\documentclass{article}
\usepackage{emulateapj}
\usepackage{apjfonts}
\usepackage[american]{babel}

\def\phn{\phantom{0}}
\def\phd{\phantom{.}}
\newcommand{\kms}{\ensuremath{\mathrm{km\ s^{-1}}}}
\newcommand{\fe}{\ensuremath{\langle\mathrm{Fe}\rangle}}
\newcommand{\femg}{\ensuremath{\langle\mathrm{Fe(Mg)}\rangle}}
\newcommand{\mgb}{\ensuremath{\mathrm{Mg}\,b}}
\newcommand{\mgone}{Mg$_1$}
\newcommand{\mgtwo}{Mg$_2$}

\newcommand{\bmv}{\ensuremath{(B-V)}}
\newcommand{\hbeta}{\ensuremath{\mathrm{H}\beta}}
\newcommand{\logt}{\ensuremath{\log t}}
\newcommand{\z}{\ensuremath{\mathrm{[Z/H]}}}
\newcommand{\feh}{\ensuremath{\mathrm{[Fe/H]}}}

\newcommand{\enh}{\ensuremath{\mathrm{[E/Fe]}}}

\newcommand{\afe}{\ensuremath{[\alpha/\mathrm{Fe}]}}
\newcommand{\reo}[1]{\ensuremath{r_{e}/#1}}

\newcommand{\XX}{\ensuremath{[X_{\mathrm{HPE}}/X_{\mathrm{LPE}}]}}
\newcommand{\roseca}{\ion{Ca}{2} H$+$H$\epsilon$/\ion{Ca}{2} K}
\newcommand{\othree}{[\ion{O}{3}]}

\submitted{Accepted for publication in the Astronomical Journal}
 
\lefthead{TRAGER ET AL.}
\righthead{STELLAR POPULATIONS OF EARLY-TYPE GALAXIES. I.}
 
\begin{document}

\title{The Stellar Population Histories of Local Early-Type Galaxies.
I. Population Parameters}

\author{S. C. Trager\altaffilmark{1}}
\affil{The Observatories of the Carnegie Institution of
Washington\\813 Santa Barbara St., Pasadena, CA 91106\\
sctrager{\@@}ociw.edu}
\authoraddr{813 Santa Barbara St., Pasadena, CA 91106; sctrager{\@@}ociw.edu}
\altaffiltext{1}{Carnegie Starr Fellow}

\author{S. M. Faber}
\affil{UCO/Lick Observatory and Board of Studies in Astronomy and
Astrophysics,\\University of California, Santa Cruz\\Santa Cruz, CA
95064\\faber{\@@}ucolick.org}
\authoraddr{University of California, Santa Cruz, CA 95064;
faber{\@@}ucolick.org}

\author{Guy Worthey}
\affil{Department of Physics and Astronomy, St.~Ambrose
University\\Davenport, IA 52803-2829\\
gworthey{\@@}saunix.sau.edu}
\authoraddr{Davenport, IA 52803-2829; gworthey{\@@}saunix.sau.edu}

\author{J. Jes\'us Gonz\'alez}
\affil{Instituto de Astronom{\'\i}a---UNAM\\Apdo Postal 70-264,
M\'exico D.F., Mexico\\jesus{\@@}astroscu.unam.mx}
\authoraddr{Apdo Postal 70-264, M\'exico D.F., Mexico;
jesus{\@@}astroscu.unam.mx}

\begin{abstract}

This paper commences a series of investigations into the stellar
populations of local elliptical galaxies as determined from their
integrated spectra.  The goal of the series is to determine the star
formation and chemical evolution histories of present-day elliptical
galaxies.  The primary galaxy sample analyzed is that of Gonz{\'a}lez
(1993, G93), which consists of 39 ellipticals drawn primarily from the
local field and nearby groups, plus the bulge of Messier 31.
Single-stellar-population (SSP) equivalent ages, metallicities, and
abundance ratios are derived from \hbeta, \mgb, and \fe\ line
strengths using an extension of the Worthey (1994) models that
incorporates non-solar line-strength ``response functions'' by Tripicco
\& Bell (1995).  These functions account for changes in the Lick/IDS
indices caused by non-solar abundance ratios, allowing us to correct
the Worthey (1994) models for the enhancements of Mg and other
$\alpha$-like elements relative to the Fe-peak elements.

SSP-equivalent ages of the G93 ellipticals are found to vary widely,
$1.5\la t\la18$ Gyr, while metallicities \z\ and enhancement ratios,
\enh\, are strongly peaked around $\langle\z\rangle=+0.26$ and
$\langle\enh\rangle=+0.20$ (in an aperture of radius \reo{8}).  The
enhancement ratios \enh\ are milder than previous estimates, owing to
the application of non-solar abundance corrections to \emph{both}
\mgb\ and \fe\ for the first time.  While \enh\ is usually $>0$, it is
not the ``E'' elements that are actually enhanced but rather the
Fe-peak elements that are depressed; this serves not only to weaken
\fe\ but also to \emph{strengthen} \mgb, accounting for the overall
generally mild enhancements.  Based on index strengths from the
Lick/IDS galaxy library (\cite{TWFBG98}), C is not depressed with Fe
but rather seems to be on a par with other elements such as Mg in the
``E'' group.  Gradients in stellar populations within galaxies are
found to be mild, with SSP-equivalent age decreasing by 25\%,
metallicity decreasing by $\langle\z\rangle=0.20$ dex, and \enh\
remaining nearly constant out to an aperture of radius \reo{2} for
nearly all systems.

Our ages have an overall zeropoint uncertainty of at least $\sim25\%$
due to uncertainties in the stellar evolution prescription, the oxygen
abundance, the effect of $\enh\neq0$ on the isochrones, and other
unknowns.  However, the \emph{relative} age rankings of stellar
populations should be largely unaffected by these errors.  In
particular, \emph{the large spread in ages appears to be real and
cannot be explained by contamination of \hbeta\ by blue stragglers or
hot horizontal branch stars, or by fill-in of \hbeta\ by emission.}
Correlations between these derived SSP-equivalent parameters and other
galaxy observables will be discussed in future papers.

\end{abstract}


\section{Introduction}

This paper is the first in a series on the stellar populations of
local field and group elliptical galaxies based on the high-quality
spectral data of Gonz{\'a}lez (1993; G93).  The present paper
concentrates on deriving improved stellar population parameters by
correcting existing population models for the effects of
\emph{non-solar abundance ratios}.  The major roadblock to population
synthesis models of elliptical galaxies is the fact that the effects
of age and metallicity are \emph{nearly degenerate} in the spectra of
old stellar populations (\cite{Faber72}, 1973; \cite{O'Connell80};
\cite{Rose85}; \cite{Renzini86}).  However, it was early noted that
certain spectral features are more sensitive to age than metallicity
(e.g., the Balmer lines [\cite{O'Connell80}; \cite{Rabin82};
\cite{BFGK84}; \cite{Rose85}], and \ion{Sr}{2} $\lambda$4077
[\cite{Rose85}]), and hope grew that such features might be able to
break the degeneracy if accurately calibrated.  (At about the same
time, several workers were also using Balmer lines to discover strong
bursts of star formation in so-called ``E+A'' or ``post-starburst''
galaxies [\cite{DG83}; \cite{CS87}; \cite{Schweizer90}], but these
applications always implicitly assumed solar metallicity.)

Our ability to decouple age and metallicity in integrated spectra has
greatly improved over the last decade, due to three developments.  In
the late 1980's, interior models of super-solar-metallicity stellar
evolution became available (e.g., \cite{VandenBerg85}, \cite{VB85},
\cite{VL87}; \cite{Padova}).  Next, the Lick/IDS stellar
absorption-line survey provided empirical polynomial fitting functions
for a set of standardized absorption-line indices as a function of
stellar temperature, gravity, and metallicity (\cite{Gorgas93};
\cite{WFGB94}).  Finally, an extensive grid of theoretical model
atmospheres and stellar flux distributions was provided by Kurucz
(1992) for stars over a wide range of temperatures and metallicities.
With these three ingredients, it finally became possible to compute
absorption-line strengths from first principles for single-burst
stellar populations (SSPs) of a given age and metallicity
(\cite{Worthey92}, 1994).

Using such models, Worthey showed that the age-metallicity degeneracy
was actually worse than suspected: a factor of \emph{two} uncertainty
in the metallicity of a galaxy mimics a factor of \emph{three}
uncertainty in its age at fixed color or metal-line strength, the
so-called ``3/2 law''.  The law implies that such commonly used
``age'' indicators as colors and metal-line strengths are by
themselves useless (although they still are widely used).  At the same
time, the Worthey models also provided a quantitative tool to break
the degeneracy (see also \cite{WO97}).  A Balmer index plotted versus
a metal line (or color) yields a two-dimensional theoretical grid; the
equivalent single-burst age and metallicity for a population can be
read off from its location in this grid.  Tests of the method on
composite stellar populations will be demonstrated in Trager et
al.~(1999; Paper II), where it is shown that these
single-stellar-population (SSP) equivalent parameters correspond
approximately to the \emph{luminosity-weighted} vector addition of
populations in the index diagrams.  A galaxy's age determined from its
integrated spectrum is thus quite sensitive to \emph{recent} star
formation, and hence to the epoch and strength of its last major
dissipative merger or accretion event.

While Worthey models validated use of the Balmer lines, they also
showed that extremely accurate Balmer data would be needed.  To our
knowledge, the line-strength data of Gonz{\'a}lez (1993) are still the
only published data on a \emph{diversified} sample of local E galaxies
that are adequate for this purpose.  Applying early Worthey models to
his data, Gonz{\'a}lez found that blue, weak-lined ellipticals in his
sample tended to have young ages, while red, strong-lined ellipticals
had older ages.  In contrast, the metallicity spread was fairly small,
less than a few tenths of a dex.  This result seemed to imply (G93,
\cite{FTGW95}) that age was the major cause of the well known
color/line-strength relation in the G93 sample, not metallicity as in
the classic picture (\cite{Baum59}; \cite{McV68}; \cite{ST69};
\cite{Faber72}, 1973).  The large age spread in G93 galaxies was later
confirmed by Trager (1997) and by Tantalo, Chiosi \& Bressan (1998a;
TCB98) using later stellar population models.

Excellent line strengths have also been measured for E and S0 galaxies
in the Fornax cluster by Kuntschner \& Davies (1998) and Kuntschner
(1998).  Fornax turns out to be the reverse of the G93 sample in
showing a larger spread in metallicity than age; the dense cluster
environment of Fornax may be the key difference.  A goal of the
present series of papers is to explore the relative importance of age
versus metallicity in driving the color and line strength relations of
ellipticals in different environments (see Paper II).

Although the data of G93 shed hope on solving the age vs. metallicity
problem, they brought another simmering problem to the fore, namely,
non-solar abundance ratios.  Enhancement of Mg relative to Fe had been
suggested by O'Connell (1976) and Peletier (1989) and shown to be
widespread in the Lick/IDS ellipticals by Worthey, Faber \& Gonz\'alez
(1992).  However, the high-quality data of G93 offered a great
improvement for hard-to-measure weak Fe lines, and, using them, Trager
(1997) showed that metallicities deduced from Mg were indeed
considerably higher than those deduced from Fe.  Other elements such
as Na, C, N, and possibly O are also probably enhanced in giant
ellipticals (\cite{Worthey98}).  Because the Worthey models are not
designed for non-solar abundance ratios, applying them to different
metal line features in elliptical spectra gives inconsistent ages and,
especially, abundances.  The progress promised by the G93 data thus
suddenly came to a full stop.

The present paper addresses the problem of non-solar abundance ratios
in a rough but hopefully satisfactory way.  On the one hand, the
general effects of non-solar ratios on evolutionary isochrones are now
beginning to be understood (\cite{SCS93}; \cite{WPM95}; \cite{SW98};
Bressan, priv.~comm.; see \cite{TCB98}).  Second, the responses of
nearly all the Lick/IDS indices to non-solar element ratios have been
modeled by Tripicco and Bell (1995; TB95).  The latter prove crucial,
and it is really these responses that open the way forward.  Using
both inputs, reasonable corrections to the W94 indices for non-solar
ratios can be estimated for the first time.  The corrected models are
used here to derive three SSP-equivalent population parameters for
each galaxy---age, mean metallicity, and mean element ``enhancement
ratio.''  Future papers will use these parameters to study stellar
populations as a function of galaxy type, determine correlations among
age, metallicity, enhancement, and other variables, and measure radial
population gradients.

Other groups (\cite{WPM95}; \cite{Greggio97}; \cite{Trager97}) have
also attempted to interpret G93 data in terms of non-solar abundance
ratios, but their approaches were more {\it ad hoc}.  Inferred
metallicities and enhancements both tend to be larger than what we
find here.  The most similar analysis so far is by Tantalo, Chiosi \&
Bressan (1998a; TCB98), building on previous work by that group
(\cite{BCT96}).  However, these authors use different response
functions from ours (and in fact do not correct the Fe index at all
for non-solar ratios).  Their results consequently differ, and a
section is devoted to comparing our results to their work
(Sec.~\ref{sec:modeldep}).

We note briefly that Balmer-line equivalent widths might be spuriously
contaminated by light from blue horizontal-branch (BHB) stars or blue
straggler stars (BSS) (e.g., \cite{BFGK84}; \cite{Lee94};
\cite{FTGW95}; \cite{Trager97}).  These possibilities are discussed in
Section~\ref{sec:hbeta_age}.  To anticipate the conclusions, we
believe that current data do not support the existence of {\it large}
numbers of BHB and BSS stars in giant elliptical galaxies, and we thus
conclude that the SSP-equivalent ages derived here for both young and
old ellipticals must be substantially correct.  Likewise, reduction of
Balmer indices by emission fill-in, though present, cannot change the
derived ages very much.  Thus, despite efforts, we have been unable to
find any explanation for the wide range of Balmer line strengths in
the G93 galaxies other than a \emph{wide range of SSP-equivalent
ages.}  This is our principal conclusion.

The outline of this paper is as follows: Section~\ref{sec:data}
presents absorption-line data for the G93 galaxies.
Section~\ref{sec:models} presents a brief description of the Worthey
(1994) models; their extension to non-solar abundance ratios using the
results of TB95; the final choice of elements for inclusion in the
enhanced element group; the method for determining the stellar
population parameters from the models; and the final population
parameters for the G93 sample.  Section~\ref{sec:results} briefly
presents the parameters for the G93, both central and global, and
their distributions.  Section~\ref{sec:disc} discusses the
assumptions, in particular the use of \hbeta\ as an age indicator, and
examines all known uncertainties in the age, metallicity, and
abundance-ratio scales and zeropoints.  Section~\ref{sec:others}
presents evidence from other absorption-line strength studies for the
presence of intermediate-age stellar populations in elliptical
galaxies; it also compares in detail our results to those of TCB98.
Two appendices discuss the effect of changing isochrones in the models
and the effect of using different prescriptions for emission and
velocity dispersion corrections to \hbeta.

\section{Data}\label{sec:data}

\subsection{The galaxy sample}

\begin{deluxetable}{lllrrrl}
\tablecaption{The Gonz{\'a}lez (1993) Sample\tablenotemark{a}\label{tbl:sample}}
\tablewidth{0pt}
\tablehead{&\colhead{Type}&\colhead{Type}&&&\colhead{$cz$}&
\colhead{Other}
\\
\colhead{Name}&\colhead{(RC3)}&\colhead{(CA)}&
\colhead{$\alpha_{2000}$}&\colhead{$\delta_{2000}$}&
\colhead{(\kms)}&\colhead{names}}
\tablecolumns{7}
\startdata
NGC 221&cE2&cE2&00:42:41.9&$+$40:51:52&$-204\pm\phn7$&M32\nl
NGC 224&SA(s)b&Sb&00:42:44.2&$+$41:16:08&$-300\pm\phn7$&M31\nl
NGC 315&E$+$:&\nodata&00:57:48.9&$+$30:21:09&$4942\pm\phn6$&\nl
NGC 507&SA(r)0&\nodata&01:23:39.8&$+$33:15:23&$4908\pm11$&\nl
NGC 547&E1&\nodata&01:26:00.7&$-$01:20:44&$5468\pm\phn6$&\nl
NGC 584&E4&S0$_1$(3,5)&01:31:20.7&$-$06:52:06&$1866\pm\phn6$&\nl
NGC 636&E3&E1&01:39:06.5&$-$07:30:46&$1860\pm\phn6$&\nl
NGC 720&E5&E5&01:53:00.4&$-$13:44:18&$1741\pm11$&\nl
NGC 821&E6?&E6&02:08:21.0&$+$10:59:44&$1730\pm\phn7$&\nl
NGC 1453&E2-3&E0&03:46:27.2&$-$03:58:09&$3886\pm\phn6$&\nl
NGC 1600&E3&E4&04:31:39.9&$-$05:05:10&$4688\pm\phn8$&\nl
NGC 1700&E4&E3&04:56:56.3&$-$04:51:52&$3895\pm\phn7$&\nl
NGC 2300&SA0&E3&07:32:22.0&$+$85:42:27&$1938\pm\phn7$&\nl
NGC 2778&E&\nodata&09:12:24.4&$+$35:01:38&$2060\pm\phn7$&\nl
NGC 3377&E5--6&E6&10:47:41.6&$+$13:59:00&$ 724\pm\phn7$&\nl
NGC 3379&E1&E0&10:47:49.5&$+$12:34:57&$ 945\pm\phn7$&M105\nl
NGC 3608&E2&E1&11:16:58.7&$+$18:08:57&$1222\pm\phn7$\nl
NGC 3818&E5&E5&11:41:57.5&$-$06:09:21&$1708\pm10$&\nl
NGC 4261&E2-3&E3&12:19:23.2&$+$05:49:31&$2238\pm\phn7$&\nl
NGC 4374&E1&E1&12:25:03.7&$+$12:53:14&$1060\pm\phn6$&M84\nl
NGC 4472&E2&E1/S0$_1$(1)&12:29:46.5&$+$07:59:48&$ 980\pm10$&M49\nl
NGC 4478&E2&E2&12:30:17.4&$+$12:19:44&$1365\pm\phn7$&\nl
NGC 4489&E&\nodata&12:30:52.2&$+$16:45:31&$ 970\pm10$&\nl
NGC 4552&E&S0$_1$(0)&12:35:39.9&$+$12:33:25&$ 364\pm\phn7$&M89\nl
NGC 4649&E2&S0$_1$(2)&12:43:39.7&$+$11:33:09&$1117\pm\phn6$&M60\nl
NGC 4697&E6&E6&12:48:35.8&$-$05:48:00&$1307\pm10$&\nl
NGC 5638&E1&E1&14:29:40.4&$+$03:14:04&$1649\pm\phn6$&\nl
NGC 5812&E0&E0&15:00:57.0&$-$07:27:19&$1929\pm\phn7$&\nl
NGC 5813&E1--2&E1&15:01:11.2&$+$01:42:08&$1954\pm\phn7$&\nl
NGC 5831&E3&E1&15:04:07.2&$+$01:13:15&$1655\pm\phn5$&\nl
NGC 5846&E0-1&E4/S0$_1$(4)&15:06:29.3&$+$01:36:21&$1714\pm\phn5$&\nl
NGC 6127&E&\nodata&16:19:11.9&$+$57:59:03&$4700\pm10$&NGC 6125/6128\nl
NGC 6702&E:&\nodata&18:46:57.6&$+$45:42:20&$4728\pm\phn5$&\nl
NGC 6703&SA0$-$&\nodata&18:47:18.9&$+$45:33:02&$2403\pm\phn7$&\nl
NGC 7052&E&\nodata&21:18:32.9&$+$26:26:48&$4672\pm\phn8$&B2 2116$+$26\nl
NGC 7454&E4&\nodata&23:01:06.6&$+$16:23:24&$2051\pm\phn7$&\nl
NGC 7562&E2--3&\nodata&23:15:57.4&$+$06:41:16&$3608\pm\phn5$&\nl
NGC 7619&E&E3&23:20:14.7&$+$08:12:23&$3762\pm\phn5$&\nl
NGC 7626&E pec:&E1&23:20:42.4&$+$08:13:02&$3405\pm\phn4$&\nl
NGC 7785&E5-6&S0$_1$(5)/E5&23:55:19.1&$+$05:54:53&$3808\pm\phn5$&\nl
\enddata
\tablenotetext{a}{NGC 4278 has been removed because of its strong
emission lines.}
\tablecomments{Col.~(1): Galaxy name. Col.~(2): Morphological type
from RC3.  Col.~(3): Morphological type from Carnegie Atlas (Sandage
\& Bedke 1994) or RSA (Sandage \& Tammann 1987).  Cols.~(4)--(5):
Position (J2000.0) from NED.  Col.~(6): Heliocentric radial velocity
from Gonz{\'a}lez (1993).  Col.~(7): Other galaxy names.}
\end{deluxetable}

The G93 galaxy sample was not selected according to quantitative
criteria but was rather chosen with the aim of covering
\emph{relatively uniformly} the full range of color, line strength,
and velocity dispersion shown by local elliptical galaxies.  As such,
it contains more dim, blue, weak-lined, low-dispersion galaxies than
would be found in a magnitude-limited sample.  In that sense the G93
sample may more closely resemble a \emph{volume}-limited sample, but
this has not been established quantitatively.

The original sample in G93 consisted of 41 galaxies, of which 40 are
included here.  NGC 4278 has been discarded because of its strong
emission.  Table~\ref{tbl:sample} presents morphologies, positions,
and heliocentric redshifts.  All galaxies are classified as elliptical
(or compact elliptical) in the RC3 (\cite{RC3}), the RSA (\cite{RSA}),
or the Carnegie Atlas (\cite{CA}) except for NGC 507 and NGC 6703,
both classified as SA0 in the RC3 but not cataloged in the RSA or the
Carnegie Atlas.  NGC 224 (the bulge of M~31) is also included.

The environmental distribution of the G93 sample bears comment.  Group
assignments and approximate group richnesses may be found for nearly
all galaxies in Faber et al. (1989).  Most of the G93 galaxies are in
poor groups, a few are quite isolated (there are no other galaxies in
the RC3 within 1 Mpc projected distance and $\pm2000\;\kms$ of NGC
6702, for example [Colbert, Mulchaey \& Zabludoff, in prep.]), and six
are members of the Virgo cluster.  Only one is in a rich cluster (NGC
547, in Abell 194).  We therefore refer to the galaxies in this sample
as local ``field'' ellipticals, given the low-density environments of
most of them.  Environmental effects are discussed in more detail in
Paper II.

\subsection{G93 indices: calibrations and corrections}\label{sec:index}

The Lick/IDS indices were introduced by Burstein et al.\ (1984) to
measure prominent absorption features in the spectra of old stellar
populations in the 4100--6300 \AA\ region.  A large and homogeneous
database of stellar and galaxy spectra was assembled (\cite{WFGB94};
\cite{TWFBG98}, hereafter TWFBG98) with the Image Dissector Scanner at
Lick Observatory (IDS; \cite{RW72}).  A description of the Lick/IDS
system and its application to stellar and galaxy spectra is given in
those papers.

Gonz\'alez (1993) measured Lick/IDS indices with a different
spectrograph setup, at higher dispersion, and over a restricted
spectral range (4700--5500 \AA).  The four best indices in his
wavelength interval are \hbeta, \mgb, Fe5270, and Fe5335, which we use
in this paper.  The bandpasses of these four indices are given in
Table~\ref{tbl:index}, and the precise index definitions are given in
G93, Worthey et al.\ (1994), and TWFBG98.

\begin{deluxetable}{llrrcl}
\tablefontsize{\normalsize}
\tablewidth{0pt}
\tablecaption{Lick/IDS Indices Used in This Study\label{tbl:index}}
\tablehead{\colhead{$j$}&\colhead{Name}&\colhead{Index Bandpass}&
\colhead{Pseudocontinua}&\colhead{Units}&
\colhead{Measures\tablenotemark{a}}}
\startdata
09&H$\beta$&4847.875--4876.625&4827.875--4847.875&\AA&H$\beta$,(Mg)\nl
&&&4876.625--4891.625&\nl
13&Mg$b$&5160.125--5192.625&5142.625--5161.375&\AA&Mg,(C),(Cr),(Fe)\nl
&&&5191.375--5206.375&\nl
14&Fe5270&5245.650--5285.650&5233.150--5248.150&\AA&Fe,C,(Mg)\nl
&&&5285.650--5318.150&\nl
15&Fe5335&5312.125--5352.125&5304.625--5315.875&\AA&Fe,(C),(Mg),Cr\nl
&&&5353.375--5363.375&\nl
\enddata
\tablenotetext{a}{Dominant species; species in parentheses control
index in a negative sense (index weakens as abundance grows).  See
Tripicco \& Bell (1995) and Worthey (1998).}
\end{deluxetable}

We use a combined ``iron'' index, \fe, in this work, which has smaller
errors than either Fe index separately and is defined as follows:
\begin{equation}
\fe\equiv\frac{\rm Fe5270 + Fe5335}{2}.
\end{equation}
It has the convenient property of being sensitive primarily to [Fe/H]
(see Sec.~\ref{sec:tb95}).  Although \mgtwo\ has also become a
standard ``metallicity'' indicator for the integrated spectra of
galaxies, we do not use it to determine stellar population parameters.
G93 was unable to transform his observations of this broad index (or
of \mgone) accurately onto the Lick/IDS system due to chromatic focus
variations in his spectrograph, coupled with the steep light gradient
in the central regions of most ellipticals (Fisher et al.\ 1995
avoided \mgtwo\ for the same reason).  We prefer to use the narrower
index \mgb, which is not affected by this problem.

\subsubsection{Velocity-dispersion corrections}
\label{sec:sigmacorr}

The observed spectrum of a galaxy is a convolution of the integrated
spectrum of its stellar population with the line-of-sight velocity
distribution function of its stars.  Indices measured for broad-line
galaxies are therefore too weak compared to unbroadened standard
stars.  TWFBG98 statistically corrected the Lick/IDS indices for this
effect in the following way: individual stellar spectra of a variety
of spectral types (plus M~32) were convolved with Gaussian broadening
functions of increasing widths and their indices were remeasured.  A
smooth multiplicative correction as a function of velocity dispersion
was determined separately for each index and applied to the galaxy
data.

G93 used a more sophisticated technique, taking advantage of the
higher resolution and signal-to-noise of his data.  His stellar
library was used to synthesize a summed stellar template representing
a best fit to the the spectrum of each galaxy.  Indices were measured
from the unbroadened template and again from the broadened template,
generating a velocity dispersion correction for each galaxy that was
tuned to its spectral type.  For \mgb, Fe5270, and Fe5335, the mean
multiplicative corrections of G93 are very similar to those of TWFBG98
(compare his Figure 4.1 with Figure 3 of TWFBG98).  However, for
\hbeta, the correction of G93 is flat or even \emph{negative}, whereas
the correction of TWFBG98 is always positive and reaches the value
1.07 at $\sigma=300\ \kms$.  Use of the TWFBG98 correction increases
\hbeta\ over G93 and leads to slightly younger ages.  In what follows,
we use the G93 correction to remain consistent with his published data
but explore the effects of the TWFBG98 correction in
Appendix~\ref{app:emsigcorr}.  The data marginally appear to favor
TWFBG98, but the differences are not large.

\subsubsection{Emission corrections}\label{sec:emission}

G93 noted that \othree\ $\lambda\lambda4959,5007$ are clearly
detectable in about half of the nuclei in his sample and that most of
these galaxies also have detectable \hbeta\ emission (see his Figure
4.10).  For galaxies in his sample with strong emission, \hbeta\ is
fairly tightly correlated with \othree\ such that EW($\hbeta_{\rm
em}$)/EW(\othree$\lambda5007$) $\;\sim0.7$.  A statistical correction
of
\begin{equation}
\Delta\hbeta=0.7\;\mathrm{EW}([\mathrm{O\;III}]\lambda5007)
\label{eq:hbcorr}
\end{equation}
was therefore added to \hbeta\ to correct for this residual emission.

We have examined the accuracy of this correction by studying
\hbeta/\othree\ among the G93 galaxies, supplemented by additional
early-type galaxies from the emission-line catalog of Ho, Filipenko \&
Sargent (1997).  The sample was restricted to include only normal,
non-AGN Hubble types E through S0$-$, and to well measured objects
with $\mathrm{EW(H\alpha)} > 1.0$ \AA.  For 27 galaxies meeting these
criteria, \hbeta/\othree\ varies from 0.33 to 1.25, with a median
value of 0.60. This suggests that a better correction coefficient in
Equation~\ref{eq:hbcorr} might be 0.6 rather than 0.7, and thus that
the average galaxy in G93 is slightly overcorrected.  For a median
\othree\ strength through the G93 \reo{8}\ aperture of $0.17$ \AA, the
error would be about $0.02$ \AA, or 3\% in age.  This systematic error
for a typical galaxy is negligible compared to other sources of error
in the ages (see Table~\ref{tbl:syserr}).  Random errors due to
scatter in the ratio are about three times larger but are still small.

Carrasco et al. (1996) report no correlation between \hbeta\ and
\othree\ emission in their sample of early-type galaxies, but give no
data.  Their claim is explored in Appendix B, which repeats our
calculations but with no \hbeta\ correction.  The ages of a few
strong-\othree\ galaxies are increased, as expected, but the broad
conclusions of this work are unaffected.

No correction for [\ion{N}{2}] emission has been made to \mgb,
although this has been suggested as a sometimes significant
contributor to this index (by increasing the flux in the red sideband;
\cite{GE96}).  Only NGC 315 and NGC 1453 would be affected (see G93).

\begin{deluxetable}{lrrrrrrrrrrrrrr}
\tablenum{3a}
\tablewidth{0pt}
\tablecaption{Fully Corrected Index Strengths in the Central $r_e/8$
Aperture\tablenotemark{a}\label{tbl:re8i}}
\tablehead{\colhead{Name}&\colhead{[\ion{O}{3}]}&\colhead{$\sigma$}&
\colhead{H$\beta$}&\colhead{$\sigma$}&\colhead{Mg $b$}&\colhead{$\sigma$}&
\colhead{Fe52}&\colhead{$\sigma$}&\colhead{Fe53}&\colhead{$\sigma$}&
\colhead{\fe}&\colhead{$\sigma$}}
\startdata
NGC 221&$0.11$&0.05&2.31&0.05&2.96&0.03&2.88&0.04&2.61&0.04&2.75&0.03\nl
NGC 224&$0.11$&0.07&1.67&0.07&4.85&0.05&2.88&0.04&2.61&0.04&3.09&0.04\nl
NGC 315&$0.45$&0.06&1.74&0.06&4.84&0.05&2.92&0.06&2.85&0.07&2.88&0.05\nl
NGC 507&$0.05$&0.09&1.73&0.09&4.52&0.11&2.95&0.12&2.60&0.15&2.78&0.09\nl
NGC 547&$0.39$&0.06&1.58&0.07&5.02&0.05&2.97&0.07&2.66&0.08&2.81&0.05\nl
NGC 584&$0.27$&0.05&2.08&0.05&4.33&0.04&3.03&0.04&2.77&0.04&2.90&0.03\nl
NGC 636&$0.07$&0.06&1.89&0.04&4.20&0.04&3.19&0.05&2.87&0.05&3.03&0.04\nl
NGC 720&$0.14$&0.09&1.77&0.12&5.17&0.11&2.94&0.12&2.80&0.14&2.87&0.09\nl
NGC 821&$0.03$&0.05&1.66&0.04&4.53&0.04&3.08&0.05&2.81&0.05&2.94&0.04\nl
NGC 1453&$0.89$&0.06&1.60&0.06&4.95&0.05&2.96&0.06&2.99&0.07&2.98&0.05\nl
NGC 1600&$0.15$&0.06&1.55&0.07&5.13&0.06&3.01&0.07&3.10&0.09&3.05&0.06\nl
NGC 1700&$0.22$&0.05&2.11&0.05&4.15&0.04&3.17&0.05&2.83&0.05&3.00&0.04\nl
NGC 2300&$0.08$&0.06&1.68&0.06&4.98&0.05&3.04&0.06&2.89&0.07&2.97&0.05\nl
NGC 2778&$0.69$&0.07&1.77&0.08&4.70&0.06&3.01&0.07&2.69&0.08&2.85&0.05\nl
NGC 3377&$0.39$&0.05&2.09&0.05&3.99&0.03&2.77&0.04&2.44&0.04&2.61&0.03\nl
NGC 3379&$0.18$&0.05&1.62&0.05&4.78&0.03&2.98&0.04&2.73&0.04&2.85&0.03\nl
NGC 3608&$0.18$&0.06&1.69&0.06&4.61&0.04&3.13&0.05&2.75&0.06&2.94&0.04\nl
NGC 3818&$0.33$&0.08&1.71&0.08&4.88&0.07&3.09&0.08&2.85&0.08&2.97&0.06\nl
NGC 4261&$0.21$&0.06&1.34&0.06&5.11&0.04&3.14&0.05&2.88&0.06&3.01&0.04\nl
NGC 4374&$0.37$&0.05&1.51&0.04&4.78&0.03&2.94&0.04&2.69&0.04&2.82&0.03\nl
NGC 4472&$0.01$&0.08&1.62&0.06&4.85&0.06&2.97&0.07&2.84&0.08&2.90&0.05\nl
NGC 4478&$0.06$&0.06&1.84&0.06&4.33&0.05&3.03&0.06&2.84&0.06&2.93&0.04\nl
NGC 4489&$0.11$&0.08&2.39&0.07&3.21&0.06&2.89&0.07&2.44&0.07&2.67&0.05\nl
NGC 4552&$0.25$&0.05&1.47&0.05&5.15&0.03&3.02&0.04&2.95&0.04&2.98&0.03\nl
NGC 4649&$0.09$&0.05&1.40&0.05&5.33&0.04&3.01&0.04&3.01&0.05&3.01&0.03\nl
NGC 4697&$0.10$&0.07&1.75&0.07&4.08&0.05&2.97&0.06&2.57&0.06&2.77&0.04\nl
NGC 5638&$0.00$&0.06&1.65&0.04&4.64&0.04&3.02&0.05&2.66&0.05&2.84&0.04\nl
NGC 5812&$0.05$&0.06&1.70&0.04&4.81&0.04&3.09&0.05&3.02&0.06&3.06&0.04\nl
NGC 5813&$0.27$&0.06&1.42&0.07&4.65&0.05&2.83&0.06&2.52&0.07&2.67&0.04\nl
NGC 5831&$0.18$&0.05&2.00&0.05&4.38&0.04&3.17&0.04&2.92&0.04&3.05&0.03\nl
NGC 5846&$0.39$&0.08&1.45&0.07&4.93&0.05&2.95&0.06&2.77&0.06&2.86&0.04\nl
NGC 6127&$0.04$&0.08&1.50&0.05&4.96&0.06&2.90&0.07&2.79&0.08&2.85&0.06\nl
NGC 6702&$0.40$&0.06&2.46&0.06&3.80&0.04&3.02&0.05&2.97&0.06&2.99&0.04\nl
NGC 6703&$0.36$&0.05&1.88&0.06&4.30&0.04&3.06&0.05&2.79&0.05&2.92&0.04\nl
NGC 7052&$0.43$&0.06&1.48&0.07&5.02&0.06&2.89&0.07&2.78&0.08&2.83&0.05\nl
NGC 7454&$0.11$&0.06&2.15&0.06&3.27&0.05&2.68&0.06&2.27&0.06&2.47&0.04\nl
NGC 7562&$0.09$&0.05&1.69&0.05&4.54&0.04&3.08&0.05&2.65&0.05&2.87&0.03\nl
NGC 7619&$-0.02$&0.05&1.36&0.04&5.06&0.04&3.03&0.05&3.08&0.06&3.06&0.04\nl
NGC 7626&$0.11$&0.05&1.46&0.05&5.05&0.04&2.85&0.05&2.80&0.05&2.83&0.03\nl
NGC 7785&$0.15$&0.06&1.63&0.06&4.60&0.04&2.88&0.05&2.94&0.06&2.91&0.04\nl
\enddata
\tablenotetext{a}{From Gonz\'alez (1993), Table 4.7.}
\end{deluxetable}

\begin{deluxetable}{lrrrrrrrrrrrrrr}
\tablenum{3b}
\tablewidth{0pt}
\tablecaption{Fully Corrected Index Strengths in the $r_e/2$
Aperture\tablenotemark{a}\label{tbl:re2i}}
\tablehead{\colhead{Name}&\colhead{[\ion{O}{3}]}&\colhead{$\sigma$}&
\colhead{H$\beta$}&\colhead{$\sigma$}&\colhead{Mg $b$}&\colhead{$\sigma$}&
\colhead{Fe52}&\colhead{$\sigma$}&\colhead{Fe53}&\colhead{$\sigma$}&
\colhead{\fe}&\colhead{$\sigma$}}
\startdata
NGC 221&$0.14$&0.07&2.15&0.07&2.96&0.07&2.79&0.06&2.47&0.06&2.63&0.04\nl
NGC 315&$0.25$&0.07&1.80&0.08&4.52&0.09&2.62&0.09&2.72&0.11&2.67&0.07\nl
NGC 507&$0.27$&0.13&2.06&0.17&4.69&0.17&2.72&0.17&2.18&0.19&2.45&0.13\nl
NGC 547&$0.19$&0.08&1.42&0.10&4.80&0.09&2.69&0.09&2.60&0.11&2.65&0.07\nl
NGC 584&$0.19$&0.07&2.06&0.07&4.13&0.07&2.77&0.06&2.55&0.06&2.66&0.04\nl
NGC 636&$0.07$&0.07&1.87&0.06&3.98&0.08&2.93&0.07&2.56&0.07&2.75&0.05\nl
NGC 720&$0.32$&0.11&2.28&0.16&4.98&0.16&2.78&0.16&2.92&0.18&2.85&0.12\nl
NGC 821&$0.04$&0.07&1.82&0.06&4.11&0.08&2.83&0.08&2.68&0.08&2.75&0.06\nl
NGC 1453&$0.78$&0.07&1.69&0.09&4.43&0.08&2.84&0.08&2.81&0.09&2.83&0.06\nl
NGC 1600&$0.15$&0.08&1.74&0.09&5.21&0.10&3.03&0.10&3.07&0.12&3.05&0.08\nl
NGC 1700&$0.20$&0.07&2.11&0.07&3.90&0.08&2.94&0.07&2.71&0.07&2.83&0.05\nl
NGC 2300&$0.05$&0.07&1.63&0.06&4.70&0.09&2.78&0.08&2.72&0.09&2.75&0.06\nl
NGC 2778&$0.60$&0.08&1.56&0.10&4.44&0.09&2.88&0.09&2.40&0.10&2.64&0.07\nl
NGC 3377&$0.50$&0.07&2.13&0.07&3.46&0.07&2.51&0.06&2.06&0.06&2.29&0.04\nl
NGC 3379&$0.11$&0.07&1.59&0.06&4.44&0.07&2.80&0.06&2.55&0.06&2.67&0.04\nl
NGC 3608&$0.11$&0.07&1.73&0.09&4.04&0.08&3.09&0.07&2.61&0.08&2.85&0.05\nl
NGC 3818&$0.42$&0.10&1.81&0.11&4.19&0.12&2.73&0.11&2.54&0.11&2.63&0.08\nl
NGC 4261&$-0.03$&0.07&1.30&0.06&4.75&0.08&3.05&0.07&2.53&0.08&2.79&0.05\nl
NGC 4374&$0.22$&0.07&1.56&0.06&4.50&0.07&2.78&0.06&2.57&0.06&2.67&0.04\nl
NGC 4472&$0.03$&0.10&1.67&0.09&4.60&0.11&2.83&0.10&2.80&0.11&2.81&0.07\nl
NGC 4478&$0.04$&0.07&1.73&0.06&4.17&0.08&2.74&0.07&2.52&0.07&2.63&0.05\nl
NGC 4489&$0.03$&0.10&2.27&0.08&2.83&0.12&2.77&0.11&2.14&0.11&2.46&0.08\nl
NGC 4552&$0.15$&0.07&1.52&0.06&4.80&0.07&2.83&0.06&2.70&0.06&2.77&0.04\nl
NGC 4649&$0.15$&0.07&1.38&0.07&5.13&0.07&2.62&0.07&2.75&0.07&2.69&0.05\nl
NGC 4697&$0.01$&0.10&1.66&0.07&3.62&0.10&2.55&0.09&2.28&0.09&2.42&0.06\nl
NGC 5638&$-0.02$&0.07&1.68&0.06&4.12&0.08&2.79&0.07&2.50&0.07&2.65&0.05\nl
NGC 5812&$0.08$&0.07&1.71&0.05&4.54&0.08&2.96&0.07&2.82&0.07&2.89&0.05\nl
NGC 5813&$0.25$&0.08&1.23&0.10&4.26&0.09&2.90&0.09&2.43&0.10&2.67&0.07\nl
NGC 5831&$0.18$&0.07&2.03&0.07&3.88&0.07&2.89&0.06&2.55&0.06&2.72&0.04\nl
NGC 5846&$0.11$&0.10&1.27&0.10&4.55&0.11&2.63&0.10&2.67&0.10&2.65&0.07\nl
NGC 6127&$0.05$&0.10&1.50&0.07&4.67&0.11&2.76&0.10&2.55&0.11&2.65&0.07\nl
NGC 6702&$0.46$&0.07&2.49&0.09&3.71&0.08&2.91&0.08&2.84&0.09&2.88&0.06\nl
NGC 6703&$0.26$&0.07&1.83&0.07&4.00&0.08&2.82&0.07&2.53&0.07&2.67&0.05\nl
NGC 7052&$0.24$&0.08&1.77&0.09&4.66&0.09&2.78&0.09&2.70&0.10&2.74&0.07\nl
NGC 7454&$0.05$&0.07&2.08&0.06&2.88&0.09&2.39&0.08&2.10&0.08&2.25&0.06\nl
NGC 7562&$0.13$&0.07&1.72&0.07&4.42&0.08&2.92&0.07&2.54&0.07&2.73&0.05\nl
NGC 7619&$-0.03$&0.07&1.47&0.05&4.70&0.08&2.78&0.07&2.84&0.08&2.81&0.05\nl
NGC 7626&$-0.03$&0.07&1.44&0.06&4.64&0.08&2.70&0.07&2.53&0.08&2.62&0.05\nl
NGC 7785&$0.05$&0.07&1.52&0.06&4.30&0.08&2.81&0.07&2.82&0.08&2.81&0.05\nl
\enddata
\tablenotetext{a}{From Gonz\'alez (1993), Table 6.1.}
\end{deluxetable}

Table~\ref{tbl:re8i} presents final corrected index strengths,
velocity dispersion corrections, and emission corrections for
measurements through a central \reo{8}\ aperture; Table~\ref{tbl:re2i}
presents similar data for a global \reo{2}\ aperture.  All values are
taken directly from G93.  The aperture index strengths are weighted
averages of the major and minor axis profile data, computed so as to
mimic what would be observed through the indicated circular aperture
(see G93 for details).

\section{SSP-equivalent stellar population parameters}\label{sec:models}

\subsection{Method}

\subsubsection{Solar-abundance ratio models of Worthey (1994)}\label{sec:w94}

SSP-equivalent population parameters have been derived by matching
observed line strengths of \mgb, \fe, and \hbeta\ to updated
single-burst stellar population (SSP) models of W94 (available at
http://astro.sau.edu/$\sim$worthey/; ``Padova" isochrones by Bertelli
et al.\ (1994) are explored in Appendix~\ref{app:padova}).  The models
of W94 depend on two adjustable parameters---metallicity and
single-burst age---and one fixed parameter, the initial mass function
exponent (IMF), here chosen to have the Salpeter value.  For reasons
stated below, we believe that the basic models of W94 have essentially
solar abundance ratios; we will presently adjust these models to allow
for non-solar ratios and, in the process, derive a third adjustable
parameter, the non-solar enhancement ratio, \enh.  The W94 models are
reviewed briefly here, and the reader is referred to Worthey (1994)
for more details.

The models incorporate three ingredients: stellar evolutionary
isochrones, a stellar SED library, and absorption-line strengths.
From the bottom of the main sequence to the base of the red-giant
branch (RGB), the models use the isochrones of VandenBerg and
collaborators (\cite{VandenBerg85}; \cite{VB85}; \cite{VL87}).  These
are mated to red giant branches from the Revised Yale Isochrones
(\cite{RYI}) by shifting the latter in $\Delta\log L$ and $\Delta\log
T_e$ to match at the base of the RGB.  Extrapolations are made to
cover a wide range of ($Z$, $Y$, age) assuming that $Z_{\odot}=0.0169$
and $Y=0.228+2.7Z$.

The SED library was constructed using the model atmospheres and SEDs
of Kurucz (1992) for stars hotter than 3750 K, and model SEDs of
Bessel et al.\ (1989, 1991) and observed SEDs from Gunn \& Stryker
(1983) for cooler M giants.\footnote{There is a systematic color
offset in the Kurucz (1992) models when compared with the empirical
colors of Johnson (1966), the Kurucz (1992) models being too red by
$0.06$ mag in \bmv\ (but not in other colors; W94).  All model \bmv\
colors in this series are corrected for this offset.}

Polynomial fitting functions from Worthey et al.\ (1994) for the
Lick/IDS indices are used as the basis of the model absorption-line
strengths.  Metal-rich stars in the Lick/IDS library are a random
sample of metal-rich stars in the solar neighborhood; since evidence
suggests that such stars have essentially solar ratios of O, Mg, Na,
and other key elements relative to Fe (\cite{EAGLNT93}), we assume
that the line-strengths produced by the metal-rich models of W94
reflect \emph{solar-abundance ratios}.

To construct a model of a given age and metallicity, the appropriate
stellar isochrone is first selected.  Each star on the isochrone is
assigned an SED from the flux library and a set of absorption-line
strengths from the Lick/IDS fitting functions.  Final model outputs
are the integrated fluxed SED (from which colors and magnitudes can be
derived) and absorption-line strengths on the Lick/IDS system.

\begin{figure*}
\plotone{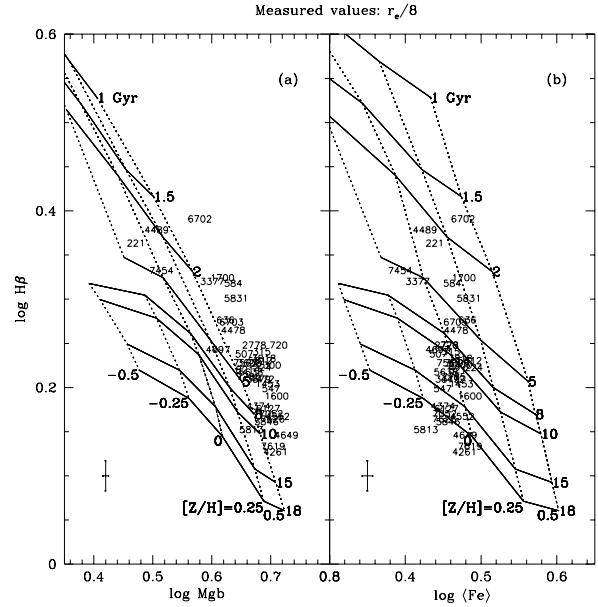}
\caption{Line strengths of early-type galaxies in the Gonz{\'a}lez
(1993) sample in the central \reo{8}\ aperture.  Model grids from
Worthey (1994) are superimposed: solid lines are contours of constant
age (from top, 1, 1.5, 2, 5, 8, 10, 15, 18 Gyr), and dotted lines are
contours of constant \z\ (from left, $\z=-0.5$, $-0.25$, 0., 0.25,
0.5, except at ages younger than 8 Gyr, where from left $\z=-0.225$,
0., 0.25, 0.5). (a) \mgb\ versus \hbeta. (b) \fe\ versus \hbeta.
Differences in the ages and metallicities inferred from these two
diagrams result from the non-solar abundance ratios of giant
elliptical galaxies.\label{fig:re8m}}
\end{figure*}

\begin{figure*}
\plotone{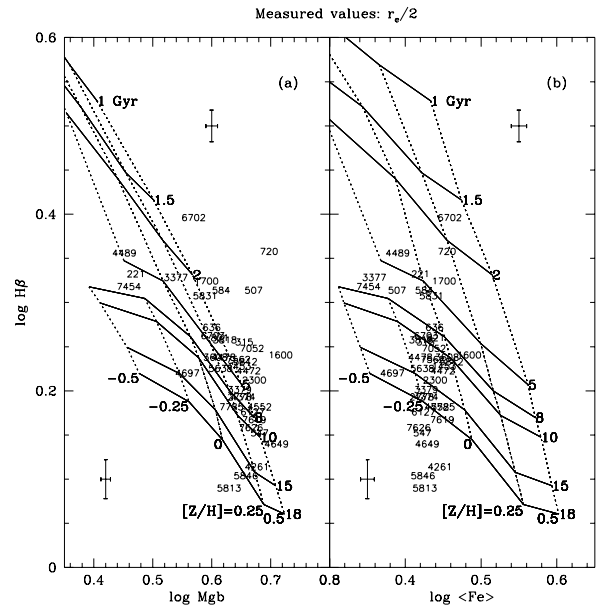}
\caption{Line strengths of early-type galaxies in the Gonz{\'a}lez
(1993) sample through the global \reo{2}\ aperture.  The remainder of
the figure is as in Figure~\ref{fig:re8m}.  The global loci here are
shifted horizontally to the left of the central loci in
Figure~\ref{fig:re8m}, reflecting the fact that the outer parts are
both older and more metal-poor.\label{fig:re2m}}
\end{figure*}

The ability of the W94 models to break the age-metallicity degeneracy
is illustrated in Figures~\ref{fig:re8m} and \ref{fig:re2m}, which
plot \mgb\ and \fe\ versus \hbeta\ for the G93 galaxies.  Model grids
from W94 are overplotted.  Both line strength pairs break the
degeneracy, but \fe-\hbeta\ does so more than \mgb-\hbeta\ because
\fe\ is less temperature sensitive than \mgb.  Metallicities inferred
from \mgb\ are clearly higher than from \fe, reflecting the probable
element enhancement in [Mg/Fe].  Low-\hbeta\ galaxies tend to fall off
the grid to high ages in the \fe-\hbeta\ diagram, especially through
the $r_e/2$ aperture (Figure 2b).  Most of this effect is removed when
\fe\ is corrected for depressed [Fe/H] (see below), and any small
remainder can be attributed to use of the G93 \hbeta\ velocity
corrections (instead of TWFBG98) and, in a few galaxies, to possible
residual, uncorrected \hbeta\ emission (see
Appendix~\ref{app:emsigcorr}).

\subsubsection{Non-solar abundance ratio models}\label{sec:tb95}

Adjusting the W94 models for non-solar ratios involves two steps.
First, one must compute new evolutionary tracks in a fully
self-consistent manner using new interior opacities, reaction rates,
and atmospheric boundary conditions that faithfully reflect the
altered compositions.  Second, one must compute new absorption-line
indices.  Part one is less developed in the literature but also proves
to be less important; we discuss it below.  Part two, the indices,
could in principle be handled by observing populations of stars with
known non-solar ratios and deriving empirical fitting functions for
them.  For example, Borges et al.\ (1995) derived a fitting function
for \mgtwo\ versus [Mg/Fe] using local dwarf and subgiant stars (this
was the function adopted by TCB98 for their population models); and
Weiss, Peletier \& Mateucci (1995) attempted to correct \mgtwo\ and
\fe\ using Galactic Bulge stars studied by Rich (1988).

However, it is hard to identify groups of stars with exactly the same
(known) enhancements, and it is even more difficult to \emph{vary} the
pattern of element abundance enhancements in a controlled way using
real stars.  For these reasons, a theoretical approach is recommended,
and we have chosen to utilize the computations of Tripicco \& Bell
(1995), who re-computed all of the Lick/IDS spectral indices from a
grid of theoretical stellar SEDs and atmospheres with varying
abundance ratios.  For three sample locations on an old stellar
isochrone, TB95 tabulate the response of each Lick/IDS index to
separate enhancements $\mathrm{[X/H]}=+0.3$ dex for the elements
$\mathrm{X=C, N, O, Mg, Fe, Ca, Na, Si, Cr}$ and $\mathrm{Ti}$.  These
response functions are the basis for our corrections to the indices
for non-solar abundance ratios.  Note that, because we use the
response functions \emph{differentially}, we are insensitive to any
zeropoint uncertainties that the TB95 indices may have (which are in
any case known to be small, as TB95 showed by comparing to real
stars).

\begin{deluxetable}{ccccccccccc}
\setcounter{table}{3}
\tablecaption{Element Mass Fractions in Non-solar Abundance Ratio
Models\tablenotemark{a}\label{tbl:tbmodels}}
\tablewidth{0pt}
\tablehead{\colhead{Model}&\colhead{}&\colhead{$A$}&\colhead{$A/(1-A)$}&
\colhead{$\Delta\enh$}&\colhead{$\Delta\mathrm{[E/H]}$}&
\colhead{$f(\mathrm{C})$}&\colhead{$f(\mathrm{O})$}&
\colhead{$f(\mathrm{Fe})$}&\colhead{$f(\mathrm{E})$}&
\colhead{\XX}}
\startdata
1&C$\circ$O$+$&0.914&$10.6\phd$&0.323&0.023&0.173&0.509&0.041&0.267&0.061\nl
2&C$\circ$O$\circ$&0.774&$\phd3.42$&0.365&0.065&0.173&0.482&0.041&0.294&0.020\nl
3&C$-$O$+$&0.759&$\phd3.15$&0.369&0.069&0.087&0.565&0.041&0.297&0.001\nl
4&C$+$O$+$&0.929&$11.7\phd$&0.319&0.019&0.181&0.504&0.041&0.265&0.067\nl
Solar&\nodata&\nodata&\nodata&\nodata&\nodata&0.172&0.482&0.082&0.253&0.000\nl
\enddata
\tablenotetext{a}{At fixed $\Delta\feh=-0.3$ dex and $\z=0$.}
\tablecomments{Cols.~(1)--(2): Enhancement model.  All models have
enhanced N, Ne, Na, Mg, Si, and S.  Ca and the Fe-peak elements (Cr,
Mn, Fe, Co, Ni, Cu, Zn) are depressed.  All other elements are fixed
at their solar abundances except that model 1 has C fixed at its solar
abundance and O enhanced (C$\circ$O$+$); model 2 has both C and O
fixed at their solar abundances (C$\circ$O$\circ$); model 3 has C
depressed like the Fe group and O enhanced (C$-$O$+$; this is the
enhancement pattern favored by TCB98 and by Salaris \& Weiss 1998);
and model 4 has both C and O enhanced (C$+$O$+$).  ``Solar'' is taken
from the solar photospheric abundances of Grevesse et al.\ (1996).
Cols.~(3)--(4): $A$ is the response of the ``enhanced'' elements to
changes in \feh\ at fixed \z: $\Delta\feh = -A\Delta\enh =
-A/(1-A)\Delta\mathrm{[E/H]}$.  Col.~(5): $\Delta\enh$ at $\Delta{\rm
[Fe/H]}=-0.3$ dex.  Col.~(6): $\Delta\mathrm{[E/H]}$ at $\Delta{\rm
[Fe/H]}=-0.3$ dex.  Col.~(7): Mass fraction of metals in C.  Col.~(8):
Mass fraction of metals in O.  Col.~(9): Mass fraction of metals in Fe
peak elements.  Col.~(10): Mass fraction of metals in enhanced
elements (except for C and O).  Note that Cols.~(7)--(10) do not add
up precisely to 1, as elements fixed at their solar abundances (other
than C and O) are not included.  Col.~(11): The ratio
$\XX\equiv[X_{\rm C} + X_{\rm N} + X_{\rm O} + X_{\rm Ne} / X_{\rm Mg}
+ X_{\rm Si} + X_{\rm S} + X_{\rm Ca} + X_{\rm Fe}]$ (Salaris, Chieffi
\& Straniero 1993); see text.}
\end{deluxetable}

Following previous practice, we adopt the convention that a certain
group of elements is ``enhanced'' in elliptical galaxies (more is said
on this below).  Precisely which elements are enhanced, and by how
much, is poorly known.  From an intercomparison of absorption-line
strengths in the Lick/IDS galaxy sample (TWFBG98), Worthey (1998)
suggested that Mg, Na, and N are enhanced in giant ellipticals but
that Ca tracks Fe (cf.~\cite{O'Connell76}; \cite{Vazdekis96}).
Comparing to additional galaxy data from TWFBG98 below, we suggest
that C also belongs to the enhanced group.  Unfortunately, the
Lick/IDS system has no indices that are capable of directly probing
oxygen in elliptical galaxies (\cite{Worthey98}).  Oxygen is important
because it dominates \z\ on account of its high masss fraction.

Because O (and perhaps C) are uncertain, we have considered four
models for the enhancement pattern in elliptical galaxies, as
described in Table~\ref{tbl:tbmodels}.  In each model there are three
groups of elements: enhanced, depressed, and fixed.  The assignment of
elements to the three groups is always the same except for C and O,
whose assignments vary.  Elements in the fixed group have their solar
(photospheric) abundances (\cite{GNS96}), while elements in each of
the enhanced and depressed groups are all scaled up or down by the
same factor.  After the amount of enhancement is chosen and C and O
are assigned to their proper groups, the depression of the depressed
elements is calculated so as to preserve \emph{constant} \z.

In the present work, we generally take the enhanced group to include
the abundant elements that are nucleosynthetically related to Mg,
several of which are actually seen to be overabundant in giant
elliptical galaxies (\cite{Worthey98}).  Elements placed in the
enhanced group include N, Ne, Na, Mg, Si, and S (plus sometimes C
and/or O).\footnote{In retrospect it would have made more sense to
group N with C since they are nucleosynthetically related
(\cite{WW95}), but making this change would negligibly affect the
conclusions.}  The iron-peak elements Cr, Mn, Fe, Co, Ni, Cu and Zn
constitute the depressed group.  All other elements (including those
heavier than Zn) are in the fixed group, with the exception of Ca (in
the depressed group), and C and O (which vary).

As noted, the four models differ in their treatment of C and O: model
1 has C fixed, O up; model 2 has C fixed, O fixed; model 3 has C down,
O up; and model 4 has C up, O up.  Because O is produced in massive
stars like Mg, it is probable that it, too, is routinely enhanced in
giant ellipticals; hence model 2 is unlikely on nucleosynthetic
grounds.  Model 3, with C down and O up, is similar to the models of
Weiss, Peletier \& Mateucci (1995), TCB98, and Salaris \& Weiss
(1998).  We show below that depression of C does not match the
Lick/IDS indices and that this model is also therefore unlikely.
Model 4, with C and O both enhanced, is our preferred model based on
McWilliam \& Rich (1990) and Rich \& McWilliam (priv.~comm.), who find
that O and C are enhanced in lockstep with Mg in stars in the Galactic
Bulge.  However model 1 (with C fixed) is very hard to distinguish
observationally from model 4 (see below).

Because the enhanced elements are not exactly the same as the
$\alpha$-elements (e.g., Ca is nominally an $\alpha$-element but
apparently tracks Fe in elliptical spectra; Worthey 1998; TWFBG98), we
use the notation \enh, where ``E'' refers to the mass fraction of
elements that are specifically enhanced in each model, in preference
to the more common notation \afe\ used by previous authors.  Following
TCB98, we write
\begin{equation}
\feh = \z - A\enh, \label{eq:z}
\end{equation}
or
\begin{equation}
\Delta\feh = -A\Delta\enh = -\frac{A}{1-A}\Delta\mathrm{[E/H]}
\label{eq:enh}
\end{equation}
at constant \z, where their very small second-order term in \enh\ has
been ignored.  Table~\ref{tbl:tbmodels} gives values of $A$ and
illustrative heavy-element fractions (C, O, E-group, Fe-peak) for the
four models, all at $\Delta\feh = -0.3$ dex and solar \z; values of
\enh\ and $\mathrm{[E/H]}$ for other values of $\Delta$\feh\ can be
calculated using Eqs.~\ref{eq:z} and \ref{eq:enh}.  For reference,
TCB98's model has $A_{\rm TCB98}=0.8$.

Table~\ref{tbl:tbmodels} reveals an important fact---because the
Fe-peak contribution to \z\ is so small (only 8\% for solar
abundance), reducing it by even 0.3 dex frees up only minimal room for
the so-called ``enhanced'' elements.  Hence, what really happens in
enhanced models is that the enhanced (and fixed) elements remain
\emph{nearly at their solar values, whereas Fe (and related elements)
are depressed}.  In short, we should think of giant ellipticals as
\emph{failing} to make Fe-peak elements rather than making too much of
certain other elements.  Likewise, the quantity \enh\ is not really an
enhancement of the E-elements but rather a \emph{depression of Fe}.

Other authors, including ourselves (e.g., \cite{WFG92}; \cite{WPM95};
\cite{Greggio97}; \cite{Vazdekis97}) have said this, but the
contradictory notion nevertheless persists that strong Mg indices are
due to an ``overabundance'' of Mg---this is not mathematically
possible if Mg, O, and the $\alpha$-elements track one another
closely, as these elements together dominate \z\ by mass.  We show
below that the TB95 response functions provide an alternative means of
strengthening \mgb\ and \mgtwo, namely, via \emph{weak} Fe-peak
elements (see below).  This unanticipated \emph{anti}-correlation
between \mgb\ and the Fe-peak elements is one of the major new
features of our treatment and the cause of our relatively small
derived values of \enh\ (compared with previous authors; see
Sec.~\ref{sec:results}).

We return next to the problem of the stellar evolutionary isochrones.
Since a full library of isochrones is not available for all abundance
ratios, we follow the lead of TCB98, who suggest from examining their
unpublished isochrones that models with varying \enh\ are ``virtually
indistinguishable in the CMD'' from models at the same \z\ with
$\enh=0$.  Earlier, Salaris et al.~(1993) had shown (at sub-solar
metallicities) that $\alpha$-enhanced isochrones are identical to
scaled-solar abundance isochrones at the same $Z$ provided that the
quantity
\begin{equation}
\left[\frac{X_{\rm HPE}}{X_{\rm LPE}}\right]\equiv 
\left[\frac{X_{\rm C} + X_{\rm N} + X_{\rm O} + X_{\rm Ne}}{X_{\rm Mg}
+ X_{\rm Si} + X_{\rm S} + X_{\rm Ca} + X_{\rm Fe}}\right]
\label{eq:XX}
\end{equation}
remains constant at the solar value ($=0$).  Here $X_i$ is the mass
fraction in element $i$, and brackets indicate the usual logarithm
relative to solar.  The elements in $X_{\rm HPE}$ have high ionization
potentials and their opacity governs the mean turnoff temperature; the
elements in $X_{\rm LPE}$ have low ionization potentials and their
opacity governs the temperature of the giant branch.  Preserving the
ratio [${X_{\rm HPE}}/{X_{\rm LPE}}$] thus preserves the \emph{shape}
of the track, they say, and the new track is found to fit neatly into
the old sequence at the same value of $Z$.  Values of [${X_{\rm
HPE}}/{X_{\rm LPE}}$] are given in Table~\ref{tbl:tbmodels} for our
four models.  Models 2 and 3 are nearly solar, while models 1 and 4
are about 15\% overabundant in HPE elements.  These small deviations
prove to be relatively unimportant, as shown in
Section~\ref{sec:uncenh}.

\begin{deluxetable}{cclrrrrrrrrr}
\tablefontsize{\footnotesize}
\tablecaption{Fractional index responses to non-solar abundance
ratios\tablenotemark{a}\label{tbl:tbind}}
\tablewidth{0pt}
\tablehead{\colhead{Model}&\colhead{}&
\colhead{Star}&\colhead{\%$F$}&
\colhead{$\delta\mathrm{C_2}$}&
\colhead{$\delta\mathrm{H}\beta$}&
\colhead{$\delta\mathrm{Mg_1}$}&
\colhead{$\delta\mathrm{Mg_2}$}&
\colhead{$\delta\mathrm{Mg}b$}&
\colhead{$\delta\mathrm{Fe52}$}&
\colhead{$\delta\mathrm{Fe53}$}&
\colhead{$\delta\fe$}}
\startdata
1&C$\circ$O$+$&cg&53&$-0.037$&0.000&0.128&0.088&0.295&$-0.113$&$-0.191$&$-0.147$\nl
&&to&44&$-0.130$&0.026&0.617&0.025&0.124&$-0.226$&$-0.260$&$-0.240$\nl
&&cd&3&$-0.193$&$-0.361$&0.183&0.086&0.153&$-0.177$&$-0.210$&$-0.192$\nl
&&total&&$-0.046$&0.027&0.146&0.080&0.249&$-0.136$&$-0.203$&$-0.165$\nl
\nl
2&C$\circ$O$\circ$&cg&53&$-0.043$&0.000&0.179&0.129&0.366&$-0.124$&$-0.205$&$-0.159$\nl
&&to&44&$-0.176$&0.031&0.541&0.049&0.184&$-0.226$&$-0.261$&$-0.241$\nl
&&cd&3&$-0.242$&$-0.317$&0.238&0.118&0.192&$-0.191$&$-0.228$&$-0.208$\nl
&&total&&$-0.056$&0.031&0.193&0.118&0.315&$-0.145$&$-0.215$&$-0.176$\nl
\nl
3&C$-$O$+$&cg&53&$-0.703$&0.000&$-0.342$&0.010&0.767&$-0.198$&$-0.138$&$-0.172$\nl
&&to&44&$-0.727$&0.027&$-0.558$&$-0.062$&0.296&$-0.232$&$-0.281$&$-0.252$\nl
&&cd&3&$-0.864$&0.314&$-0.021$&0.023&0.168&$-0.199$&$-0.208$&$-0.203$\nl
&&total&&$-0.708$&0.026&$-0.330$&0.001&0.625&$-0.204$&$-0.165$&$-0.187$\nl
\nl
4&C$+$O$+$&cg&53&0.055&0.000&0.175&0.095&0.265&$-0.107$&$-0.195$&$-0.145$\nl
&&to&44&$-0.049$&0.026&0.786&0.032&0.111&$-0.226$&$-0.259$&$-0.240$\nl
&&cd&3&$-0.075$&$-0.395$&0.200&0.091&0.152&$-0.175$&$-0.210$&$-0.191$\nl
&&total&&0.045&0.027&0.195&0.086&0.225&$-0.131$&$-0.206$&$-0.164$\nl
\nl
&TCB98&total\tablenotemark{b}&&\nodata&$-0.050$&\nodata&0.172&\nodata&\nodata&\nodata&$-0.125$\nl
\enddata
\tablenotetext{a}{For a W94 model with age 12 Gyr, fixed
$\Delta\feh=-0.3$ dex, and $\z=0$.}
\tablenotetext{b}{For the TCB98 model with an age of 10 Gyr and
$\z=0$.}
\tablecomments{Cols.~(1)--(2): Enhancement model from Table 6;
``TCB98'' is close to the model of Tantalo et al.~(1998a; TCB98); model
4 is our preferred model.  Col.~(3): Stellar component on isochrone,
following TB95.  ``cg'' refers to cool giants ($T_e=4255$ K, $\log
g=1.9$); ``to'' refers to turn-off stars ($T_e=6200$ K, $\log g=4.1$);
``cd'' refers to cool dwarfs ($T_e=4575$ K, $\log g=4.6$); and
``total'' refers to the composite spectrum.  Col.~(4): Typical
fractional light contribution from each stellar component to total
flux, in percent (C$_2$ has a slightly higher contribution from
turn-off stars than the other indices).  Col.~(5)--(12): Fractional
index responses for $\Delta\feh=-0.3$ dex, in the sense $\delta I =
\Delta I/I$, where $\Delta I$ is the index change and $I$ is the
original value of the index. These are computed by inserting the
response functions of Tripicco \& Bell (1995) for each element into
Eq.~\ref{eq:index}.  (The three stellar entries do not sum to the
total value because they are weighted by the unequal stellar indices
of each type.)  Fractional responses for the TCB98 model are based on
element abundances computed from their Eqs.~(3)--(5) and their Table
1.}
\end{deluxetable}

More recently, Salaris \& Weiss (1998) have suggested that, at higher
metallicities near solar, track constancy may break down and that
increasing \afe\ both shifts the track to the blue and changes its
shape.  It is not clear whether these effects are due to high \enh, to
$\XX\neq0$, or both.  However, the motions are small, and we show in
Section~\ref{sec:uncenh} that their impact on the indices is probably
slight.

If isochrones do not shift (at fixed metallicity), we can assume that
$\log g$, $\log T_e$, $\log L$, and the SED of each star on the track
are also constant.  Hence, it is necessary only to calculate the
changes in each spectral feature using the index response functions of
TB95, by perturbing each element up or down according to the model.
TB95 tabulate fractional index changes for three typical stars, one on
the lower main sequence, one at the turnoff, and one on the RGB, at
solar metallicity.  We assume the same fractional changes at all
metallicities and combine these responses by weighting by the
fractional light contributions of each type of star at each
index.\footnote{We have ignored the dependence of the line strength
indices on Ti, as TB95 make contradictory statements about its
inclusion in their model atmospheres.  Although their tables include
the effects of varying Ti, they clearly state that they have not
included TiO lines in their line lists.  This will affect the line
strengths in the coolest giants.  However, \hbeta, \mgb, and \fe\ are
little affected by Ti in their models; see their Tables 4--6.}
Details are given in the notes to Table~\ref{tbl:tbind}.

Note that the TB95 response functions are for enhancement values
corresponding to $\mathrm{[X/H]} = +0.3$ dex.  Response functions for
arbitrary values of \enh\ are calculated via Eq.~\ref{eq:enh} to get
$\Delta\feh$ and $\Delta\enh$ and then by exponentially scaling the
response functions in Tables 4--6 of TB95 by the appropriate element
abundance.  The fractional response of index $I$ is therefore
\begin{equation}
\frac{\Delta I}{I_0} = \left\{\prod_i[1 +
R_{0.3}(X_i)]^{(\mathrm{[X_i/H]}/0.3)}\right\} - 1,\label{eq:index}
\end{equation}
where $R_{0.3}(X_i)$ is the TB95 response function for element $i$ at
$\mathrm{[X_{\it i}/H]}=+0.3$ dex.\footnote{This equation assumes that
the percentage index change is constant for each step of 0.3 dex in
abundance.  This assures that index values approach zero gracefully at
low abundances but predicts infinite indices at high abundances, which
is impossible.  The scaling law should therefore probably not be
applied at levels much above $\mathrm{[X_i/H]} = +0.6$.}

Table~\ref{tbl:tbind} shows changes in the indices corresponding to
the four models in Table~\ref{tbl:tbmodels}, all of which have
$\Delta\enh\ \sim$ 0.3.  \hbeta\ is virtually unaffected by non-solar
abundance ratios, even at substantial \enh; all changes are less than
3\%, which translates to $\la8$\% in age.  Changes in \fe\ are roughly
the same in all models and amount to a decrease of about 20\%, driven
mostly by the decrease in \feh\ (of 0.3 dex).  However, C$_2$4668,
\mgone, \mgtwo, and \mgb\ are all different, owing to the presence (or
not) of C$_2$ bands in the passband or sidebands of these indices;
C$_2$ and \mgone\ increase greatly with increasing C, \mgb\ declines
with increasing C, while \mgtwo\ stays about the same independent of
C.  These changes all reflect the different abundance of C in the
models since the abundance of Mg (and other elements in the E group)
is always about constant (cf. Table~\ref{tbl:tbmodels}).

Finally, we note that \mgb\ increases in all models, in apparent
contradiction to the near constancy of \enh.  This increase is due
mostly to the decrease in Fe and Cr (see TB95), which has the effect
of increasing \mgb.  In fact, changes in all the Mg indices are driven
more by the Fe-peak deficit than by any actual increase in Mg, proving
once again that a more correct way of looking at elliptical galaxies
is to regard them as Fe-poor rather than $\alpha$-enhanced.

\subsection{Ages, metallicities, and abundance ratios}\label{sec:params}

SSP-equivalent parameters are derived for each G93 galaxy by choosing,
for each model 1--4, the best-fitting age $t$, metallicity \z, and
enhancement ratio \enh.  Solving for three free parameters requires
three indices, for which we use \hbeta, \fe, and \mgb.  First, an
expanded model grid of line strengths as a function of $t$, \z, and
(now) \enh\ is generated by applying the TB95 response functions to
the base W94 models at each ($t$, \z).  These new grids (one for each
model 1--4) are created by interpolating the W94 models at intervals
of $\Delta t=0.1$ Gyr and $\Delta\z=0.01$ and then interpolating the
TB95 results at intervals of $\Delta\enh=0.01$ at each $(t,\z)$.  The
process is then inverted to derive $(t,\z,\enh)$ for each galaxy by
searching in the grid to find that point with minimum distance from
the observed parameters $(\hbeta,\mgb,\fe)$.  It was necessary to
linearly extrapolate the W94 models to slightly higher ages and to
both lower and higher \z\ values to cover the full range of
$(\hbeta,\mgb,\fe)$-space populated by the observations.  The range of
$(t,\z,\enh)$ space covered by the final grids is
\begin{displaymath}
\begin{array}{ll}
\phn1\leq t\; (\mathrm{Gyr})\leq 22,&-0.5\leq\z\leq 1.25,\\
&\qquad\qquad-0.3\leq\enh\leq 0.75\\
\\
22 < t\; (\mathrm{Gyr})\leq 30,&-0.5\leq\z\leq 0.5,\phn\\
&\qquad\qquad-0.3\leq\enh\leq 0.75.
\end{array}
\end{displaymath}

Tables~\ref{tbl:re8tza} and \ref{tbl:re2tza} give derived
$(t,\z,\enh)$ values and associated uncertainties in the \reo{8}\ and
\reo{2}\ apertures, respectively.  Errors were derived by searching
the grid at $(\hbeta \pm\sigma_{\hbeta},\mgb,\fe)$, $(\hbeta,\mgb
\pm\sigma_{\mgb},\fe)$, and $(\hbeta,\mgb,\fe \pm\sigma_{\fe})$ and
taking the maximum deviations $\mathrm{max}(\Delta t)$,
$\mathrm{max}(\Delta\z)$, and ${\rm max}(\Delta\enh)$ as the
associated uncertainties.\footnote{These errors faithfully relect the
magnitude of the uncertainties but not their correlations.  Correlated
errors in \z\ and $t$ can be important, driven jointly by
observational errors in \hbeta\ (\cite{Trager97}).  Fortunately the
G93 errors are so small that observationally driven correlations in
the output parameters are not important.}

The derived SSP-equivalent parameters should be treated with caution
for the extrapolated solutions ($t>18$ Gyr at all metallicities,
$\z>0.5$ at all ages, and $t<8$ Gyr at $\z<-0.225$).  However, in the
\reo{8}\ aperture, which we concentrate on in this and the following
paper, only one galaxy (NGC 5813) has $t>18$ Gyr, and only a few more
have $\z>0.5$ for any enhancement model.  The extrapolations are more
significant for the stellar population parameters in the \reo{2}\
aperture (Table~\ref{tbl:re2tza}).  However, many of these would also
lessen or disappear if TWFBG98 velocity corrections to \hbeta\ were
substituted for those of G93, or if small \hbeta\ emission fill-in
errors were corrected (see Appendix~\ref{app:emsigcorr}).

\begin{figure*}
\plotone{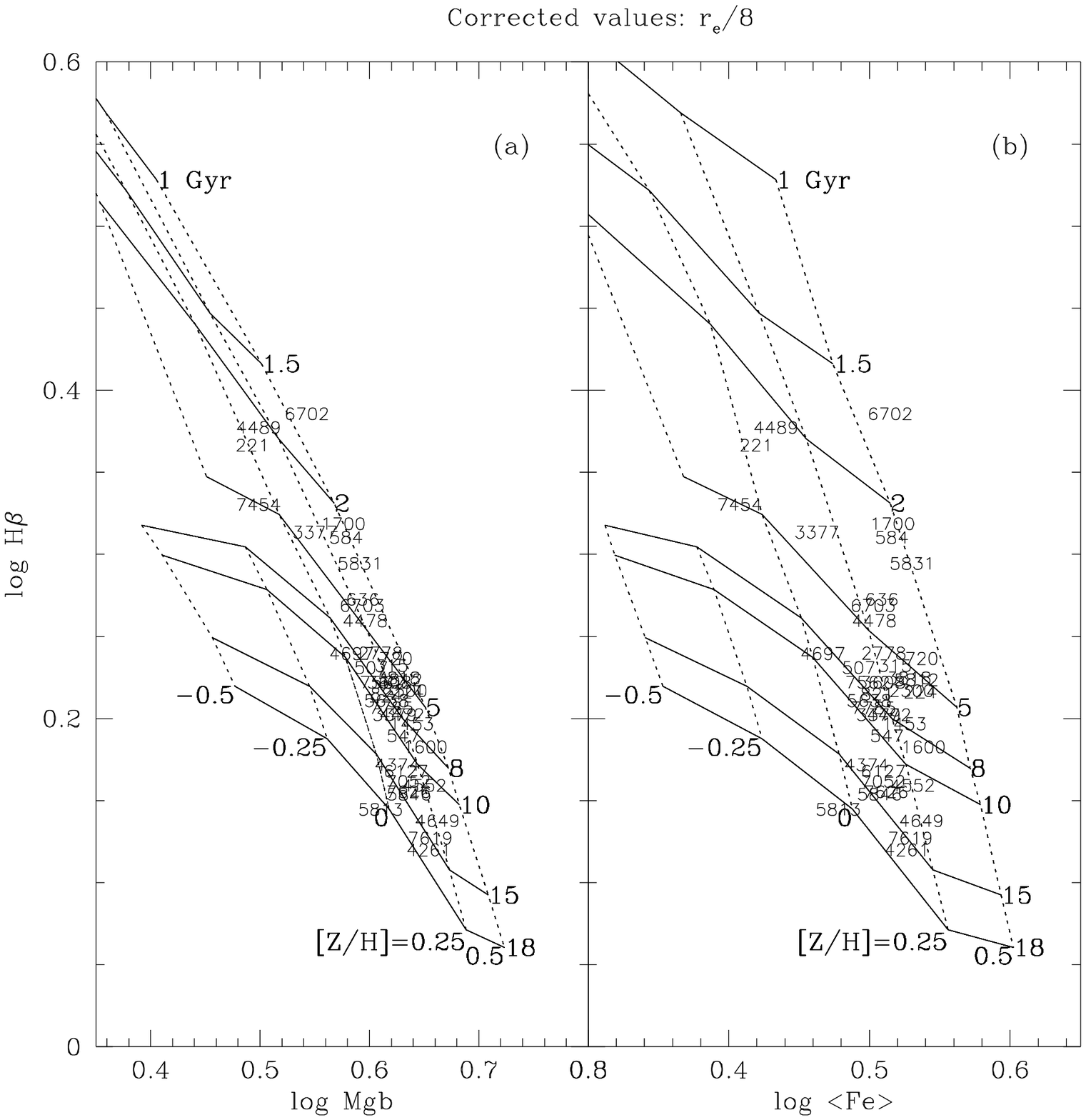}
\caption{Line strengths of early-type galaxies in the Gonz{\'a}lez
(1993) sample in the central \reo{8}\ aperture, corrected to solar
abundance ratios ($\enh=0$) using the method described in
Section~\ref{sec:tb95} and enhancement model 4.  Solar abundance model
grids from Worthey (1994) are again superimposed as in
Figure~\ref{fig:re8m}.  The corrections to solar abundance ratios
bring inferred SSP ages and metallicities into good agreement between
the \mgb--\hbeta\ and the \fe--\hbeta\ diagrams, as expected.  These
figures indicate the final central SSP values for $t$ and
\z.\label{fig:re8c}}
\end{figure*}

\begin{figure*}
\plotone{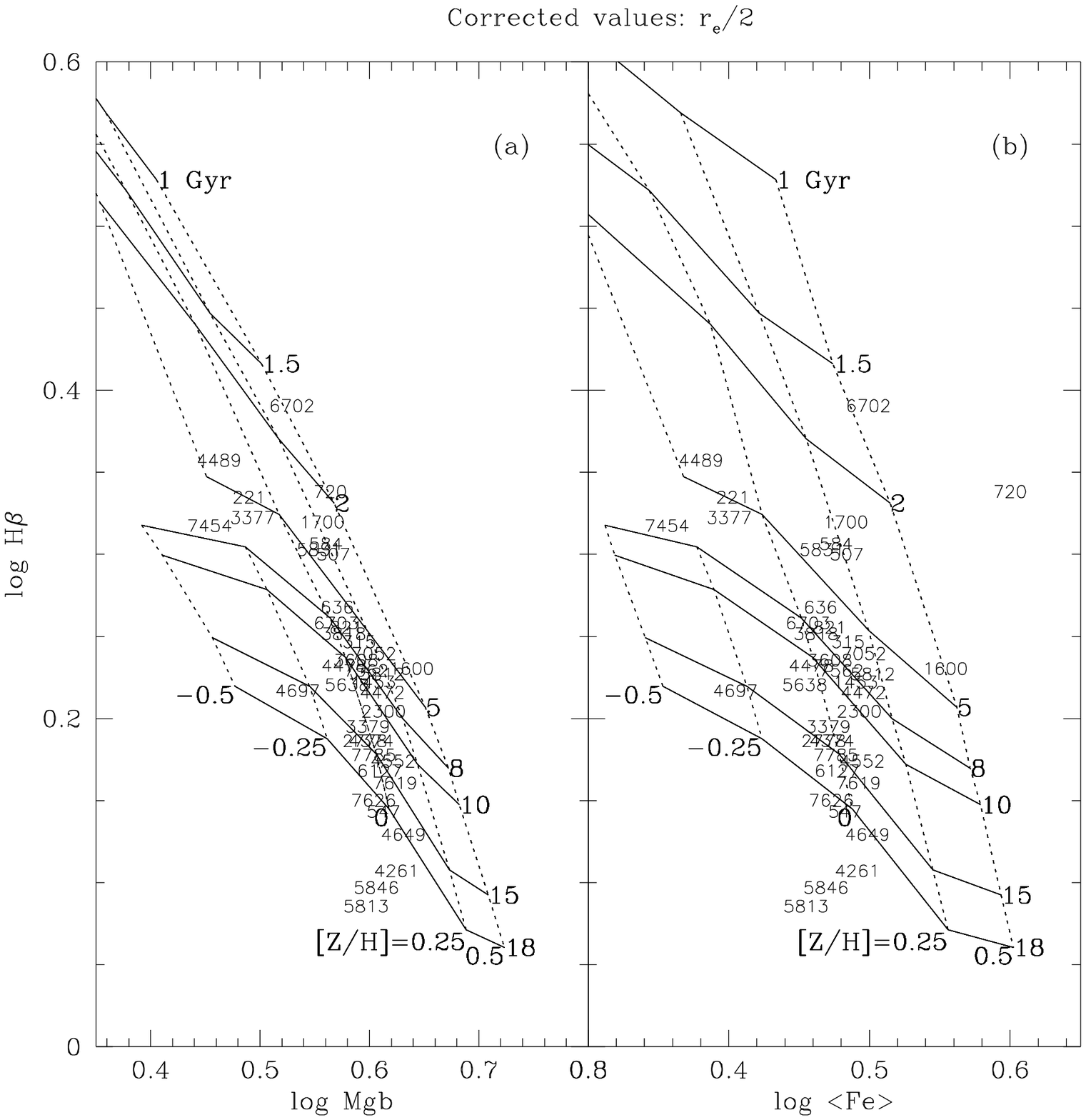}
\caption{Line strengths of early-type galaxies in the Gonz{\'a}lez
(1993) sample in the global \reo{2}\ aperture, corrected to solar
abundance ratios ($\enh=0$) using the method described in
Section~\ref{sec:tb95} and enhancement model 4.  Model grids from
Worthey (1994) are again superimposed as in Figure~\ref{fig:re8m}.
Ages and metallicities in the two panels agree, as they do for the
central values in Figure~\ref{fig:re8c} (with the exception of NGC
720, which lies far off the grid).  These figures indicate the final
global SSP values for $t$ and \z. \label{fig:re2c}}
\end{figure*}

We have checked our fitting procedure by correcting the observed line
strengths back to solar abundance ratios using the TB95 response
functions for the solved-for values of \z\ and \enh.  The resulting
corrected line strengths are presented in Figures~\ref{fig:re8c} (for
\reo{8}) and \ref{fig:re2c} (for \reo{2}) with the W94 models
overplotted.  These are the predicted line strengths that would be
seen if the populations had the same $t$ and \z\ but \enh\ = 0.
Metallicities and ages inferred from \mgb\ and \fe\ now agree,
suggesting that our method for finding for the best-fitting parameters
by searching in the three-dimensional grid is working correctly.
These corrected points show graphically our final values of $t$ and
\z.

\begin{figure*}
\plotone{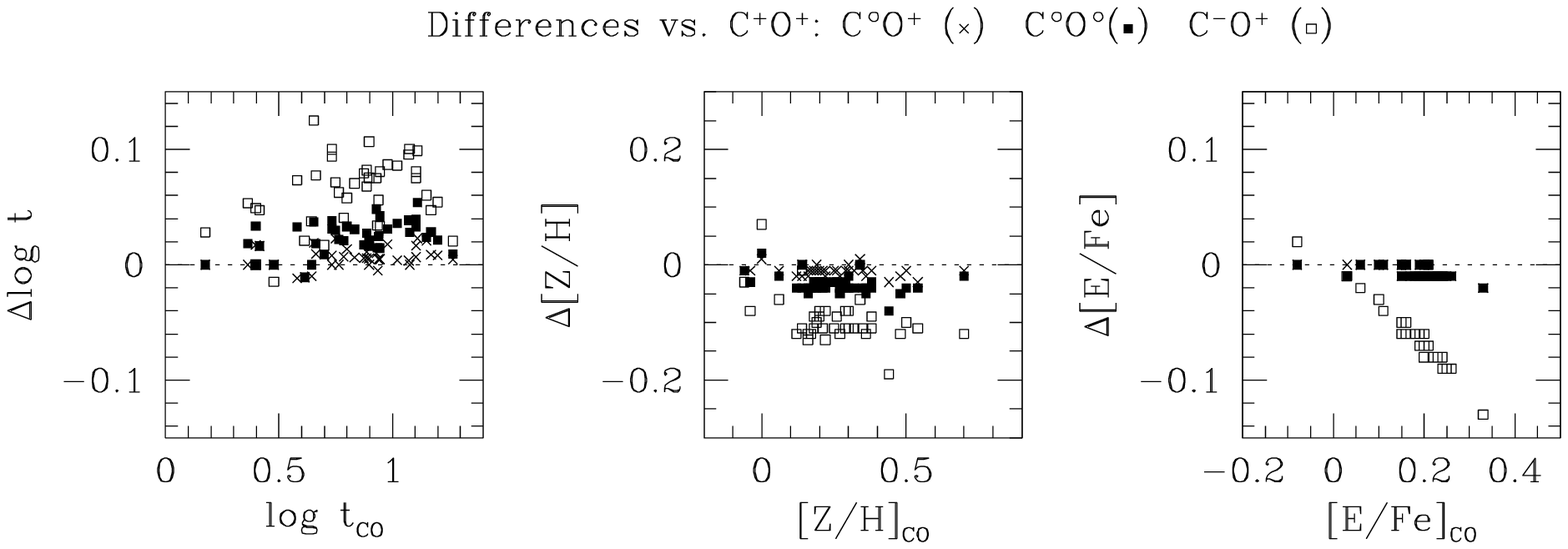}
\caption{SSP ages, metallicities, and enhancement ratios in models 1,
2, and 3 compared to model 4 (which has C$+$O$+$).  (Data are for the
central \reo{8}\ aperture.)  Models 1, 2, and 4 give nearly identical
population parameters because the C abundance changes little among
them (cf.~Table~\ref{tbl:tbmodels}); model 3 deviates strongly from
the others (especially in age and enhancement) because of the strong
dependence of \mgb\ on C abundance (TB95;
cf.~Table~\ref{tbl:tbind}).\label{fig:comp_enhmods}}
\end{figure*}

Derived stellar parameters from the four enhancement models are
compared in Figure~\ref{fig:comp_enhmods}.  The most notable
difference is between model 3 (C down, O up) versus all other models:
galaxies are older, more metal-poor, and less enhanced in model 3 than
in the others.  These differences are driven entirely by the low C
abundance in model 3; reducing C increases \mgb\ but has little effect
on \fe.  Models with low C (like model 3) therefore result in lower
overall metallicities, smaller \enh, and older ages, as may be seen by
following through the consequences of a higher \mgb\ response function
in Figure~\ref{fig:re8m}.

\begin{figure*}
\plotone{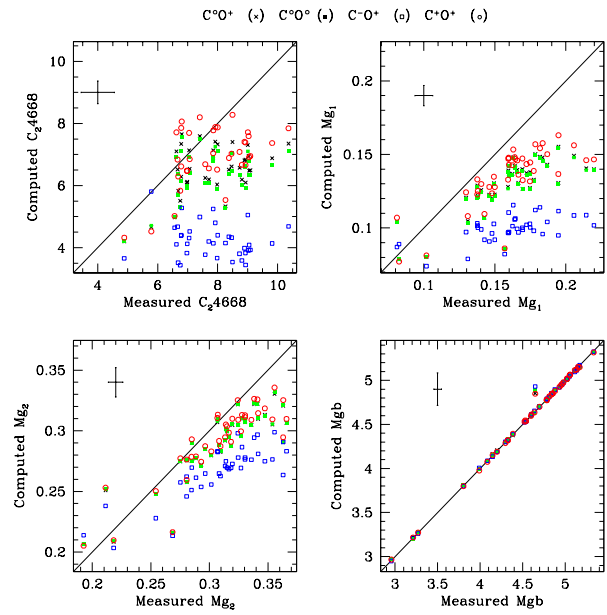}
\caption{A comparison of measured and model-derived indices for the
four element enhancement models.  Shown here are three new C-sensitive
line indices for the central regions (from Lick/IDS data), plus \mgb,
shown for reference.  The three new indices were not used to compute
the original SSP parameters and therefore provide an independent check
on them; ``computed'' indices are those predicted by the SSP
parameters.  Model 3 (open squares) systematically underestimates the
C-sensitive indices compared to models 1, 2, and 4. This is because
its C abundance is too low.  The good agreement for \mgb\ shows that
we have the proper solutions.
\label{fig:comp4index}}
\end{figure*}

Is model 3 in fact compatible with observed galaxy line strengths?  To
test this, we augment the G93 indices with data on the C-sensitive
feature C$_2$4668 from the Lick/IDS sample of TWFBG98.  For each
population model, we use the response functions of TB95 to compute
predicted line strengths for three new features---C$_2$4668, \mgone,
and \mgtwo---none of which were used in the original fits.  Observed
versus predicted indices are shown in Figure~\ref{fig:comp4index}.
Enhancement model 3, in which C is depressed, clearly fails
systematically to reproduce the strengths of the new indices,
especially C$_2$4668.  Models 1, 2, and 4 are nearly
indistinguishable, as expected since the C abundance hardly varies
among them (cf.  Table~\ref{tbl:tbmodels}).  Model 4 is marginally the
best ($\sim1\sigma$) on account of its slightly higher C abundance, a
further slight boost for our preferred model.  Although model 4 fits
best, it still fails systematically to reproduce the highest values of
C$_2$4668, \mgone\, and, especially, \mgtwo.  This may indicate that C
(and perhaps Mg) are actually \emph{over}-enhanced compared to the
E-group generally and may signal a breakdown in our assumption that
all E-group elements scale in lockstep.  Specific element abundance
ratios will be explored using the full set of Lick indices in future
papers.

\begin{deluxetable}{lcrrrrr}
\tablefontsize{\footnotesize}
\tablenum{6a}
\tablecaption{Ages, metallicities and enhancement ratios through the
central $r_e/8$ aperture\label{tbl:re8tza}}
\tablewidth{0pt}
\tablehead{\colhead{Name}&\colhead{Model}&\colhead{Age (Gyr)}&
\colhead{\z}&\colhead{\enh}&\colhead{\feh}&\colhead{$[\mathrm{E/H}]$}}
\startdata
NGC  221&1&$ 3.0\pm0.7$&$ 0.01\pm0.04$&$-0.08\pm0.01$&$ 0.08\pm0.04$&$ 0.00\pm0.04$\nl
&2&$ 3.0\pm0.5$&$ 0.02\pm0.04$&$-0.08\pm0.01$&$ 0.08\pm0.04$&$ 0.00\pm0.04$\nl
&3&$ 2.9\pm0.5$&$ 0.07\pm0.04$&$-0.06\pm0.01$&$ 0.12\pm0.04$&$ 0.06\pm0.04$\nl
&4&$ 3.0\pm0.7$&$ 0.00\pm0.05$&$-0.08\pm0.01$&$ 0.07\pm0.05$&$-0.01\pm0.05$\nl
\tablevspace{2pt}
NGC  224&1&$ 6.2\pm1.5$&$ 0.37\pm0.06$&$ 0.18\pm0.02$&$ 0.21\pm0.06$&$ 0.39\pm0.06$\nl
&2&$ 6.4\pm1.6$&$ 0.35\pm0.05$&$ 0.18\pm0.02$&$ 0.21\pm0.05$&$ 0.39\pm0.05$\nl
&3&$ 6.7\pm1.5$&$ 0.29\pm0.04$&$ 0.12\pm0.01$&$ 0.20\pm0.04$&$ 0.32\pm0.04$\nl
&4&$ 6.1\pm1.5$&$ 0.38\pm0.07$&$ 0.19\pm0.02$&$ 0.20\pm0.07$&$ 0.39\pm0.07$\nl
\tablevspace{2pt}
NGC  315&1&$ 5.8\pm1.1$&$ 0.31\pm0.04$&$ 0.23\pm0.02$&$ 0.10\pm0.04$&$ 0.33\pm0.04$\nl
&2&$ 6.1\pm1.3$&$ 0.28\pm0.05$&$ 0.23\pm0.02$&$ 0.10\pm0.05$&$ 0.33\pm0.05$\nl
&3&$ 6.7\pm1.5$&$ 0.21\pm0.04$&$ 0.16\pm0.01$&$ 0.09\pm0.04$&$ 0.25\pm0.04$\nl
&4&$ 5.8\pm1.2$&$ 0.32\pm0.06$&$ 0.24\pm0.02$&$ 0.10\pm0.06$&$ 0.34\pm0.06$\nl
\tablevspace{2pt}
NGC  507&1&$ 7.6\pm2.9$&$ 0.17\pm0.07$&$ 0.19\pm0.03$&$ 0.00\pm0.08$&$ 0.19\pm0.07$\nl
&2&$ 7.8\pm3.0$&$ 0.15\pm0.07$&$ 0.19\pm0.02$&$ 0.00\pm0.07$&$ 0.19\pm0.07$\nl
&3&$ 9.0\pm2.3$&$ 0.07\pm0.07$&$ 0.13\pm0.02$&$-0.03\pm0.07$&$ 0.10\pm0.07$\nl
&4&$ 7.5\pm2.5$&$ 0.18\pm0.06$&$ 0.20\pm0.03$&$-0.01\pm0.07$&$ 0.19\pm0.06$\nl
\tablevspace{2pt}
NGC  547&1&$ 9.9\pm2.3$&$ 0.20\pm0.05$&$ 0.25\pm0.01$&$-0.03\pm0.05$&$ 0.22\pm0.05$\nl
&2&$10.2\pm2.5$&$ 0.18\pm0.05$&$ 0.25\pm0.02$&$-0.01\pm0.05$&$ 0.24\pm0.05$\nl
&3&$11.6\pm2.7$&$ 0.09\pm0.05$&$ 0.17\pm0.01$&$-0.04\pm0.05$&$ 0.13\pm0.05$\nl
&4&$ 9.5\pm2.2$&$ 0.22\pm0.05$&$ 0.26\pm0.01$&$-0.02\pm0.05$&$ 0.24\pm0.05$\nl
\tablevspace{2pt}
NGC  584&1&$ 2.6\pm0.3$&$ 0.46\pm0.03$&$ 0.21\pm0.01$&$ 0.27\pm0.03$&$ 0.48\pm0.03$\nl
&2&$ 2.7\pm0.3$&$ 0.43\pm0.03$&$ 0.21\pm0.01$&$ 0.27\pm0.03$&$ 0.48\pm0.03$\nl
&3&$ 2.8\pm0.4$&$ 0.36\pm0.03$&$ 0.14\pm0.01$&$ 0.25\pm0.03$&$ 0.39\pm0.03$\nl
&4&$ 2.5\pm0.3$&$ 0.48\pm0.03$&$ 0.22\pm0.01$&$ 0.28\pm0.03$&$ 0.50\pm0.03$\nl
\tablevspace{2pt}
NGC  636&1&$ 4.0\pm0.4$&$ 0.35\pm0.04$&$ 0.11\pm0.01$&$ 0.25\pm0.04$&$ 0.36\pm0.04$\nl
&2&$ 4.0\pm0.6$&$ 0.34\pm0.05$&$ 0.11\pm0.02$&$ 0.25\pm0.05$&$ 0.36\pm0.05$\nl
&3&$ 4.3\pm0.5$&$ 0.28\pm0.04$&$ 0.07\pm0.01$&$ 0.23\pm0.04$&$ 0.30\pm0.04$\nl
&4&$ 4.1\pm0.7$&$ 0.34\pm0.07$&$ 0.11\pm0.02$&$ 0.24\pm0.07$&$ 0.35\pm0.07$\nl
\tablevspace{2pt}
NGC  720&1&$ 4.7\pm2.4$&$ 0.41\pm0.14$&$ 0.31\pm0.04$&$ 0.13\pm0.14$&$ 0.44\pm0.14$\nl
&2&$ 4.9\pm2.5$&$ 0.36\pm0.14$&$ 0.31\pm0.04$&$ 0.12\pm0.14$&$ 0.43\pm0.14$\nl
&3&$ 6.0\pm2.4$&$ 0.25\pm0.10$&$ 0.20\pm0.03$&$ 0.10\pm0.10$&$ 0.30\pm0.10$\nl
&4&$ 4.5\pm2.7$&$ 0.44\pm0.15$&$ 0.33\pm0.04$&$ 0.13\pm0.15$&$ 0.46\pm0.15$\nl
\tablevspace{2pt}
NGC  821&1&$ 7.8\pm1.5$&$ 0.21\pm0.03$&$ 0.14\pm0.01$&$ 0.08\pm0.03$&$ 0.22\pm0.03$\nl
&2&$ 8.2\pm1.0$&$ 0.19\pm0.03$&$ 0.14\pm0.01$&$ 0.08\pm0.03$&$ 0.22\pm0.03$\nl
&3&$ 9.0\pm1.3$&$ 0.14\pm0.03$&$ 0.10\pm0.01$&$ 0.06\pm0.03$&$ 0.16\pm0.03$\nl
&4&$ 7.7\pm1.3$&$ 0.22\pm0.03$&$ 0.15\pm0.01$&$ 0.08\pm0.03$&$ 0.23\pm0.03$\nl
\tablevspace{2pt}
NGC 1453&1&$ 7.9\pm1.4$&$ 0.29\pm0.04$&$ 0.20\pm0.02$&$ 0.11\pm0.04$&$ 0.31\pm0.04$\nl
&2&$ 8.2\pm1.8$&$ 0.26\pm0.04$&$ 0.20\pm0.02$&$ 0.11\pm0.04$&$ 0.31\pm0.04$\nl
&3&$ 9.4\pm2.0$&$ 0.19\pm0.04$&$ 0.14\pm0.01$&$ 0.08\pm0.04$&$ 0.22\pm0.04$\nl
&4&$ 7.9\pm1.4$&$ 0.30\pm0.04$&$ 0.21\pm0.02$&$ 0.10\pm0.04$&$ 0.31\pm0.04$\nl
\tablevspace{2pt}
NGC 1600&1&$ 8.5\pm1.7$&$ 0.34\pm0.05$&$ 0.21\pm0.02$&$ 0.15\pm0.05$&$ 0.36\pm0.05$\nl
&2&$ 8.9\pm1.7$&$ 0.31\pm0.05$&$ 0.21\pm0.02$&$ 0.15\pm0.05$&$ 0.36\pm0.05$\nl
&3&$ 9.3\pm2.2$&$ 0.24\pm0.05$&$ 0.14\pm0.01$&$ 0.13\pm0.05$&$ 0.27\pm0.05$\nl
&4&$ 8.6\pm1.7$&$ 0.35\pm0.05$&$ 0.22\pm0.02$&$ 0.15\pm0.05$&$ 0.37\pm0.05$\nl
\tablevspace{2pt}
NGC 1700&1&$ 2.3\pm0.3$&$ 0.49\pm0.03$&$ 0.16\pm0.01$&$ 0.34\pm0.03$&$ 0.50\pm0.03$\nl
&2&$ 2.4\pm0.4$&$ 0.46\pm0.04$&$ 0.16\pm0.01$&$ 0.34\pm0.04$&$ 0.50\pm0.04$\nl
&3&$ 2.6\pm0.4$&$ 0.40\pm0.04$&$ 0.10\pm0.01$&$ 0.32\pm0.04$&$ 0.42\pm0.04$\nl
&4&$ 2.3\pm0.3$&$ 0.50\pm0.03$&$ 0.16\pm0.01$&$ 0.35\pm0.03$&$ 0.51\pm0.03$\nl
\tablevspace{2pt}
NGC 2300&1&$ 6.5\pm1.6$&$ 0.34\pm0.05$&$ 0.23\pm0.02$&$ 0.13\pm0.05$&$ 0.36\pm0.05$\nl
&2&$ 6.8\pm1.8$&$ 0.31\pm0.05$&$ 0.23\pm0.02$&$ 0.13\pm0.05$&$ 0.36\pm0.05$\nl
&3&$ 7.2\pm1.5$&$ 0.24\pm0.04$&$ 0.15\pm0.01$&$ 0.13\pm0.04$&$ 0.28\pm0.04$\nl
&4&$ 6.3\pm1.4$&$ 0.36\pm0.04$&$ 0.24\pm0.02$&$ 0.14\pm0.04$&$ 0.38\pm0.04$\nl
\tablevspace{2pt}
NGC 2778&1&$ 5.4\pm2.1$&$ 0.28\pm0.06$&$ 0.22\pm0.02$&$ 0.08\pm0.06$&$ 0.30\pm0.06$\nl
&2&$ 5.8\pm1.9$&$ 0.25\pm0.06$&$ 0.22\pm0.02$&$ 0.08\pm0.06$&$ 0.30\pm0.06$\nl
&3&$ 6.8\pm1.8$&$ 0.18\pm0.05$&$ 0.15\pm0.01$&$ 0.07\pm0.05$&$ 0.22\pm0.05$\nl
&4&$ 5.4\pm1.7$&$ 0.29\pm0.07$&$ 0.23\pm0.02$&$ 0.08\pm0.07$&$ 0.31\pm0.07$\nl
\tablevspace{2pt}
NGC 3377&1&$ 3.7\pm0.9$&$ 0.19\pm0.05$&$ 0.19\pm0.02$&$ 0.02\pm0.05$&$ 0.21\pm0.05$\nl
&2&$ 4.1\pm0.8$&$ 0.15\pm0.05$&$ 0.19\pm0.01$&$ 0.00\pm0.05$&$ 0.19\pm0.05$\nl
&3&$ 4.5\pm0.5$&$ 0.09\pm0.02$&$ 0.13\pm0.01$&$-0.01\pm0.02$&$ 0.12\pm0.02$\nl
&4&$ 3.8\pm0.9$&$ 0.19\pm0.06$&$ 0.19\pm0.02$&$ 0.01\pm0.06$&$ 0.20\pm0.06$\nl
\tablevspace{2pt}
NGC 3379&1&$ 8.9\pm1.9$&$ 0.20\pm0.04$&$ 0.20\pm0.01$&$ 0.02\pm0.04$&$ 0.22\pm0.04$\nl
&2&$ 9.7\pm1.9$&$ 0.17\pm0.04$&$ 0.20\pm0.01$&$ 0.02\pm0.04$&$ 0.22\pm0.04$\nl
&3&$10.6\pm1.6$&$ 0.10\pm0.03$&$ 0.14\pm0.01$&$-0.01\pm0.03$&$ 0.13\pm0.03$\nl
&4&$ 8.8\pm1.6$&$ 0.21\pm0.04$&$ 0.21\pm0.01$&$ 0.01\pm0.04$&$ 0.22\pm0.04$\nl
\tablevspace{2pt}
NGC 3608&1&$ 6.9\pm1.5$&$ 0.25\pm0.04$&$ 0.16\pm0.02$&$ 0.10\pm0.04$&$ 0.26\pm0.04$\nl
&2&$ 7.3\pm1.6$&$ 0.23\pm0.04$&$ 0.16\pm0.02$&$ 0.11\pm0.04$&$ 0.27\pm0.04$\nl
&3&$ 8.0\pm1.8$&$ 0.17\pm0.04$&$ 0.11\pm0.01$&$ 0.09\pm0.04$&$ 0.20\pm0.04$\nl
&4&$ 6.8\pm1.3$&$ 0.26\pm0.04$&$ 0.17\pm0.02$&$ 0.10\pm0.04$&$ 0.27\pm0.04$\nl
\tablevspace{2pt}
NGC 3818&1&$ 5.9\pm1.5$&$ 0.34\pm0.07$&$ 0.22\pm0.02$&$ 0.14\pm0.07$&$ 0.36\pm0.07$\nl
&2&$ 6.0\pm1.5$&$ 0.32\pm0.05$&$ 0.22\pm0.02$&$ 0.15\pm0.05$&$ 0.37\pm0.05$\nl
&3&$ 6.6\pm1.7$&$ 0.24\pm0.05$&$ 0.15\pm0.01$&$ 0.13\pm0.05$&$ 0.28\pm0.05$\nl
&4&$ 5.6\pm1.8$&$ 0.36\pm0.06$&$ 0.23\pm0.02$&$ 0.15\pm0.06$&$ 0.38\pm0.06$\nl
\tablevspace{2pt}
NGC 4261&1&$16.1\pm3.2$&$ 0.17\pm0.04$&$ 0.19\pm0.01$&$ 0.00\pm0.04$&$ 0.19\pm0.04$\nl
&2&$16.6\pm2.3$&$ 0.15\pm0.03$&$ 0.19\pm0.01$&$ 0.00\pm0.03$&$ 0.19\pm0.03$\nl
&3&$17.9\pm2.6$&$ 0.09\pm0.03$&$ 0.13\pm0.01$&$-0.01\pm0.03$&$ 0.12\pm0.03$\nl
&4&$15.8\pm2.9$&$ 0.18\pm0.04$&$ 0.20\pm0.01$&$-0.01\pm0.04$&$ 0.19\pm0.04$\nl
\tablevspace{2pt}
NGC 4374&1&$13.2\pm1.6$&$ 0.10\pm0.03$&$ 0.20\pm0.01$&$-0.08\pm0.03$&$ 0.12\pm0.03$\nl
&2&$13.7\pm1.8$&$ 0.08\pm0.03$&$ 0.20\pm0.01$&$-0.07\pm0.03$&$ 0.13\pm0.03$\nl
&3&$15.1\pm1.6$&$ 0.00\pm0.03$&$ 0.14\pm0.01$&$-0.11\pm0.03$&$ 0.03\pm0.03$\nl
&4&$12.7\pm2.0$&$ 0.12\pm0.03$&$ 0.20\pm0.01$&$-0.07\pm0.03$&$ 0.13\pm0.03$\nl
\tablevspace{2pt}
NGC 4472&1&$ 8.0\pm2.0$&$ 0.24\pm0.04$&$ 0.20\pm0.02$&$ 0.06\pm0.04$&$ 0.26\pm0.04$\nl
&2&$ 8.3\pm2.1$&$ 0.22\pm0.04$&$ 0.21\pm0.02$&$ 0.06\pm0.04$&$ 0.27\pm0.04$\nl
&3&$10.1\pm2.2$&$ 0.14\pm0.04$&$ 0.14\pm0.01$&$ 0.03\pm0.04$&$ 0.17\pm0.04$\nl
&4&$ 7.9\pm2.1$&$ 0.25\pm0.05$&$ 0.21\pm0.02$&$ 0.05\pm0.05$&$ 0.26\pm0.05$\nl
\tablevspace{2pt}
NGC 4478&1&$ 4.7\pm2.0$&$ 0.28\pm0.09$&$ 0.14\pm0.03$&$ 0.15\pm0.09$&$ 0.29\pm0.09$\nl
&2&$ 4.8\pm2.3$&$ 0.26\pm0.09$&$ 0.14\pm0.03$&$ 0.15\pm0.09$&$ 0.29\pm0.09$\nl
&3&$ 5.5\pm2.1$&$ 0.21\pm0.07$&$ 0.09\pm0.02$&$ 0.14\pm0.07$&$ 0.23\pm0.07$\nl
&4&$ 4.6\pm2.3$&$ 0.29\pm0.10$&$ 0.15\pm0.03$&$ 0.15\pm0.10$&$ 0.30\pm0.10$\nl
\tablevspace{2pt}
NGC 4489&1&$ 2.5\pm0.4$&$ 0.14\pm0.06$&$ 0.03\pm0.02$&$ 0.11\pm0.06$&$ 0.14\pm0.06$\nl
&2&$ 2.5\pm0.4$&$ 0.14\pm0.06$&$ 0.02\pm0.02$&$ 0.12\pm0.06$&$ 0.14\pm0.06$\nl
&3&$ 2.5\pm0.4$&$ 0.13\pm0.06$&$ 0.02\pm0.02$&$ 0.11\pm0.06$&$ 0.13\pm0.06$\nl
&4&$ 2.5\pm0.4$&$ 0.14\pm0.06$&$ 0.03\pm0.02$&$ 0.11\pm0.06$&$ 0.14\pm0.06$\nl
\tablevspace{2pt}
NGC 4552&1&$10.6\pm1.5$&$ 0.25\pm0.04$&$ 0.22\pm0.01$&$ 0.05\pm0.04$&$ 0.27\pm0.04$\nl
&2&$11.4\pm2.1$&$ 0.22\pm0.04$&$ 0.22\pm0.01$&$ 0.05\pm0.04$&$ 0.27\pm0.04$\nl
&3&$12.8\pm2.1$&$ 0.15\pm0.03$&$ 0.15\pm0.01$&$ 0.04\pm0.03$&$ 0.19\pm0.03$\nl
&4&$10.5\pm1.4$&$ 0.27\pm0.04$&$ 0.23\pm0.01$&$ 0.06\pm0.04$&$ 0.29\pm0.04$\nl
\tablevspace{2pt}
NGC 4649&1&$11.9\pm2.8$&$ 0.25\pm0.04$&$ 0.23\pm0.01$&$ 0.04\pm0.04$&$ 0.27\pm0.04$\nl
&2&$12.7\pm2.3$&$ 0.23\pm0.03$&$ 0.23\pm0.01$&$ 0.05\pm0.03$&$ 0.28\pm0.03$\nl
&3&$15.0\pm2.9$&$ 0.15\pm0.03$&$ 0.16\pm0.01$&$ 0.03\pm0.03$&$ 0.19\pm0.03$\nl
&4&$11.9\pm2.6$&$ 0.27\pm0.04$&$ 0.24\pm0.01$&$ 0.05\pm0.04$&$ 0.29\pm0.04$\nl
\tablevspace{2pt}
NGC 4697&1&$ 8.9\pm2.0$&$ 0.05\pm0.05$&$ 0.10\pm0.01$&$-0.04\pm0.05$&$ 0.06\pm0.05$\nl
&2&$ 9.1\pm2.0$&$ 0.04\pm0.05$&$ 0.10\pm0.01$&$-0.04\pm0.05$&$ 0.06\pm0.05$\nl
&3&$ 9.5\pm1.9$&$ 0.00\pm0.05$&$ 0.07\pm0.01$&$-0.05\pm0.05$&$ 0.02\pm0.05$\nl
&4&$ 8.8\pm2.0$&$ 0.06\pm0.05$&$ 0.10\pm0.02$&$-0.03\pm0.05$&$ 0.07\pm0.05$\nl
\tablevspace{2pt}
NGC 5638&1&$ 8.7\pm1.3$&$ 0.18\pm0.03$&$ 0.18\pm0.01$&$ 0.02\pm0.03$&$ 0.20\pm0.03$\nl
&2&$ 9.5\pm1.5$&$ 0.15\pm0.03$&$ 0.18\pm0.01$&$ 0.01\pm0.03$&$ 0.19\pm0.03$\nl
&3&$10.1\pm1.4$&$ 0.09\pm0.03$&$ 0.13\pm0.01$&$-0.01\pm0.03$&$ 0.12\pm0.03$\nl
&4&$ 8.5\pm1.5$&$ 0.19\pm0.03$&$ 0.19\pm0.01$&$ 0.01\pm0.03$&$ 0.20\pm0.03$\nl
\tablevspace{2pt}
NGC 5812&1&$ 5.5\pm1.0$&$ 0.37\pm0.03$&$ 0.19\pm0.01$&$ 0.20\pm0.03$&$ 0.39\pm0.03$\nl
&2&$ 5.9\pm1.0$&$ 0.34\pm0.03$&$ 0.19\pm0.01$&$ 0.19\pm0.03$&$ 0.38\pm0.03$\nl
&3&$ 6.7\pm1.1$&$ 0.27\pm0.03$&$ 0.12\pm0.01$&$ 0.18\pm0.03$&$ 0.30\pm0.03$\nl
&4&$ 5.4\pm1.0$&$ 0.38\pm0.04$&$ 0.20\pm0.01$&$ 0.19\pm0.04$&$ 0.39\pm0.04$\nl
\tablevspace{2pt}
NGC 5813&1&$18.7\pm2.3$&$-0.05\pm0.03$&$ 0.20\pm0.01$&$-0.23\pm0.03$&$-0.03\pm0.03$\nl
&2&$18.9\pm2.4$&$-0.07\pm0.03$&$ 0.21\pm0.02$&$-0.23\pm0.03$&$-0.02\pm0.03$\nl
&3&$19.4\pm2.0$&$-0.12\pm0.03$&$ 0.14\pm0.01$&$-0.23\pm0.03$&$-0.09\pm0.03$\nl
&4&$18.5\pm2.3$&$-0.04\pm0.03$&$ 0.21\pm0.01$&$-0.24\pm0.03$&$-0.03\pm0.03$\nl
\tablevspace{2pt}
NGC 5831&1&$ 2.7\pm0.4$&$ 0.51\pm0.05$&$ 0.18\pm0.01$&$ 0.35\pm0.05$&$ 0.53\pm0.05$\nl
&2&$ 2.7\pm0.3$&$ 0.50\pm0.03$&$ 0.18\pm0.01$&$ 0.36\pm0.03$&$ 0.54\pm0.03$\nl
&3&$ 2.9\pm0.5$&$ 0.43\pm0.05$&$ 0.12\pm0.01$&$ 0.34\pm0.05$&$ 0.46\pm0.05$\nl
&4&$ 2.6\pm0.3$&$ 0.54\pm0.04$&$ 0.19\pm0.01$&$ 0.36\pm0.04$&$ 0.55\pm0.04$\nl
\tablevspace{2pt}
NGC 5846&1&$14.8\pm3.3$&$ 0.12\pm0.05$&$ 0.20\pm0.02$&$-0.06\pm0.05$&$ 0.14\pm0.05$\nl
&2&$14.9\pm3.3$&$ 0.10\pm0.05$&$ 0.21\pm0.01$&$-0.06\pm0.05$&$ 0.15\pm0.05$\nl
&3&$16.2\pm2.9$&$ 0.03\pm0.04$&$ 0.14\pm0.01$&$-0.08\pm0.04$&$ 0.06\pm0.04$\nl
&4&$14.1\pm2.9$&$ 0.14\pm0.05$&$ 0.21\pm0.02$&$-0.06\pm0.05$&$ 0.15\pm0.05$\nl
\tablevspace{2pt}
NGC 6127&1&$11.9\pm2.9$&$ 0.16\pm0.04$&$ 0.22\pm0.01$&$-0.04\pm0.04$&$ 0.18\pm0.04$\nl
&2&$12.9\pm2.0$&$ 0.13\pm0.03$&$ 0.22\pm0.02$&$-0.04\pm0.03$&$ 0.18\pm0.03$\nl
&3&$14.7\pm2.6$&$ 0.05\pm0.04$&$ 0.15\pm0.01$&$-0.06\pm0.04$&$ 0.09\pm0.04$\nl
&4&$11.8\pm2.3$&$ 0.17\pm0.03$&$ 0.23\pm0.01$&$-0.04\pm0.03$&$ 0.19\pm0.03$\nl
\tablevspace{2pt}
NGC 6702&1&$ 1.5\pm0.2$&$ 0.69\pm0.11$&$ 0.15\pm0.03$&$ 0.55\pm0.11$&$ 0.70\pm0.11$\nl
&2&$ 1.5\pm0.2$&$ 0.68\pm0.11$&$ 0.15\pm0.02$&$ 0.56\pm0.11$&$ 0.71\pm0.11$\nl
&3&$ 1.6\pm0.1$&$ 0.58\pm0.05$&$ 0.10\pm0.01$&$ 0.50\pm0.05$&$ 0.60\pm0.05$\nl
&4&$ 1.5\pm0.1$&$ 0.70\pm0.06$&$ 0.15\pm0.03$&$ 0.56\pm0.07$&$ 0.71\pm0.06$\nl
\tablevspace{2pt}
NGC 6703&1&$ 4.3\pm0.6$&$ 0.30\pm0.06$&$ 0.15\pm0.02$&$ 0.16\pm0.06$&$ 0.31\pm0.06$\nl
&2&$ 4.4\pm0.9$&$ 0.28\pm0.05$&$ 0.15\pm0.02$&$ 0.16\pm0.05$&$ 0.31\pm0.05$\nl
&3&$ 4.8\pm1.2$&$ 0.22\pm0.06$&$ 0.10\pm0.01$&$ 0.14\pm0.06$&$ 0.24\pm0.06$\nl
&4&$ 4.4\pm0.7$&$ 0.30\pm0.06$&$ 0.15\pm0.02$&$ 0.16\pm0.06$&$ 0.31\pm0.06$\nl
\tablevspace{2pt}
NGC 7052&1&$12.9\pm3.4$&$ 0.15\pm0.05$&$ 0.23\pm0.01$&$-0.06\pm0.05$&$ 0.17\pm0.05$\nl
&2&$13.9\pm2.9$&$ 0.12\pm0.05$&$ 0.23\pm0.02$&$-0.06\pm0.05$&$ 0.17\pm0.05$\nl
&3&$15.3\pm3.2$&$ 0.04\pm0.04$&$ 0.16\pm0.01$&$-0.08\pm0.04$&$ 0.08\pm0.04$\nl
&4&$12.7\pm3.0$&$ 0.16\pm0.05$&$ 0.24\pm0.01$&$-0.06\pm0.05$&$ 0.18\pm0.05$\nl
\tablevspace{2pt}
NGC 7454&1&$ 5.1\pm0.9$&$-0.07\pm0.03$&$ 0.06\pm0.02$&$-0.12\pm0.04$&$-0.06\pm0.03$\nl
&2&$ 5.1\pm1.0$&$-0.07\pm0.04$&$ 0.06\pm0.02$&$-0.12\pm0.04$&$-0.06\pm0.04$\nl
&3&$ 5.2\pm1.0$&$-0.09\pm0.03$&$ 0.04\pm0.01$&$-0.12\pm0.03$&$-0.08\pm0.03$\nl
&4&$ 5.0\pm1.0$&$-0.06\pm0.04$&$ 0.06\pm0.02$&$-0.12\pm0.04$&$-0.06\pm0.04$\nl
\tablevspace{2pt}
NGC 7562&1&$ 7.9\pm1.3$&$ 0.19\pm0.03$&$ 0.16\pm0.01$&$ 0.04\pm0.03$&$ 0.20\pm0.03$\nl
&2&$ 8.2\pm1.4$&$ 0.17\pm0.04$&$ 0.16\pm0.02$&$ 0.05\pm0.04$&$ 0.21\pm0.04$\nl
&3&$ 9.3\pm1.7$&$ 0.11\pm0.04$&$ 0.11\pm0.01$&$ 0.03\pm0.04$&$ 0.14\pm0.04$\nl
&4&$ 7.7\pm1.4$&$ 0.20\pm0.04$&$ 0.17\pm0.01$&$ 0.04\pm0.04$&$ 0.21\pm0.04$\nl
\tablevspace{2pt}
NGC 7619&1&$15.1\pm2.6$&$ 0.19\pm0.03$&$ 0.17\pm0.01$&$ 0.03\pm0.03$&$ 0.20\pm0.03$\nl
&2&$15.8\pm2.4$&$ 0.17\pm0.03$&$ 0.17\pm0.01$&$ 0.04\pm0.03$&$ 0.21\pm0.03$\nl
&3&$16.5\pm1.5$&$ 0.12\pm0.03$&$ 0.12\pm0.01$&$ 0.03\pm0.03$&$ 0.15\pm0.03$\nl
&4&$14.8\pm2.3$&$ 0.20\pm0.03$&$ 0.18\pm0.01$&$ 0.03\pm0.03$&$ 0.21\pm0.03$\nl
\tablevspace{2pt}
NGC 7626&1&$13.6\pm2.3$&$ 0.14\pm0.03$&$ 0.24\pm0.01$&$-0.08\pm0.03$&$ 0.16\pm0.03$\nl
&2&$14.6\pm2.6$&$ 0.11\pm0.04$&$ 0.24\pm0.01$&$-0.08\pm0.04$&$ 0.16\pm0.04$\nl
&3&$16.2\pm2.4$&$ 0.03\pm0.03$&$ 0.16\pm0.01$&$-0.09\pm0.03$&$ 0.07\pm0.03$\nl
&4&$12.9\pm2.9$&$ 0.16\pm0.04$&$ 0.25\pm0.01$&$-0.07\pm0.04$&$ 0.18\pm0.04$\nl
\tablevspace{2pt}
NGC 7785&1&$ 8.8\pm1.9$&$ 0.19\pm0.04$&$ 0.16\pm0.01$&$ 0.04\pm0.04$&$ 0.20\pm0.04$\nl
&2&$ 9.2\pm1.9$&$ 0.17\pm0.04$&$ 0.16\pm0.01$&$ 0.05\pm0.04$&$ 0.21\pm0.04$\nl
&3&$ 9.9\pm1.9$&$ 0.12\pm0.04$&$ 0.11\pm0.01$&$ 0.04\pm0.04$&$ 0.15\pm0.04$\nl
&4&$ 8.7\pm1.9$&$ 0.20\pm0.04$&$ 0.16\pm0.01$&$ 0.05\pm0.04$&$ 0.21\pm0.04$\nl
\enddata
\end{deluxetable}

\begin{deluxetable}{lcrrrrr}
\tablenum{6b}
\tablecaption{Ages, metallicities and enhancement ratios through the
global $r_e/2$ aperture\label{tbl:re2tza}}
\tablewidth{0pt}
\tablehead{\colhead{Name}&\colhead{Model}&\colhead{Age (Gyr)}&
\colhead{\z}&\colhead{\enh}&\colhead{\feh}&\colhead{$[\mathrm{E/H}]$}}
\startdata
NGC  221&1&$ 4.9\pm1.1$&$-0.08\pm0.04$&$-0.07\pm0.02$&$-0.02\pm0.04$&$-0.09\pm0.04$\nl
&2&$ 4.9\pm1.0$&$-0.07\pm0.04$&$-0.07\pm0.02$&$-0.02\pm0.04$&$-0.09\pm0.04$\nl
&3&$ 4.9\pm1.0$&$-0.06\pm0.03$&$-0.05\pm0.01$&$-0.02\pm0.03$&$-0.07\pm0.03$\nl
&4&$ 4.9\pm1.3$&$-0.08\pm0.05$&$-0.07\pm0.02$&$-0.01\pm0.05$&$-0.08\pm0.05$\nl
\tablevspace{2pt}
NGC  315&1&$ 6.9\pm2.4$&$ 0.16\pm0.07$&$ 0.23\pm0.02$&$-0.05\pm0.07$&$ 0.18\pm0.07$\nl
&2&$ 7.3\pm2.2$&$ 0.13\pm0.06$&$ 0.23\pm0.02$&$-0.05\pm0.06$&$ 0.18\pm0.06$\nl
&3&$ 8.3\pm2.4$&$ 0.04\pm0.06$&$ 0.15\pm0.02$&$-0.07\pm0.06$&$ 0.08\pm0.06$\nl
&4&$ 6.9\pm1.9$&$ 0.17\pm0.06$&$ 0.24\pm0.02$&$-0.05\pm0.06$&$ 0.19\pm0.06$\nl
\tablevspace{2pt}
NGC  507&1&$ 3.5\pm3.2$&$ 0.27\pm0.15$&$ 0.37\pm0.04$&$-0.07\pm0.15$&$ 0.30\pm0.15$\nl
&2&$ 4.1\pm3.1$&$ 0.20\pm0.13$&$ 0.37\pm0.04$&$-0.09\pm0.13$&$ 0.28\pm0.13$\nl
&3&$ 4.9\pm3.0$&$ 0.08\pm0.12$&$ 0.25\pm0.02$&$-0.11\pm0.12$&$ 0.14\pm0.12$\nl
&4&$ 3.5\pm2.7$&$ 0.28\pm0.13$&$ 0.39\pm0.04$&$-0.08\pm0.14$&$ 0.31\pm0.13$\nl
\tablevspace{2pt}
NGC  547&1&$18.7\pm3.8$&$-0.04\pm0.06$&$ 0.24\pm0.02$&$-0.26\pm0.06$&$-0.02\pm0.06$\nl
&2&$18.6\pm3.3$&$-0.05\pm0.04$&$ 0.24\pm0.02$&$-0.24\pm0.04$&$ 0.00\pm0.04$\nl
&3&$19.7\pm3.2$&$-0.13\pm0.05$&$ 0.17\pm0.02$&$-0.26\pm0.05$&$-0.09\pm0.05$\nl
&4&$18.1\pm3.2$&$-0.02\pm0.05$&$ 0.25\pm0.02$&$-0.25\pm0.05$&$ 0.00\pm0.05$\nl
\tablevspace{2pt}
NGC  584&1&$ 3.4\pm1.2$&$ 0.25\pm0.07$&$ 0.20\pm0.02$&$ 0.07\pm0.07$&$ 0.27\pm0.07$\nl
&2&$ 3.8\pm0.9$&$ 0.21\pm0.06$&$ 0.20\pm0.02$&$ 0.06\pm0.06$&$ 0.26\pm0.06$\nl
&3&$ 4.4\pm1.1$&$ 0.13\pm0.06$&$ 0.14\pm0.02$&$ 0.02\pm0.06$&$ 0.16\pm0.06$\nl
&4&$ 3.4\pm1.1$&$ 0.26\pm0.07$&$ 0.21\pm0.02$&$ 0.06\pm0.07$&$ 0.27\pm0.07$\nl
\tablevspace{2pt}
NGC  636&1&$ 6.7\pm1.3$&$ 0.10\pm0.06$&$ 0.11\pm0.02$&$ 0.00\pm0.06$&$ 0.11\pm0.06$\nl
&2&$ 6.8\pm1.3$&$ 0.09\pm0.06$&$ 0.11\pm0.02$&$ 0.00\pm0.06$&$ 0.11\pm0.06$\nl
&3&$ 7.2\pm1.3$&$ 0.05\pm0.05$&$ 0.07\pm0.01$&$ 0.00\pm0.05$&$ 0.07\pm0.05$\nl
&4&$ 6.8\pm1.4$&$ 0.10\pm0.05$&$ 0.11\pm0.02$&$ 0.00\pm0.05$&$ 0.11\pm0.05$\nl
\tablevspace{2pt}
NGC  720&1&$ 1.5\pm0.4$&$ 0.90\pm0.23$&$ 0.41\pm0.16$&$ 0.53\pm0.27$&$ 0.94\pm0.23$\nl
&2&$ 1.7\pm0.4$&$ 0.75\pm0.21$&$ 0.40\pm0.08$&$ 0.44\pm0.22$&$ 0.84\pm0.21$\nl
&3&$ 1.8\pm0.7$&$ 0.59\pm0.15$&$ 0.27\pm0.03$&$ 0.39\pm0.15$&$ 0.66\pm0.15$\nl
&4&$ 1.1\pm0.8$&$ 1.13\pm0.42$&$ 0.56\pm0.18$&$ 0.61\pm0.45$&$ 1.17\pm0.42$\nl
\tablevspace{2pt}
NGC  821&1&$ 7.4\pm1.4$&$ 0.10\pm0.05$&$ 0.12\pm0.02$&$-0.01\pm0.05$&$ 0.11\pm0.05$\nl
&2&$ 7.6\pm1.4$&$ 0.08\pm0.05$&$ 0.12\pm0.02$&$-0.01\pm0.05$&$ 0.11\pm0.05$\nl
&3&$ 7.9\pm1.4$&$ 0.04\pm0.04$&$ 0.08\pm0.02$&$-0.02\pm0.04$&$ 0.06\pm0.04$\nl
&4&$ 7.1\pm1.2$&$ 0.12\pm0.05$&$ 0.13\pm0.02$&$ 0.00\pm0.05$&$ 0.13\pm0.05$\nl
\tablevspace{2pt}
NGC 1453&1&$ 8.7\pm2.6$&$ 0.14\pm0.06$&$ 0.15\pm0.02$&$ 0.00\pm0.06$&$ 0.15\pm0.06$\nl
&2&$ 9.1\pm2.5$&$ 0.12\pm0.06$&$ 0.15\pm0.02$&$ 0.00\pm0.06$&$ 0.15\pm0.06$\nl
&3&$ 9.9\pm2.6$&$ 0.07\pm0.06$&$ 0.10\pm0.02$&$-0.01\pm0.06$&$ 0.09\pm0.06$\nl
&4&$ 8.5\pm2.7$&$ 0.15\pm0.07$&$ 0.16\pm0.02$&$ 0.00\pm0.07$&$ 0.16\pm0.07$\nl
\tablevspace{2pt}
NGC 1600&1&$ 4.4\pm1.8$&$ 0.50\pm0.14$&$ 0.28\pm0.03$&$ 0.24\pm0.14$&$ 0.52\pm0.14$\nl
&2&$ 4.6\pm1.9$&$ 0.46\pm0.13$&$ 0.28\pm0.03$&$ 0.24\pm0.13$&$ 0.52\pm0.13$\nl
&3&$ 5.2\pm1.8$&$ 0.35\pm0.08$&$ 0.18\pm0.02$&$ 0.21\pm0.08$&$ 0.39\pm0.08$\nl
&4&$ 4.3\pm1.5$&$ 0.53\pm0.13$&$ 0.29\pm0.04$&$ 0.26\pm0.14$&$ 0.55\pm0.13$\nl
\tablevspace{2pt}
NGC 1700&1&$ 2.8\pm0.6$&$ 0.32\pm0.06$&$ 0.13\pm0.02$&$ 0.20\pm0.06$&$ 0.33\pm0.06$\nl
&2&$ 2.9\pm0.5$&$ 0.29\pm0.04$&$ 0.13\pm0.02$&$ 0.19\pm0.04$&$ 0.32\pm0.04$\nl
&3&$ 3.0\pm0.9$&$ 0.25\pm0.05$&$ 0.08\pm0.02$&$ 0.19\pm0.05$&$ 0.27\pm0.05$\nl
&4&$ 2.8\pm0.5$&$ 0.32\pm0.05$&$ 0.13\pm0.02$&$ 0.20\pm0.05$&$ 0.33\pm0.05$\nl
\tablevspace{2pt}
NGC 2300&1&$10.2\pm1.9$&$ 0.13\pm0.04$&$ 0.21\pm0.02$&$-0.06\pm0.04$&$ 0.15\pm0.04$\nl
&2&$10.6\pm1.8$&$ 0.10\pm0.04$&$ 0.22\pm0.02$&$-0.07\pm0.04$&$ 0.15\pm0.04$\nl
&3&$11.5\pm2.3$&$ 0.02\pm0.04$&$ 0.15\pm0.02$&$-0.09\pm0.04$&$ 0.06\pm0.04$\nl
&4&$10.1\pm2.0$&$ 0.14\pm0.04$&$ 0.22\pm0.02$&$-0.06\pm0.04$&$ 0.16\pm0.04$\nl
\tablevspace{2pt}
NGC 2778&1&$15.1\pm3.6$&$-0.05\pm0.06$&$ 0.18\pm0.02$&$-0.21\pm0.06$&$-0.03\pm0.06$\nl
&2&$15.1\pm3.4$&$-0.06\pm0.05$&$ 0.18\pm0.02$&$-0.20\pm0.05$&$-0.02\pm0.05$\nl
&3&$15.7\pm3.6$&$-0.12\pm0.06$&$ 0.13\pm0.02$&$-0.22\pm0.06$&$-0.09\pm0.06$\nl
&4&$14.9\pm3.5$&$-0.04\pm0.06$&$ 0.19\pm0.02$&$-0.22\pm0.06$&$-0.03\pm0.06$\nl
\tablevspace{2pt}
NGC 3377&1&$ 6.0\pm1.2$&$-0.13\pm0.05$&$ 0.16\pm0.02$&$-0.28\pm0.05$&$-0.12\pm0.05$\nl
&2&$ 6.1\pm1.1$&$-0.14\pm0.04$&$ 0.16\pm0.02$&$-0.26\pm0.04$&$-0.10\pm0.04$\nl
&3&$ 6.3\pm1.3$&$-0.18\pm0.04$&$ 0.11\pm0.01$&$-0.26\pm0.04$&$-0.15\pm0.04$\nl
&4&$ 5.9\pm1.2$&$-0.12\pm0.04$&$ 0.17\pm0.02$&$-0.28\pm0.04$&$-0.11\pm0.04$\nl
\tablevspace{2pt}
NGC 3379&1&$13.4\pm2.4$&$-0.01\pm0.04$&$ 0.17\pm0.02$&$-0.17\pm0.04$&$ 0.00\pm0.04$\nl
&2&$13.7\pm2.1$&$-0.03\pm0.03$&$ 0.18\pm0.02$&$-0.17\pm0.03$&$ 0.01\pm0.03$\nl
&3&$14.5\pm2.5$&$-0.08\pm0.03$&$ 0.12\pm0.01$&$-0.17\pm0.03$&$-0.05\pm0.03$\nl
&4&$13.2\pm2.4$&$ 0.00\pm0.04$&$ 0.18\pm0.02$&$-0.17\pm0.04$&$ 0.01\pm0.04$\nl
\tablevspace{2pt}
NGC 3608&1&$ 8.9\pm2.8$&$ 0.07\pm0.07$&$ 0.07\pm0.02$&$ 0.01\pm0.07$&$ 0.08\pm0.07$\nl
&2&$ 9.1\pm2.5$&$ 0.06\pm0.07$&$ 0.07\pm0.02$&$ 0.01\pm0.07$&$ 0.08\pm0.07$\nl
&3&$ 9.2\pm2.5$&$ 0.04\pm0.06$&$ 0.05\pm0.02$&$ 0.00\pm0.06$&$ 0.05\pm0.06$\nl
&4&$ 9.0\pm2.5$&$ 0.07\pm0.07$&$ 0.07\pm0.02$&$ 0.00\pm0.07$&$ 0.07\pm0.07$\nl
\tablevspace{2pt}
NGC 3818&1&$ 8.2\pm3.0$&$ 0.04\pm0.09$&$ 0.16\pm0.03$&$-0.11\pm0.09$&$ 0.05\pm0.09$\nl
&2&$ 8.3\pm2.9$&$ 0.02\pm0.09$&$ 0.17\pm0.03$&$-0.11\pm0.09$&$ 0.06\pm0.09$\nl
&3&$ 9.2\pm2.4$&$-0.03\pm0.07$&$ 0.11\pm0.02$&$-0.11\pm0.07$&$ 0.00\pm0.07$\nl
&4&$ 8.0\pm3.1$&$ 0.05\pm0.09$&$ 0.17\pm0.03$&$-0.11\pm0.09$&$ 0.06\pm0.09$\nl
\tablevspace{2pt}
NGC 4261&1&$21.4\pm2.2$&$-0.03\pm0.03$&$ 0.19\pm0.02$&$-0.20\pm0.04$&$-0.01\pm0.03$\nl
&2&$21.4\pm1.7$&$-0.04\pm0.02$&$ 0.19\pm0.01$&$-0.19\pm0.02$&$ 0.00\pm0.02$\nl
&3&$21.8\pm1.6$&$-0.09\pm0.03$&$ 0.13\pm0.01$&$-0.19\pm0.03$&$-0.06\pm0.03$\nl
&4&$21.0\pm1.7$&$-0.01\pm0.03$&$ 0.19\pm0.02$&$-0.19\pm0.04$&$ 0.00\pm0.03$\nl
\tablevspace{2pt}
NGC 4374&1&$14.2\pm2.3$&$-0.01\pm0.04$&$ 0.18\pm0.02$&$-0.17\pm0.04$&$ 0.01\pm0.04$\nl
&2&$14.4\pm2.4$&$-0.03\pm0.04$&$ 0.19\pm0.02$&$-0.18\pm0.04$&$ 0.01\pm0.04$\nl
&3&$15.3\pm2.3$&$-0.09\pm0.04$&$ 0.13\pm0.01$&$-0.19\pm0.04$&$-0.06\pm0.04$\nl
&4&$14.4\pm2.8$&$-0.01\pm0.05$&$ 0.19\pm0.01$&$-0.19\pm0.05$&$ 0.00\pm0.05$\nl
\tablevspace{2pt}
NGC 4472&1&$ 8.6\pm2.9$&$ 0.17\pm0.07$&$ 0.18\pm0.02$&$ 0.01\pm0.07$&$ 0.19\pm0.07$\nl
&2&$ 9.1\pm2.7$&$ 0.14\pm0.06$&$ 0.19\pm0.03$&$-0.01\pm0.06$&$ 0.18\pm0.06$\nl
&3&$ 9.9\pm2.7$&$ 0.08\pm0.06$&$ 0.13\pm0.02$&$-0.02\pm0.06$&$ 0.11\pm0.06$\nl
&4&$ 8.4\pm2.7$&$ 0.18\pm0.06$&$ 0.19\pm0.02$&$ 0.00\pm0.06$&$ 0.19\pm0.06$\nl
\tablevspace{2pt}
NGC 4478&1&$10.5\pm1.8$&$-0.02\pm0.04$&$ 0.15\pm0.02$&$-0.16\pm0.04$&$-0.01\pm0.04$\nl
&2&$10.6\pm1.6$&$-0.03\pm0.03$&$ 0.15\pm0.02$&$-0.15\pm0.03$&$ 0.00\pm0.03$\nl
&3&$11.1\pm1.6$&$-0.07\pm0.04$&$ 0.10\pm0.02$&$-0.15\pm0.04$&$-0.05\pm0.04$\nl
&4&$10.3\pm1.7$&$-0.01\pm0.04$&$ 0.16\pm0.02$&$-0.16\pm0.04$&$ 0.00\pm0.04$\nl
\tablevspace{2pt}
NGC 4489&1&$ 4.6\pm0.5$&$-0.15\pm0.04$&$-0.04\pm0.03$&$-0.11\pm0.05$&$-0.15\pm0.04$\nl
&2&$ 4.6\pm0.5$&$-0.15\pm0.04$&$-0.04\pm0.04$&$-0.12\pm0.05$&$-0.16\pm0.04$\nl
&3&$ 4.6\pm0.6$&$-0.14\pm0.05$&$-0.03\pm0.02$&$-0.12\pm0.05$&$-0.15\pm0.05$\nl
&4&$ 4.6\pm0.5$&$-0.15\pm0.04$&$-0.04\pm0.04$&$-0.11\pm0.05$&$-0.15\pm0.04$\nl
\tablevspace{2pt}
NGC 4552&1&$13.3\pm2.3$&$ 0.09\pm0.04$&$ 0.21\pm0.02$&$-0.10\pm0.04$&$ 0.11\pm0.04$\nl
&2&$13.9\pm2.5$&$ 0.06\pm0.04$&$ 0.22\pm0.02$&$-0.11\pm0.04$&$ 0.11\pm0.04$\nl
&3&$15.3\pm2.3$&$-0.02\pm0.04$&$ 0.15\pm0.01$&$-0.13\pm0.04$&$ 0.02\pm0.04$\nl
&4&$13.1\pm2.7$&$ 0.10\pm0.04$&$ 0.22\pm0.02$&$-0.10\pm0.04$&$ 0.12\pm0.04$\nl
\tablevspace{2pt}
NGC 4649&1&$18.2\pm2.5$&$ 0.04\pm0.04$&$ 0.28\pm0.02$&$-0.22\pm0.04$&$ 0.06\pm0.04$\nl
&2&$19.2\pm2.8$&$ 0.00\pm0.04$&$ 0.29\pm0.02$&$-0.22\pm0.04$&$ 0.07\pm0.04$\nl
&3&$20.0\pm2.5$&$-0.08\pm0.03$&$ 0.20\pm0.01$&$-0.23\pm0.03$&$-0.03\pm0.03$\nl
&4&$18.3\pm2.8$&$ 0.05\pm0.04$&$ 0.29\pm0.02$&$-0.22\pm0.04$&$ 0.07\pm0.04$\nl
\tablevspace{2pt}
NGC 4697&1&$15.6\pm2.2$&$-0.29\pm0.03$&$ 0.08\pm0.03$&$-0.36\pm0.04$&$-0.28\pm0.03$\nl
&2&$15.5\pm2.2$&$-0.29\pm0.03$&$ 0.08\pm0.03$&$-0.35\pm0.04$&$-0.27\pm0.03$\nl
&3&$15.6\pm2.1$&$-0.31\pm0.04$&$ 0.06\pm0.02$&$-0.36\pm0.04$&$-0.30\pm0.04$\nl
&4&$15.6\pm2.5$&$-0.29\pm0.04$&$ 0.09\pm0.03$&$-0.37\pm0.05$&$-0.28\pm0.04$\nl
\tablevspace{2pt}
NGC 5638&1&$11.6\pm2.0$&$-0.05\pm0.03$&$ 0.13\pm0.02$&$-0.17\pm0.04$&$-0.04\pm0.03$\nl
&2&$11.7\pm2.1$&$-0.06\pm0.03$&$ 0.13\pm0.02$&$-0.16\pm0.03$&$-0.03\pm0.03$\nl
&3&$12.1\pm2.5$&$-0.10\pm0.03$&$ 0.09\pm0.01$&$-0.17\pm0.03$&$-0.08\pm0.03$\nl
&4&$11.6\pm2.1$&$-0.04\pm0.04$&$ 0.13\pm0.02$&$-0.16\pm0.04$&$-0.03\pm0.04$\nl
\tablevspace{2pt}
NGC 5812&1&$ 7.2\pm1.4$&$ 0.22\pm0.04$&$ 0.16\pm0.02$&$ 0.07\pm0.04$&$ 0.23\pm0.04$\nl
&2&$ 7.5\pm1.1$&$ 0.20\pm0.03$&$ 0.16\pm0.02$&$ 0.08\pm0.03$&$ 0.24\pm0.03$\nl
&3&$ 8.2\pm1.3$&$ 0.14\pm0.04$&$ 0.11\pm0.01$&$ 0.06\pm0.04$&$ 0.17\pm0.04$\nl
&4&$ 7.0\pm1.3$&$ 0.23\pm0.04$&$ 0.17\pm0.02$&$ 0.07\pm0.04$&$ 0.24\pm0.04$\nl
\tablevspace{2pt}
NGC 5813&1&$24.3\pm2.1$&$-0.21\pm0.05$&$ 0.12\pm0.03$&$-0.32\pm0.06$&$-0.20\pm0.05$\nl
&2&$24.4\pm2.0$&$-0.22\pm0.05$&$ 0.12\pm0.03$&$-0.31\pm0.06$&$-0.19\pm0.05$\nl
&3&$24.7\pm2.0$&$-0.27\pm0.05$&$ 0.09\pm0.02$&$-0.34\pm0.05$&$-0.25\pm0.05$\nl
&4&$24.3\pm2.0$&$-0.20\pm0.06$&$ 0.12\pm0.03$&$-0.31\pm0.07$&$-0.19\pm0.06$\nl
\tablevspace{2pt}
NGC 5831&1&$ 4.3\pm1.2$&$ 0.17\pm0.07$&$ 0.12\pm0.02$&$ 0.06\pm0.07$&$ 0.18\pm0.07$\nl
&2&$ 4.5\pm1.2$&$ 0.15\pm0.07$&$ 0.12\pm0.02$&$ 0.06\pm0.07$&$ 0.18\pm0.07$\nl
&3&$ 4.8\pm1.3$&$ 0.11\pm0.04$&$ 0.08\pm0.01$&$ 0.05\pm0.04$&$ 0.13\pm0.04$\nl
&4&$ 4.2\pm1.0$&$ 0.18\pm0.06$&$ 0.13\pm0.02$&$ 0.06\pm0.06$&$ 0.19\pm0.06$\nl
\tablevspace{2pt}
NGC 5846&1&$23.1\pm2.5$&$-0.14\pm0.04$&$ 0.19\pm0.03$&$-0.31\pm0.05$&$-0.12\pm0.04$\nl
&2&$23.4\pm2.4$&$-0.16\pm0.04$&$ 0.19\pm0.03$&$-0.31\pm0.05$&$-0.12\pm0.04$\nl
&3&$23.6\pm2.3$&$-0.21\pm0.06$&$ 0.13\pm0.02$&$-0.31\pm0.06$&$-0.18\pm0.06$\nl
&4&$23.1\pm2.7$&$-0.13\pm0.04$&$ 0.19\pm0.03$&$-0.31\pm0.05$&$-0.12\pm0.04$\nl
\tablevspace{2pt}
NGC 6127&1&$16.2\pm2.6$&$-0.02\pm0.04$&$ 0.21\pm0.02$&$-0.21\pm0.04$&$ 0.00\pm0.04$\nl
&2&$16.3\pm2.5$&$-0.04\pm0.04$&$ 0.22\pm0.02$&$-0.21\pm0.04$&$ 0.01\pm0.04$\nl
&3&$17.2\pm2.1$&$-0.11\pm0.04$&$ 0.15\pm0.01$&$-0.22\pm0.04$&$-0.07\pm0.04$\nl
&4&$16.0\pm2.9$&$-0.01\pm0.05$&$ 0.22\pm0.02$&$-0.21\pm0.05$&$ 0.01\pm0.05$\nl
\tablevspace{2pt}
NGC 6702&1&$ 1.6\pm0.2$&$ 0.57\pm0.12$&$ 0.15\pm0.04$&$ 0.43\pm0.13$&$ 0.58\pm0.12$\nl
&2&$ 1.6\pm0.2$&$ 0.55\pm0.10$&$ 0.15\pm0.02$&$ 0.43\pm0.10$&$ 0.58\pm0.10$\nl
&3&$ 1.7\pm0.2$&$ 0.46\pm0.09$&$ 0.10\pm0.02$&$ 0.38\pm0.09$&$ 0.48\pm0.09$\nl
&4&$ 1.6\pm0.2$&$ 0.57\pm0.13$&$ 0.16\pm0.03$&$ 0.42\pm0.13$&$ 0.58\pm0.13$\nl
\tablevspace{2pt}
NGC 6703&1&$ 7.8\pm2.0$&$ 0.03\pm0.05$&$ 0.12\pm0.02$&$-0.08\pm0.05$&$ 0.04\pm0.05$\nl
&2&$ 8.2\pm1.8$&$ 0.00\pm0.07$&$ 0.12\pm0.02$&$-0.09\pm0.07$&$ 0.03\pm0.07$\nl
&3&$ 8.7\pm1.7$&$-0.03\pm0.04$&$ 0.08\pm0.01$&$-0.09\pm0.04$&$-0.01\pm0.04$\nl
&4&$ 7.9\pm2.1$&$ 0.03\pm0.06$&$ 0.12\pm0.02$&$-0.08\pm0.06$&$ 0.04\pm0.06$\nl
\tablevspace{2pt}
NGC 7052&1&$ 6.7\pm2.0$&$ 0.21\pm0.06$&$ 0.23\pm0.02$&$ 0.00\pm0.06$&$ 0.23\pm0.06$\nl
&2&$ 7.2\pm2.3$&$ 0.18\pm0.06$&$ 0.23\pm0.02$&$ 0.00\pm0.06$&$ 0.23\pm0.06$\nl
&3&$ 7.8\pm2.6$&$ 0.10\pm0.06$&$ 0.16\pm0.01$&$-0.02\pm0.06$&$ 0.14\pm0.06$\nl
&4&$ 6.6\pm2.2$&$ 0.22\pm0.07$&$ 0.24\pm0.03$&$ 0.00\pm0.08$&$ 0.24\pm0.07$\nl
\tablevspace{2pt}
NGC 7454&1&$ 7.1\pm1.3$&$-0.31\pm0.03$&$ 0.02\pm0.03$&$-0.33\pm0.04$&$-0.31\pm0.03$\nl
&2&$ 7.0\pm1.3$&$-0.31\pm0.03$&$ 0.02\pm0.03$&$-0.33\pm0.04$&$-0.31\pm0.03$\nl
&3&$ 7.0\pm1.3$&$-0.31\pm0.03$&$ 0.01\pm0.02$&$-0.32\pm0.03$&$-0.31\pm0.03$\nl
&4&$ 7.1\pm1.3$&$-0.31\pm0.03$&$ 0.02\pm0.03$&$-0.33\pm0.04$&$-0.31\pm0.03$\nl
\tablevspace{2pt}
NGC 7562&1&$ 8.9\pm2.2$&$ 0.10\pm0.06$&$ 0.18\pm0.02$&$-0.06\pm0.06$&$ 0.12\pm0.06$\nl
&2&$ 9.2\pm2.0$&$ 0.08\pm0.05$&$ 0.18\pm0.02$&$-0.06\pm0.05$&$ 0.12\pm0.05$\nl
&3&$10.2\pm2.2$&$ 0.01\pm0.05$&$ 0.12\pm0.01$&$-0.08\pm0.05$&$ 0.04\pm0.05$\nl
&4&$ 8.9\pm2.0$&$ 0.11\pm0.06$&$ 0.18\pm0.02$&$-0.06\pm0.06$&$ 0.12\pm0.06$\nl
\tablevspace{2pt}
NGC 7619&1&$15.5\pm2.3$&$ 0.05\pm0.04$&$ 0.18\pm0.02$&$-0.11\pm0.04$&$ 0.07\pm0.04$\nl
&2&$15.9\pm2.0$&$ 0.03\pm0.03$&$ 0.18\pm0.02$&$-0.11\pm0.03$&$ 0.07\pm0.03$\nl
&3&$16.5\pm1.5$&$-0.02\pm0.03$&$ 0.12\pm0.01$&$-0.11\pm0.03$&$ 0.01\pm0.03$\nl
&4&$15.1\pm2.2$&$ 0.07\pm0.04$&$ 0.18\pm0.02$&$-0.10\pm0.04$&$ 0.08\pm0.04$\nl
\tablevspace{2pt}
NGC 7626&1&$18.3\pm2.2$&$-0.07\pm0.03$&$ 0.22\pm0.02$&$-0.27\pm0.04$&$-0.05\pm0.03$\nl
&2&$18.8\pm2.0$&$-0.09\pm0.03$&$ 0.22\pm0.02$&$-0.26\pm0.03$&$-0.04\pm0.03$\nl
&3&$19.4\pm1.9$&$-0.15\pm0.03$&$ 0.15\pm0.01$&$-0.26\pm0.03$&$-0.11\pm0.03$\nl
&4&$18.2\pm2.0$&$-0.06\pm0.02$&$ 0.23\pm0.02$&$-0.27\pm0.03$&$-0.04\pm0.02$\nl
\tablevspace{2pt}
NGC 7785&1&$15.0\pm2.2$&$-0.01\pm0.04$&$ 0.10\pm0.02$&$-0.10\pm0.04$&$ 0.00\pm0.04$\nl
&2&$15.2\pm2.3$&$-0.02\pm0.04$&$ 0.10\pm0.02$&$-0.10\pm0.04$&$ 0.00\pm0.04$\nl
&3&$15.6\pm2.3$&$-0.05\pm0.04$&$ 0.07\pm0.01$&$-0.10\pm0.04$&$-0.03\pm0.04$\nl
&4&$15.0\pm2.3$&$-0.01\pm0.04$&$ 0.11\pm0.02$&$-0.11\pm0.04$&$ 0.00\pm0.04$\nl
\enddata
\end{deluxetable}

\clearpage

\section{SSP-equivalent parameters for the G93 sample}\label{sec:results}

This section presents a brief overview of the resultant SSP-equivalent
population parameters for the G93 galaxies; detailed discussion is
reserved to Papers II and III.  Our focus here is on the preferred
model 4 (C and O both up), but results from models 1 and 2 are similar
(model 3 being ruled out).

\begin{figure*}
\plotone{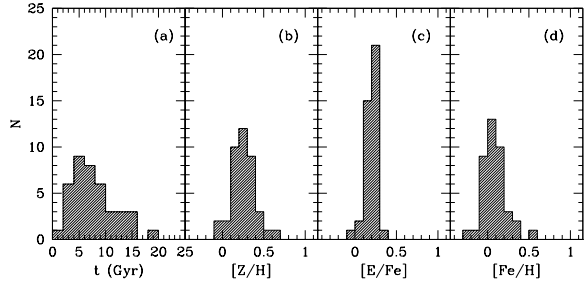}
\caption{Distributions of central (\reo{8}) stellar population
parameters for the Gonz{\'a}lez (1993) sample using enhancement model
4.  As expected from Figures 1 and 3, these local field ellipticals
span a wide range of ages but a small range ($\la0.3$ dex) of
metallicities and a very small range in abundance
enhancements.\label{fig:tza_hist_re8}}
\end{figure*}

Figure~\ref{fig:tza_hist_re8} presents histograms of $t$, \z, and
\enh\ for the G93 sample through the \reo{8}\ aperture.  The original
conclusions of G93 are confirmed using this more rigorous analysis:
the central stellar populations of galaxies in this sample span a
large range of SSP-equivalent ages, from $1.5\la t\;({\rm Gyr})\la18$
(more than 1 dex), but a relatively narrow range in \z,
$-0.1\la\z\la+0.6$, and an even smaller spread in \enh.  The
metallicity distribution has a peak at $\langle\z\rangle=+0.24$ and a
dispersion of $\sigma(\z)=0.14$, while the enhancement distribution
peaks strongly at $\langle\enh\rangle=+0.20$ with a dispersion
$\sigma$(\enh) of only 0.05 (these values vary slightly with the
model).

A striking fact to emerge from Figure~\ref{fig:tza_hist_re8} is how
\emph{mild} the mean metallicities and enhancements of ellipticals
really are.  Matching the high Mg index values of ellipticals has been
problematic in the past (e.g., \cite{Mat94}; \cite{Greggio97}), and
previous authors have typically invoked rather large enhancements in
the range $\enh=+0.3$--0.5 (\cite{WPM95}; \cite{Trager97};
\cite{Greggio97}).  With the TB95 response functions, however, the
average \z\ is only a factor of two higher than solar, and the average
\enh\ is only $+0.2$.  The latter is small compared to the maximum
value of $\afe \sim +0.5$ found in metal-poor Galactic stars
(\cite{WST89}; \cite{EAGLNT93}), which is widely regarded as an
empirical upper limit to the amount of depression in Fe that can
result from total suppression of SNae Ia.  The depression of the
Fe-peak in ellipticals appears to be much less than this and should be
easier to accommodate with reasonable galacto-nucleosynthesis models.

\begin{figure*}
\plotone{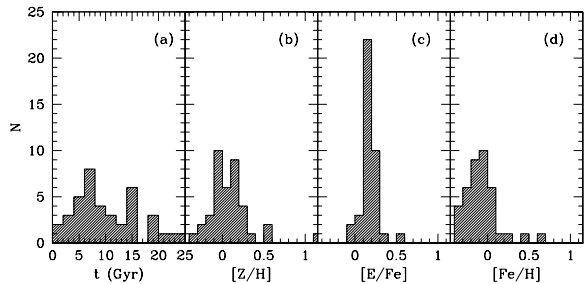}
\caption{Distributions of global (\reo{2}) stellar population
parameters for the Gonz{\'a}lez (1993) sample using enhancement model
4.  The global parameters here are slightly offset to older ages and
lower metallicities from the central parameters in Figure 7, but
otherwise the distributions are strikingly
similar.\label{fig:tza_hist_re2}}
\end{figure*}

Figure~\ref{fig:tza_hist_re2} presents similar histograms of $t$, \z,
and \enh\ for the \reo{2}\ aperture.  The global stellar populations
span a slightly larger range of ages, from 1.5 to 25 Gyr and a
slightly wider range of metallicities, $-0.3\la\z\la+0.7$ (with NGC
720 at $\z=1.05$), although some of this larger scatter is surely due
to the larger uncertainties in the \reo{2}\ line strengths.
Otherwise, the shapes of the distributions are similar.  Comparing
\reo{8}\ with \reo{2}\ shows that mean $\langle\z\rangle$ is down by
0.18 dex in the outer parts, indicating that the outer regions are
slightly more metal-poor than the centers.  The outer mean enhancement
$\langle\enh\rangle$ is lower by only 0.03 dex, however, confirming
the conclusion of Worthey et al.~(1992), Davies, Sadler \& Peletier
(1993), and G93 that enhancement gradients \emph{within} galaxies are
weak.  Ages increase slightly outwards, the outer parts being on
average roughly 25\% older.  Overall, the differences \emph{among}
galaxies are much more striking than the differences \emph{within}
galaxies, at least in the G93 sample, through these apertures.

\section{Uncertainties and systematic errors}\label{sec:disc}

This section assesses both zeropoint and scale errors in $t$, \z, and
\enh.  We begin by examining our basic assumption that the ages,
metallicities, and enhancement ratios we have derived above represent
true light-weighted ages and abundances of elliptical galaxies.  In
particular, we first ask whether the apparent large age spread among
the G93 galaxies could be due to spurious effects.

\subsection{\hbeta\ as an age indicator}\label{sec:hbeta_age}

The assumption that we are measuring real ages of stellar populations
rests on the further assumption that \hbeta\ light is coming purely
from main-sequence and red giant-branch stars.  We now discuss three
scenarios whereby \hbeta\ might be contaminated by light from other
sources.

(1) \emph{Fill-in by emission} (see Section~\ref{sec:emission}).  The
extreme form of this hypothesis says that \emph{all} ellipticals are
actually young and that the apparent large age spread is due entirely
to variable amounts of infill by emission.  This extreme view is
strictly ruled out by numerous observational studies of emission in
elliptical galaxies.  For example, G93's plot of precision
continuum-subtracted spectra (G93, Figure 4.10) shows that emission is
nearly always less than a few tenths of an \AA, not nearly large
enough to create the observed age spread.  In the same vein, Carrasco
et al.\ (1996) went so far as to suggest that no emission corrections
should be applied \emph{at all} to most ellipticals, implying that any
emission can at most be small.  The final point is that \hbeta\
correlates strongly both with Mg$_2$ and $\sigma$ (G93;
\cite{Jorgensen97}), inconsistent with emission fill-in, which varies
unpredictably from galaxy to galaxy.

A more reasonable hypothesis is that \emph{errors} in the emission
correction contribute noticeably to the age spread.  Such errors were
investigated in Section~\ref{sec:emission}, where we noted that
scatter in the \hbeta/\othree\ ratio would induce age errors of only
$\pm9$\% for typical galaxies.  An even more drastic test is presented
in Appendix~\ref{app:emsigcorr}, which shows that neglecting the
emission correction altogether affects a few strong-\othree\ galaxies
but makes at most small changes in the broad age distribution.

(2) \emph{Contamination by blue horizontal branch stars (BHBs)}. BHB
stars are not present in the standard Worthey (1994) models, which
assume red clumps for old metal-rich populations.  BHB stars might
come from an anomalous BHB population associated with the metal-rich
stars, or from contamination by a normal BHB associated with a
subordinate metal-poor population.  By ``BHB,'' we mean blue
horizontal branches similar to M~92, which would contribute
significantly to the light at 4000--5500 \AA, not the extremely hot
horizontal branches identified in populations like NGC 6791
(\cite{LSG94}) which contribute primarily to 1500 \AA\ flux and the
``UV upturn'' (e.g., \cite{Lee94}, \cite{Yi99}).

The galaxy M~32 can be used to rule out the hypothesis that BHBs alone
are responsible for the large \hbeta\ excesses seen in
\emph{high}-\hbeta\ ellipticals.  It can be shown that nearly the
\emph{entire} red clump in M~32 would have to be moved to a BHB at
approximately spectral type mid-F to explain its high \hbeta\ index
(\cite{BFGK84}); this is strictly ruled out by blue spectral indices
(\cite{Rose85}, 1994).  Moreover, the HB has actually been detected in
the outer part of M~32 by HST (\cite{Grillmair96}) and is seen to be
mostly red.\footnote{To be precise, the Grillmair et al.~(1996) data
are not deep enough to rule out a \emph{small} number of BHB stars
(\cite{Grillmair96}; C.  Gallart, priv.~comm.), but the lack of point
sources in archival F300W images suggests that any BHB must indeed be
weak.}  Extrapolating the G93 indices outward to this field and
matching to W94 models there yields an excellent fit to both the
integrated colors and the color of the RGB at this point
(\cite{Grillmair96}), supporting the assumption that the HB is indeed
red.

The existence of a dominant BHB population in metal-rich ellipticals
is not expected on astrophysical grounds.  If ellipticals were
\emph{very} old (i.e., $>18$ Gyr, as Lee 1994 has suggested), then
BHBs could conceivably be significant components, but such large ages
violate current constraints on the age of the Universe (see, e.g.,
\cite{Gratton97}).  No \emph{solar metallicity} cluster populations in
the Milky Way have BHBs (\cite{Worthey94}), although we note that Rich
et al.\ (1997) have discovered significant M3-like BHB populations in
two metal-rich Galactic bulge globular clusters, NGC 6388 and NGC 6441
($\feh\sim-0.5$).  However, these two globulars are the densest known
in the Galactic globular cluster system; the fact that BHB stars occur
precisely there (\cite{Sosin97}) suggests that dynamical interactions
are the cause.  We conclude that the occurrence of BHB stars in
low-density systems like giant elliptical galaxies is unlikely, but a
deeper understanding of their presence in these globulars is obviously
necessary.

Contamination by BHBs from a subordinate metal-poor population also
does not seem probable.  As noted, such contamination by a trace blue
BHB component cannot materially affect the indices of
\emph{high}-\hbeta\ galaxies, but perturbations in \emph{weak}-\hbeta\
galaxies should be considered.  For example, 5\% of the $V$-band light
in metal-poor BHB stars would decrease the inferred age of a galaxy
from 13 Gyr to 8 Gyr at solar metallicity.  However, Rose (1985,
1994), using a set of high-resolution spectral indices in the 4000
\AA\ region, has shown in M~32 and eight strong-lined ellipticals that
no more than $\sim5\%$ of the light in the blue region (and less than
2\% in the $V$-band) can come from very hot stars (F0 and earlier).
This falls short by a factor of two.  Moreover 5\% of $V$-band light
in BHB stars would imply that altogether $\sim$25\% of the
\emph{total} light would have to come from metal-poor stars.  This is
twenty-five times more than the amount of metal-poor ($\z<-1.5$)
$V$-band light actually found in the outer part of M~32 by Grillmair
et al.~(1996).  That a much larger quantity of metal-poor stars could
be found near the centers of \emph{more} metal-rich elliptical
galaxies seems implausible.

We conclude that contamination by metal-poor populations is a
negligible perturbation to the \emph{central} ages of the G93 galaxies
and could cause at most a $\sim$10\% reduction near $r_e$, if that.

(3) \emph{Contamination by blue straggler stars (BSSs).} A typical BSS
has $M_V\approx3$ mag, $B-V\approx0.2$ (\cite{Bailyn95}), and spectral
type A8--F0.  From Worthey et al.\ (1994), dwarf A8--F0 stars have
$\hbeta\approx5.8$--$7$ \AA.  To explain the high \hbeta\ strength of
M~32 ($\hbeta=2.4$ \AA) as arising from a population of blue
stragglers superimposed on an old (15 Gyr), solar-metallicity
population ($\hbeta=1.5$ \AA) would require that $\ga15\%$ of the
$V$-band light come from BSSs.  This implies a BSS specific frequency
of $\ga275$ per $10^4\;L_{\odot}$, which is a factor of 8 higher than
seen in the most BSS-rich Galactic globular cluster (Palomar 5) and a
factor of 28 higher than seen in the average Galactic globular cluster
(\cite{FFB95}).  We again conclude that \emph{high}-\hbeta\ galaxies
like M~32 are immune to perturbations by spurious hot components such
as BSS stars.

Consider next a trace contamination by BSSs in low-\hbeta\ galaxies.
For example, to decrease the age of an elliptical from 13 Gyr to 8 Gyr
at solar metallicity, BSSs would again need to contribute $\ga5\%$ of
the $V$-band light.  For NGC 3379, this implies a specific frequency
of $\ga 120$ per $10^4\;L_{\odot}$.  This specific frequency is a
little more than a factor of 3 higher than that seen in the most
BSS-rich Galactic globular cluster (\cite{FFB95}).  In the absence of
a complete theory of BSS formation, a factor of three increase might
not be impossible.  On the other hand, as noted, Rose (1985, 1994) has
shown that no more than $\sim2\%$ of $V$-band light can come from hot
stars F0 and earlier in M~32 and eight strong-lined ellipticals.  This
is less than half the light required and would perturb the age from 13
Gyr to only 11 Gyr, a reduction of only 15\%.

To summarize, the leading hot-star contaminants, BHBs and BSSs, both
peak at temperatures at or hotter than F0, whereas the blue
line-strength data of Rose (1985, 1994) imply that the great bulk of
Balmer absorption must be coming from cooler F and G stars, at or at
most only slightly hotter than the derived turnoff temperatures.
Barring some as-yet-undiscovered contaminating population of cooler
stars, the Rose limits imply that contamination of \hbeta\ by non-main
sequence stars can reduce the ages of even the oldest ellipticals by
at most 10--15\%.

\subsection{Errors due to theoretical model uncertainties}\label{sec:tunc}

The next three sections assess additional sources of systematic
errors; results are collected in Table~\ref{tbl:syserr}.  This section
discusses theoretical model uncertainties caused by errors in the
stellar isochrones and line-strength response functions of TB95.  The
major uncertainty in the interior models is the \emph{age scale},
which is continually being refined.  With the recent release of
parallaxes from the HIPPARCOS satellite, much effort has been spent
recalibrating the ages of Galactic globular clusters using both the
new parallaxes and up-to-date models of stellar evolution (e.g.,
\cite{Reid97}, 1998; \cite{Gratton97}; \cite{Chaboyer98};
\cite{GVA98}; \cite{Pont98}; \cite{SW98}).  This effort has brought
the ages of the oldest globular clusters down from $\sim14$--15 Gyr to
$\sim12$ Gyr, a reduction of $\sim15\%$.  At least half of this
reduction is due to corrections in the metallicity scale of globular
clusters and to the use of more up-to-date stellar evolutionary models
(\cite{Gratton97}).

These age redeterminations have so far been restricted to clusters
with metallicities $\feh\la-0.7$.  At the metallicities typical of
elliptical galaxies, the effect of the age recalibrations is not yet
known but could be as much as $\sim20\%$, just by using isochrones
from the most modern stellar evolutionary models.  This agrees with
Charlot, Worthey \& Bressan (1996), who found that absolute ages are
uncertain at the 25\% level in stellar populations with ages $>10$
Gyr, resulting almost entirely from the choice of different stellar
models.  Below and in Appendix~\ref{app:padova}, we explore the effect
of substituting ``Padova" isochrones by Bertelli et al.\ (1994) for
those of W94 and find that young ages differ by 35\% but that old ages
change by only 4\%.  As a rough rule of thumb, we assume that both the
age zeropoint and age scale of the models are uncertain at the
$\sim20$-$25\%$ level.

\begin{figure*}
\plotone{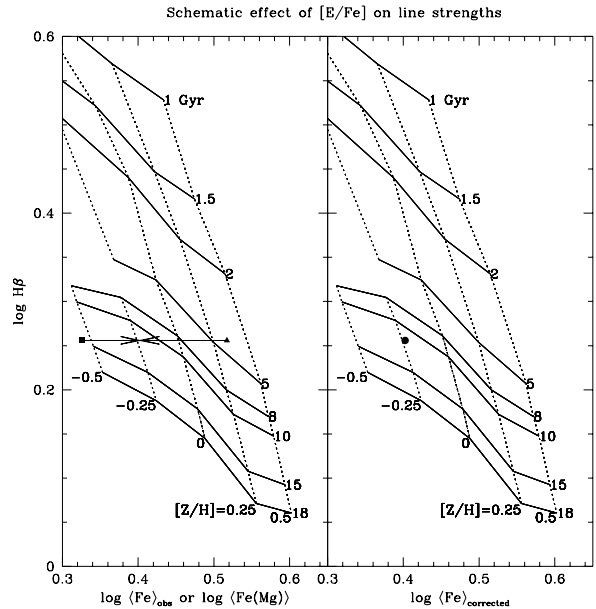}
\caption{A schematic representation showing how SSP parameters are
derived for a galaxy with non-solar abundance ratios.  A model
non-solar galaxy with $(t,\z,\enh)=(12,-0.25,0.3)$ is plotted in panel
(a) with its true value of \fe\ (solid square) and the value of \fe\
that would be inferred from its \mgb\ strength assuming solar
abundance ratios (solid triangle; called \femg\ in the text).  This
latter point lies to the right of \fe\, reflecting non-solar ratios.
The vectors show corrections to each feature for the non-solar
enhancement ($\enh\neq0$; small changes in \hbeta\ are ignored).  When
the correct enhancement is chosen, the two vector tips agree.  The sum
of the two vectors determines the size of the enhancement, while their
ratio determines the location of the final point, which gives $t$ and
\z.  The corrected point is also shown in panel (b) and coincides with
the inferred ($t$,\z) of $(12,-0.25)$.  Errors in the two correction
vectors affect the final population parameters as described in the
text.\label{fig:reserr}}
\end{figure*}

The effect of errors in the theoretical response functions of TB95 is
illustrated in Figure~\ref{fig:reserr}, which is a schematic repeat of
Figure 1(b) showing \hbeta\ versus \fe.  Figure~\ref{fig:reserr} shows
a galaxy plotted two ways, one using raw \fe, the other using the
value of \fe\ inferred from \mgb\ by assuming solar abundance ratios
(call this \femg).  \fe\ lies to the left of \femg, indicating Fe
depression.  Applying the TB95 corrections for non-solar \enh\ moves
\fe\ to the right and \femg to the left, as shown by the arrows (the
correction to \hbeta\ is small and is ignored). When the correct value
of \enh\ is reached, the two points coincide, giving final $t$, $Z$,
and \enh\ (right hand panel).

Where the solution lands is evidently governed by the \emph{relative
lengths} of the two correction vectors; for model 4, this ratio is
$\Delta \log \femg/\Delta \log \fe$ = 1.25.  The systematic errors of
the final point depend mainly on the error of this ratio.  Assuming
that the two response functions of TB95 are individually uncertain by
as much as 30\% and that the errors of their three stellar types add
in quadrature, the resultant zeropoint uncertainties are 3\% in age,
0.10 dex in \z, and 0.04 dex in \enh\ for highly enhanced galaxies.
Since this last error drops to zero for galaxies with $\enh = 0$, we
derive an overall scale uncertainty in \enh\ of $\leq$20\%.

We note that fundamental uncertainties in stellar models, for example
the use of a single-parameter mixing length theory for convection or
the detailed effects of rotation and diffusion, may induce additional,
unknown systematic errors in our absolute age estimates.  Such
uncertainties also affect the globular cluster age scale.  At present,
our estimated uncertainties in the absolute ages of galaxies therefore
should be considered to be relative to the globular cluster age scale.

\subsection{Errors due to empirical model uncertainties}\label{sec:eunc}

Errors in this category include errors in the metallicity and
temperature scales of the Lick/IDS fitting functions and errors in the
fitting accuracy of the functions themselves.

We have checked the metallicity scale of the Lick/IDS system by
comparing our assumed stellar \z\ values (summarized by \cite{WFGB94})
with the compilation of published spectroscopically-determined values
by Cayrel de Strobel et al.\ (1997).  For stars with $\log g\ge4$
(mostly dwarfs), the Lick/IDS metallicity scale is in excellent
agreement with the published values at all \feh.  For giants with
$\feh\la0$, the Lick/IDS metallicity scale is within $0.05$--$0.1$ dex
of the Cayrel de Strobel scale (systematically slightly high).
However, for giants with $\feh>0$ (``SMR'' stars), the Lick/IDS
metallicity scale deviates strongly from the Cayrel de Strobel scale,
such that the Lick/IDS giants appear to be more metal rich.  The
Lick/IDS metallicity scale for giants is based on the narrow-band
photometric metallicity scales of Hansen \& Kj{\ae}rgaard (1971) and
Gottlieb \& Bell (1971), and on the high-resolution spectral study of
Gustafsson, Kj{\ae}rgaard \& Anderson (1974) (\cite{FFBG85}).  In
contrast, the Cayrel de Strobel et al.\ (1997) catalog is populated in
the SMR giant regime by older spectroscopic abundance determinations
based on stellar atmospheres that typically (1) have a too-low solar
iron abundance (\cite{McWilliam97}) and (2) do not properly account
for molecule formation in SMR giants (\cite{Castro96}).  Correcting
the abundances of SMR giants for these two effects suggests that the
Lick/IDS scale may actually be very close (possibly $0.05$--$0.1$ dex
too high) to the modern spectroscopic metallicity scale, even at
$\feh\sim0.4$ (\cite{Castro96}; McWilliam, priv.~comm.).

The next question is whether the fitting functions are in fact good
fits to the stellar line strengths.  By inspecting the residual
diagrams in Gorgas et al.~(1993) and Worthey et al.~(1994), we
estimate that any systematic errors in the metal-line fits are less
than 3\%, which translates to zeropoint uncertainties of 0.05 in \z\
and 0.10 in \enh\ (the former averages Fe and Mg while the latter
differences them, accounting for its larger error).  The crucial
function for age is the fit to \hbeta\ versus $V-K$ for main sequence
A-F stars.  Again, we estimate that the basic line-strength
calibration level is accurate to better than 3\% in this interval,
which translates to about 10\% in age.  Finally, the temperature scale
($V-K$ vs. $T_e$) of main sequence stars is needed to attach \hbeta\
strengths to the theoretical isochrones.  An error of 100 K
(\cite{WFGB94}) again translates to about 10\% in age.  Note that all
these errors in the fitting functions affect only the absolute
zeropoints of age, metallicity, and \enh\ but not their differential
values.

\subsection{Errors due to unknown element enhancements}\label{sec:uncenh}

Our treatment of element enhancements is crude---we simply group all
elements into three categories (enhanced, depressed, and fixed) and
assume that differences within each group are nil.  The group
assignment of certain elements is also uncertain.  Unknown element
abundance ratios introduce errors in the predicted index response
functions and, to a smaller extent, in the theoretical stellar
evolutionary tracks.

According to TB95, the elements that significantly influence the
indices used here are Fe, Cr, C, and Mg.  Fe and Cr are produced in
both Type Ia and in intermediate-mass progenitor Type II SNae
(\cite{WW95}).  They should vary closely together by virtue of similar
nucleosynthesis; i.e., their relative uncertainty should be small.
Breaking the link between Fe and Cr, for example by decreasing
[Cr/Fe], would have the effect of altering the \mgb\ index strength
without significantly affecting other indices (TB95).  In our own
galaxy, however, [Cr/Fe] is solidly at the solar value until
$\feh\sim-2$ (\cite{McWilliam97}), much lower than the metallicities
of interest in elliptical galaxies.  We will discuss possible
element-to-element variations in a future paper.  Likewise we have
tested the sensitivity of the indices to C explicitly in models 1--4
and found that low-C models (like model 3) are ruled out.  With this
eliminated, remaining uncertainties due to the C abundance variations
are limited to 10\% in age, 0.05 dex in \z, and 0.01 dex in \enh\
(Figure~\ref{fig:comp_enhmods}).

A larger source of uncertainty arises from uncertain ratios
\emph{within} the Type II SNae group.  The metallicity \z\ is
controlled by O, which has little spectroscopic signature (TB95),
while a major spectral impact comes from Mg.  Our inferred values of
\z\ thus depend critically on the assumption that Mg and O track one
another.  Breaking this link, e.g., by enhancing Mg over O, could
reduce our inferred \z's substantially.  For example, suppose that
[O/H] is always solar regardless of Mg (this would place O in the
depressed group in metal-rich galaxies).  Since O contributes half the
mass in $Z$ (see Table~\ref{tbl:tbmodels}), our values of \z\ would be
overestimated by a factor of two.  Correcting for this would reduce
the typical \z\ from 0.26 to 0.13, and in so doing would increase ages
by about 20\%; enhancements \enh\ would remain unchanged.  We are thus
relying quite heavily on the notion that decoupling O and Mg is
astrophysically unreasonable.

Finally, our analysis assumes that isochrone shape and location are
unaffected by the exact value of \enh\ or by the detailed pattern of
element enhancements within \enh.  Existing models suggest that this
assumption might be tolerable.  Salaris \& Weiss (1998) have
calculated an isochrone for an old population model with $\z = -0.3$,
$\enh = +0.4$, and non-solar $\XX = 0.12$.  Log $T_e$ at the turnoff
shifts to the blue by 0.0044, while log $T_e$ on the RGB shifts to the
blue by 0.011 relative to a scaled solar model.  The blueward shifts
should scale in proportion to both \z\ and \enh, while the shape
change may also depend on \XX\ (\cite{SCS93}; \cite{SW98}).  A typical
G93 galaxy is four times more metal-rich than their model but smaller
by a factor of two in \enh.  The quantity \XX\ is also likely to be
smaller, being +0.07 in model 4 (Table~\ref{tbl:tbmodels}) versus
+0.12 in their model.  On balance, the net shifts and shape changes in
the elliptical isochrones are plausibly no more than twice those in
their model.

The effects of such motions would be small.  A shift of log $T_e =
0.0044$ at the turnoff causes a change of only 0.016 in log \hbeta,
for a change in age of 6\%.  A shift of log $T_e = 0.011$ on the RGB
causes a change in metal lines of the same amount, for a change in \z\
of about 0.05 and no change in \enh.  Even if multiplied by two, as
estimated above, these effects would still be small compared to other
errors.  On tho other hand, it should be stressed that the effect of
non-solar ratios on isochrone location is acclerating at high
metallicity, and the above models were calculated for metallicities
considerably smaller than what we require.  A failure of O to track Mg
(as mentioned above) could also introduce further shape changes that
have not yet been modeled in detail.  In sum, our assumption that
isochrone shape and location are unaffected by the value of \enh\ or
by the pattern of non-solar enhancements within \enh\ looks promising
but is in need of further validation.

\subsection{Error summary}\label{sec:errorsummary}

\begin{deluxetable}{lrl}
\setcounter{table}{6}
\tablefontsize{\normalsize}
\tablewidth{0pt}
\tablecaption{Summary of Systematic Errors
\label{tbl:syserr}}
\tablehead{\colhead{Source}&\colhead{Amount}&\colhead{Effect}}
\startdata
\sidehead{Age errors:}
Metal-poor BHB contamination&$<-10\%$\tablenotemark{a}&Weak-\hbeta\ objects
only\nl
Blue straggler contamination&$<-15\%$&Weak-\hbeta\ objects only\nl
Theoretical isochrones&20--25\%&Zeropoint\nl
Theoretical isochrones&20--25\%&Scale\nl
Theoretical response functions (TB95)&3\%&Zeropoint\nl
\hbeta\ fitting function vs. $V-K$&10\%&Mainly zeropoint\nl
$V-K$ vs.~$T_e$&10\%&Zeropoint\nl
Unknown O/Mg ratio&20\%\tablenotemark{b}&Metal-rich objects only\nl
Unknown C abundance&10\%&Zeropoint\nl
Effect of $\enh\neq0$ on isochrones&6--12\%&Zeropoint\nl
Overestimation of \hbeta\ emission correction&+3\%&Zeropoint\nl
Undercorrected emission fill-in&$<+25$\%&A few weak-\hbeta\ objects only\nl
TWFBG98 \hbeta\ velocity dispersion correction&+25\%&Weak-\hbeta\ objects
with high $\sigma$ only\nl
\sidehead{Metallicity errors, [Z/H]:}
Theoretical response functions (TB95)&0.10 dex&Zeropoint\nl
Metal-line fitting functions&0.05 dex&Zeropoint\nl
Unknown O/Mg ratio&100\%\tablenotemark{b}&Scale\nl
Unknown C abundance&0.05 dex&Zeropoint\nl
Effect of $\enh\neq0$ on tracks&0.05--0.1 dex&Zeropoint\nl
\sidehead{Enhancement errors, [E/Fe]:}
Theoretical response functions&20\%&Scale\nl
Metal-line fitting functions&0.10 dex&Zeropoint\nl
Unknown C and O abundance&0.01 dex&Zeropoint\nl
Effect of $\enh\neq0$ on tracks&0.0 dex&\nodata\nl
\enddata
\tablenotetext{a}{Outer regions only.}
\tablenotetext{b}{Assumes that [O/H] is always solar regardless of \z.}
\tablecomments{When signs are given, they are in the sense that the
standard models are wrong by that amount.  If no sign is given,
the error could have either sign.}
\end{deluxetable}

The results of the preceding sections, plus some additional
experiments in Appendices~\ref{app:padova} and \ref{app:emsigcorr},
are summarized in Table~\ref{tbl:syserr}.  Age errors are
significant---several terms amount individually to 10--25\% and their
addition is uncertain.  Some age errors are also larger for
weak-\hbeta\ objects and therefore tend to stretch or compress the age
scale.  However, most of the errors, including those in age, are
simple zeropoint shifts.  Future applications will take advantage of
the relative robustness and use the data differentially.

The galaxy M~32 offers a final check on the zeropoints of both \z\ and
\enh.  The integrated spectrum of M~32 has been modeled by many
authors, and the upper CM diagram of the outer parts has been measured
(\cite{Grillmair96}).  All spectrum modelers concur that a mix of
moderately young stars of near-solar metallicity matches every known
feature of the spectrum.  The mean turnoff spectral type within the
$r_e/2$ aperture is accurately known to be F7--8 (\cite{Faber72};
\cite{O'Connell80}; \cite{Rose94}), while the light-weighted
metallicity in the Grillmair field is $\feh = -0.25$.  The enhancement
ratio \enh\ is also known to be small based on the excellent spectral
fits using solar-neighborhood-abundance stars (e.g., \cite{Faber72}).

These independently measured parameters agree well with the
SSP-equivalent parameters.  G93 indices for the $r_e/2$ aperture yield
an SSP-equivalent age of 5 Gyr, a mean \z\ of $-0.07$, and an \enh\ of
$-0.07$.  These parameters imply a turnoff spectral type of exactly
F7--8, as the modelers have concluded, and the near-solar \z\ and
\enh\ also agree with their results.  Extrapolating the G93 indices
outward, Grillmair et al.\ (1996) found an SSP-equivalent age in their
field of 8 Gyr, a mean \z\ of $-0.25$, and \enh\ of $-0.05$.  This
metallicity coincides precisely with the metallicity distribution that
they inferred from the color locus of the RGB for that assumed age.
Putting this information together, we conclude that the actual
absolute uncertainties in both \z\ and \enh\ are $\leq0.05$, at least
for galaxies close to solar composition like M~32.

\section{Comparisons with previous studies: Evidence for
intermediate-age populations}
\label{sec:others}

Many previous studies have examined the line strengths and colors of
elliptical galaxies to determine their stellar content.  A complete
review of all previous models is beyond the scope of this paper (see
Charlot, Worthey \& Bressan 1996, Vazdekis et al.\ 1996, and Arimoto
1996 for comparison of some modern stellar population synthesis
models, and Worthey 1998 for a historical review of the metallicities
and abundance ratios of early-type galaxies).  We concentrate here on
previous investigations that derived SSP ages and models by using the
Balmer lines.  We begin with the results of TCB98 and then turn to
those of other workers.  Consideration of other methods, in particular
those using colors, is delayed to future papers.

\subsection{Model dependence of derived stellar population
parameters: comparison with TCB98}\label{sec:modeldep}

In a recent paper, TCB98 have analyzed line strengths of the G93
galaxies in the context of their own stellar population models.  These
models are based on isochrones by Bertelli et al.\ (1994) (which, like
ours, neglect the effects of $\enh\neq0$), the original fitting
functions for \hbeta\ and \fe\ by Worthey et al.\ (1994) (which also
neglect $\enh\neq0$), and a fitting function for \mgtwo\ by Borges et
al.\ (1995), who claim to take into account the effects of
$\enh\neq0$.  Like us, TCB98 assume that their isochrones depend only
on bulk metallicity \z\ but not on \enh\ (which they call \afe).  To
determine the stellar population parameters \logt, \z, and \afe, TCB98
compute {\it averaged} derivatives (from their models) of (the
logarithms of) \mgtwo, \fe, and \hbeta\ versus population parameters.
These derivatives are then inverted to derive a series of linear
equations that yield relative values of \logt, \z, and \afe\ as
functions of \mgtwo, \fe, and \hbeta.  This solution method is
equivalent to assuming that all line strengths depend {\it linearly}
on all parameters, which is marginally inconsistent with the curved
shapes of the actual grids (cf.  Figure~\ref{fig:re8m}).

\begin{figure*}
\plotone{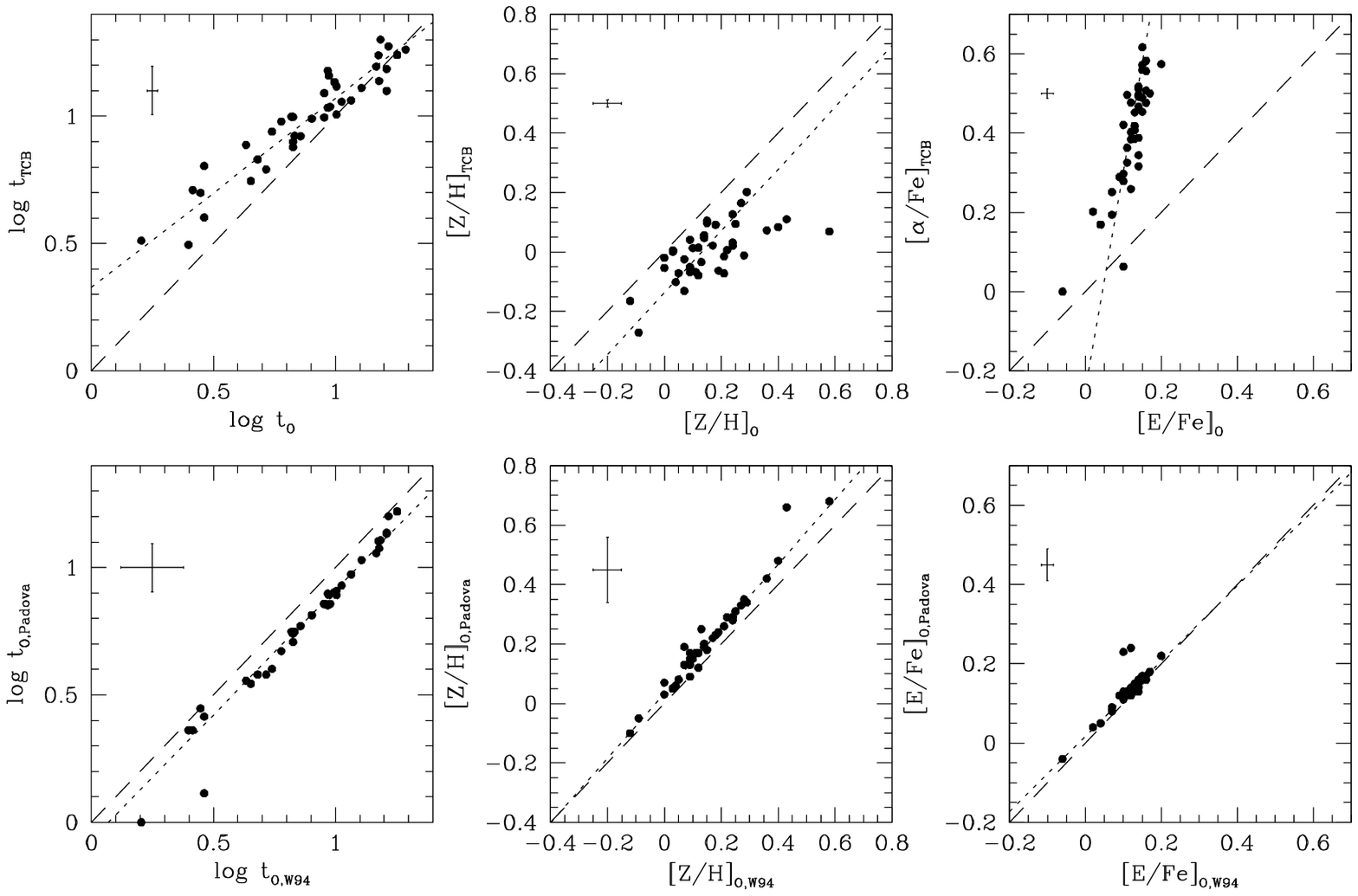}
\caption{Two tests of the model dependences of our results, for the
central \reo{8}\ aperture.  In the top row, the results of Tantalo et
al.\ (1998a; TCB98) are plotted as a function of our ages,
metallicities, and enhancement ratios, inferred here using enhancement
model 3 (C down, O up, labelled ``O''), which has an elemental mix
similar to that of Tantalo et al.  We have shifted their results so
that NGC 221 (M~32) has an age of 4 Gyr, $\z=-0.1$, and $\enh=0$, as
suggested by them.  Note the disagreement between the inferred \z\ and
\enh\ (\afe) distributions in the two studies.  This disagreement is
due mainly to their use of different response functions to correct the
line indices for \enh$\ne0$; in particular, no correction to \fe\ was
applied.  The bottom row shows the results of substituting the
Bertelli et al.\ (1994; ``Padova'') isochrones for the Worthey (1994;
``W94'') isochrones in model 4; all other ingredients (fitting
functions, response to non-solar-neighborhood abundance ratios) remain
the same.  using enhancement model 4 (``CO'').  Apart from slight
zeropoint and slope differences, the two sets of isochrones give very
similar results (see Appendix~\ref{app:padova}).  In all plots, the
long-dashed line is the line of equivalence for the two sets of
results; the short-dashed line is a linear least-squares fit
(iteratively rejecting 3-$\sigma$ outliers).\label{fig:padmodels}}
\end{figure*}

Figure~\ref{fig:padmodels} (top row) shows the stellar population
parameters derived by TCB98 as a function of our derived parameters.
For this comparison we have used our enhancement model 3 with C
depressed and O enhanced, which is closest to their model.  While the
two studies roughly agree in the distribution of ages of the G93
galaxies, they are discordant in both \z\ and \enh, for which the
slopes of their values versus ours deviate strongly from unity.  These
different inferred \z\ and \enh\ values imply different
interpretations of the star formation histories of these galaxies
(particularly in the \logt--\z\ relation; see below).

To isolate the source of the differences, we constructed new models
(``Padova'') by substituting the Bertelli et al.\ (1994) isochrones
for the RYI/VandenBerg isochrones used by W94
(Appendix~\ref{app:padova}).  The results are presented in the bottom
row of Figure~\ref{fig:padmodels}.  The match between the new Padova
models and the standard W94 models is quite good, with only slight
slope changes and mild zeropoint offsets; \logt\ decreases by from
10\% to 30\%, \z\ increases by less than $+0.08$ dex, and \enh\
increases by no more than $+0.02$ dex in the Padova models.

The differences between the present results and those of TCB98 in the
top row are therefore not caused by differences in the isochrones, but
must rather stem from one of the following other differences: (1) use
of different response functions for $\enh\neq0$ for \mgtwo, Fe5270,
and Fe5335; (2) use of \mgtwo\ instead of \mgb; and/or (3) use of a
linearized solution method for deriving ages, metallicities, and
enhancement ratios from observed line strengths.  Inspection shows
that differences (2) and (3) are most likely minor; in particular, the
G93 \mgtwo\ strengths are less reliable than the \mgb\ strengths, but
broadly the two agree fairly well.  Likewise, the linearized method
deviates at large distances from the middles of the grids owing to
grid curvature, but these differences are not large enough to cause
the global slope differences seen in Figure~\ref{fig:padmodels}.

Difference (1), the use of different response functions, dominates the
differences in \z\ and \enh.  TCB98 use the Borges et al.\ (1995)
fitting function for \mgtwo, which nominally takes $\enh\neq0$ into
account but is derived from only a small set of calibration
stars.\footnote{The stellar metallicites are most likely on a
different metallicity scale, as they are drawn directly from Cayrel de
Strobel et al.~(1997), which may be unreliable at $\feh>0$; see
Section~\ref{sec:uncenh}.  The Borges et al.\ sample is also deficient
in calibrating stars on the RGB.}  Furthermore, by using the original
Worthey et al.~(1994) fitting functions for Fe5270 and Fe5335 without
correction for $\enh\neq0$, they implicitly assume that
\emph{metallicities can be determined from \fe\ alone}.  In other
words, the ages and metallicities of the G93 galaxies are effectively
defined by the \fe--\hbeta\ line-strength diagram alone in the TCB98
scheme, and the enhancement ratios \enh\ are defined by the offset of
the galaxies in the \mgtwo--\hbeta\ line-strength diagram (scaled by
some factor from the \cite{Borges95} fitting function for \mgtwo).  By
not correcting \fe\ upwards for Fe-deficiency, TCB98
\emph{underestimate} the metallicites \z\ and \emph{overestimate} the
ages $t$ and enhancement ratios \enh.  This matches the behavior of
residuals seen in Figure~\ref{fig:padmodels}.

These systematic effects cause TCB98 to find a much narrower spread in
in \z\ in the centers of the G93 galaxies than we do, and also a much
wider spread in \enh.  The narrow spread in \z\ prevents them from
finding any age--metallicity relation, which is a major focus of our
Paper II; conversely, the broad spread in \enh\ causes them to find a
strong age--enhancement ratio relation, which we do not find in Paper
II.  Further discussion of these trends is reserved to future papers.
However, it is clear that the adopted response functions can have
far-reaching consequences for parameter correlation studies.

\subsection{Other authors}

Few other authors have fitted stellar population parameters to Balmer
line data.  Kuntschner (1998; \cite{KD98}) has studied the line
strengths of a complete, magnitude-limited ($M_B<-17$) sample of
early-type galaxies in the Fornax cluster, split evenly between
ellipticals and lenticulars.  He derives stellar population parameters
but does not correct his line strengths for non-solar abundance
ratios.  In general, Kuntschner finds old ages for ellipticals but a
wide spread in the ages of S0s.  His data are of excellent quality,
and we lump them together with the G93 sample and analyze them in
parallel in Paper II.

Using moderate-$S/N$ ($\sim30$) long-slit and fiber spectroscopy,
J{\o}rgensen (1999) has studied the line strengths of 115 early-type
galaxies in Coma.  Of these galaxies, 71 have measured \mgb\, \fe, and
\hbeta (the last with typical errors of $0.22$ \AA).  However, Trager
(1997) has shown that errors in \hbeta\ of this magnitude (typical of
the Lick/IDS galaxy sample; TWFBG98) seriously compromise the
determination of stellar population parameters through correlated
errors in age, metallicity, and enhancement ratio; errors of $\la0.1$
\AA\ in \hbeta\ are required to determine ages to 10\% or better and
to reduce the correlated errors to insignificant levels.  Further
consideration of Jorgensen's work is reserved to Paper II.

Vazdekis et al. (1997) fit their own SSP-equivalent models to three
early-type galaxies, including NGC 3379 and NGC 4472 studied here.
Their derived ages are about 50\% larger than ours, for a variety of
reasons.  Although \hbeta is included in the suite of data fitted, it
is only one among many features used.  The resultant models
significantly {\it under}-predict their own \hbeta strengths, and
matching them would yield ages as young or younger than we find.
Their high ages (and low metallicities) seem to be driven by the very
red near-IR colors of their models at high \z, which in turn may stem
from the cool giant-branch tips of the Padova isochrones used.  Since
giant-branch temperatures are still in flux, we prefer the Balmer
lines, which are less sensitive to stellar evolution uncertainties.

Fisher, Franx \& Illingworth (1995) studied the line strengths of
nearby field ellipticals and brightest cluster galaxies (BCG).  All of
the nearby ellipticals (seven galaxies) were drawn from G93; the data
are consistent, and we have therefore not added them to this series of
papers.  The BCG data are of slightly lower quality as the galaxies
are more distant.  Fisher et al.\ compare their line strengths to the
W94 models---ignoring $\enh$ variations---and generally find old
($t\ga10$ Gyr) mean stellar populations in the centers.  However, two
of nine BCGs have \hbeta\ strengths indicative of intermediate-age
populations, NGC 2329 (Abell 569) and NGC 7720 (Abell 2634).

Jones \& Worthey (1995) developed a novel H$\gamma$ index that has
lower sensitivity to metallicity, and therefore in principle better
age discrimination.  Using W94 models, they applied this index to the
center of M 32 and determined an SSP age of $t\approx5$--7 Gyr.  Our
age for this object is only $t=3.0\pm0.6$ Gyr (with $\z=0.00\pm0.05$
dex; formal errors only).  Jones and Worthey also fitted other Balmer
indices (including different versions of H$\gamma$), which gave
similarly low ages.  They were not able to identify a reason for the
discrepancy.  This disagreement among Balmer indices is an outstanding
issue.

Finally, we mention the results of Rose (1985, 1994) for a sample of
10 normal elliptical nuclei, 6 of which overlap with our sample,
including M 32.  Rose's spectra were taken around the 4000 \AA\ break,
and he developed a large number of stellar population indicators in
this wavelength region, including the Balmer index \roseca, a
sensitive measure of the presence of hot A and B stars, and
\ion{Sr}{2}/\ion{Fe}{1}, a measure of the total dwarf-to-giant light.
By balancing these and other indices, Rose found that there must be a
substantial intermediate-temperature component of dwarf light in all
of these galaxies, but that no more than 2\% of 4000 \AA\ light could
come from stars hotter than F0.  He concluded that all 10 galaxies
contained a significant component of intermediate-age main sequence
stars.  Our SSP ages range from 3 to 10 Gyr for the 6 galaxies in
common, consistent with these conclusions.

\section{Summary}\label{sec:summary}

We have presented central ($r_e/8$) and global ($r_e/2$) line
strengths for the Gonz\'alez (1993) local elliptical galaxy sample.  A
method for deriving SSP-equivalent stellar population parameters is
presented using the models of Worthey (1994), supplemented by
model-atmosphere line-strength response functions for non-solar
element abundance ratios by Tripicco \& Bell (1995).  The resultant
stellar population parameters broadly confirm the findings of G93 in
showing a wide range of ages but a fairly narrow range of
metallicities and enhancement ratios.  Differences among galaxies in
the sample are larger than radial differences within them.

Four different models are considered with different patterns of
element enhancement.  The best-fitting model (model 4) has all
elements enhanced or normal except for the Fe-peak (and Ca), which are
depressed.  The actual atomic abundance ratios of the so-called
``enhanced'' elements are in fact virtually solar---it is really the
Fe-peak elements that are depressed.  Indeed, the TB95 response
functions imply that the observed strengthening of \mgb\ is not due to
an overabundance of Mg but to an \emph{under}abundance of Fe (and Cr).
It is shown that C must also belong to the enhanced group (i.e., it
does not follow Fe, as sometimes assumed).  Hence, a more accurate
description of elliptical galaxies is that they failed to make Fe-peak
elements rather than that they made an overabundance of
$\alpha$-elements.  The element enhancement pattern of ellipticals
will be considered in more detail in a future paper.

Sources of error in the population parameters are considered.
Contamination of \hbeta\ by hot stars such as horizontal branch stars
and blue stragglers can cause small reductions in the measured ages of
the oldest galaxies but cannot noticeably affect the strong \hbeta\
lines, and thus the deduced low ages, of young ellipticals (as also
found by \cite{Rose85}, 1994 and \cite{Greggio97}).  Emission fill-in
may increase the measured ages of a few, largely old galaxies, but the
broad age distribution is unaffected by whether any emission
corrections are made or not.  Uncertainties in the theoretical tracks,
index response functions, element enhancement patterns, and the
Lick/IDS metallicity scale all affect the absolute zero points of age,
\z, and \enh\ at the level of a few tens of percent or tenths of a
dex---but not the relative age rankings among galaxies.

Finally, we have compared our population parameters to those derived
by TCB98, who apply a different modeling technique to the G93 sample.
Our values of \z\ and \enh\ correlate with theirs, but the slopes
differ significantly from unity.  This appears to stem from the use of
different response functions; in particular, TCB98 do not correct \fe\
for the underabundance of Fe.  When this is allowed for, the two
studies are consistent.

Future papers will discuss the central stellar populations of the G93
sample in detail, correlations between stellar populations and
structural parameters, scaling relation of these local ellipticals in
the context of stellar populations, and stellar population gradients
in elliptical galaxies.

\acknowledgments

We thank Drs.~M. Bolte, A. Bressan, D. Burstein, J. Dalcanton,
G. Illingworth, D. Kelson, I. King, A. McWilliam, A. Renzini, M. Rich,
M. Salaris, and A. Zabludoff for stimulating discussions.  We thank
especially the referee, Dr.~J. Rose, for a careful and thorough
reading of the manuscript which helped improve the final presentation.
We are indebted to Drs.~M. Tripicco and R. Bell for calculating
response functions for the Lick/IDS indices, without which this work
would not have been possible, and to Dr. Tripicco for sending
electronic versions of their tables. We also thank Dr.~Salaris for
sending us his and Dr.~Weiss's solar-metallicity, $\alpha$-enhanced
isochrones in advance of publication.

\appendix

\section{Stellar population parameters using Padova
isochrones}\label{app:padova}

This section provides more details on the ``Padova'' models discussed
in Section~\ref{sec:modeldep}.  The Padova models are identical to the
W94 models except that the isochrones (and opacities) are replaced by
the isochrone library of Bertelli et al.\ (1994).  This isochrone
library is based on the stellar evolutionary tracks developed by the
Padova group (see Bertelli et al.~1994 and Charlot, Worthey, \&
Bressan 1996 for more details) using the Iglesias, Rogers \& Wilson
(1992) radiative opacities.  These isochrones include all phases of
stellar evolution from the ZAMS to the remnant stage for stars of
masses in the age range $0.004\leq t\leq16$ Gyr and metallicity range
$0.0004\leq Z\leq0.1$ ($Z_{\odot}=0.02$).  The models include
convective overshooting in stars more massive than $1\;M_{\odot}$ and
an analytic prescription for the TP-AGB regime.

Figure~\ref{fig:re8m_pad} presents the inferred \hbeta, \mgb, and \fe\
line strengths for the W94 models using the Padova isochrones.
Comparing this with Figure~\ref{fig:re8m} shows small differences: the
Padova models have a higher metallicity at a given \mgb\ or \fe\
strength and a younger age at a given \hbeta\ strength.

\begin{figure*}
\setcounter{figure}{0}
\plotone{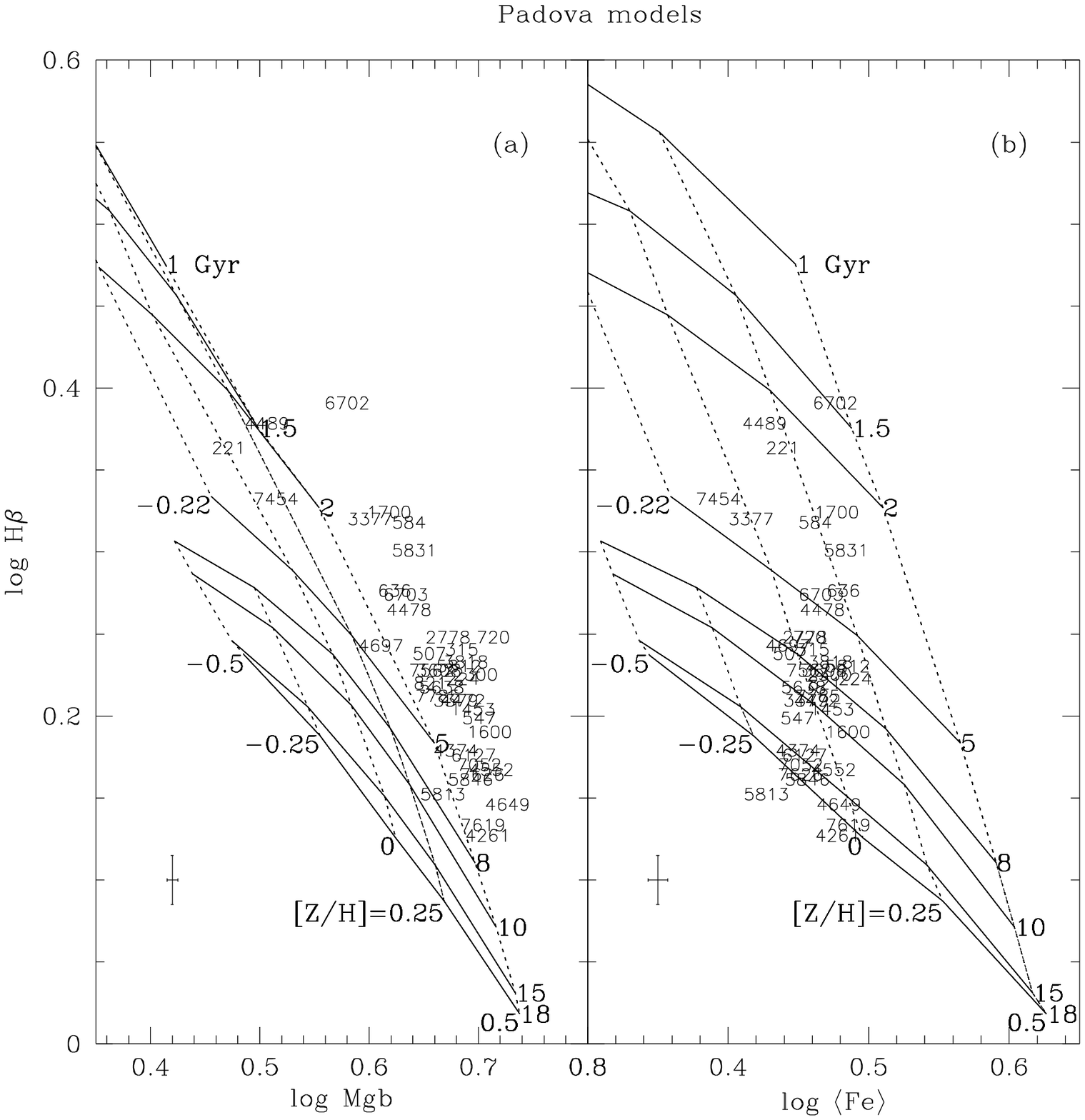}
\caption{The effect of substituting model isochrones from Bertelli et
al.\ (1994) for those of Worthey (1994).  The grid in this figure
should be compared to that in Figure 1.  The basic morphology of the
grid is unchanged.  The data are the central $r_e/8$ values as in
Figure 1.\label{fig:re8m_pad}}
\end{figure*}

Figure~\ref{fig:padmodels} (bottom row) shows the results of applying
these new models to the G93 central (\reo{8}) line strengths.  The
derived ages, metallicities, and enhancement ratios agree quite well
with the W94 models, apart from slight slope changes and zeropoint
offsets:
\begin{eqnarray}
\logt_{\rm Padova} &=& 1.02\; \logt_{\rm W94} - 0.10, \\
\z_{\rm Padova} &=& 1.09\; \z_{\rm W94} + 0.03, \\
\enh_{\rm Padova} &=& 0.97\; \enh_{\rm W94} + 0.02.
\end{eqnarray}
The above are linear least-square fits using enhancement model 4 and
rejecting 3-$\sigma$ outliers.  The fit for \logt\ is in accordance
with the results of Charlot, Worthey \& Bressan (1996): changing
isochrones can alter the inferred ages from line strengths at young
ages by as much as $\sim25\%$; agreement at old ages is within 10\%.
On average, the inferred metallicities \z\ are increased by
$\approx10\%$ in the Padova models, as expected at fixed line
strengths from the 3/2 rule ($\Delta\logt/\Delta\z\approx1.4$ between
the two sets of models).

\section{Corrections to \hbeta}
\label{app:emsigcorr}

\subsection{Emission corrections}

This section presents stellar population parameters derived by
omitting the emission fill-in correction to \hbeta\ discussed in
Section~\ref{sec:emission}.  This follows the suggestion by Carrasco
et al.~(1996) that no correction to \hbeta\ should be made for
residual \hbeta\ emission based on [\ion{O}{3}], as they find no such
correlation in their own sample of early-type galaxies.

\begin{figure*}
\setcounter{figure}{0}
\plotone{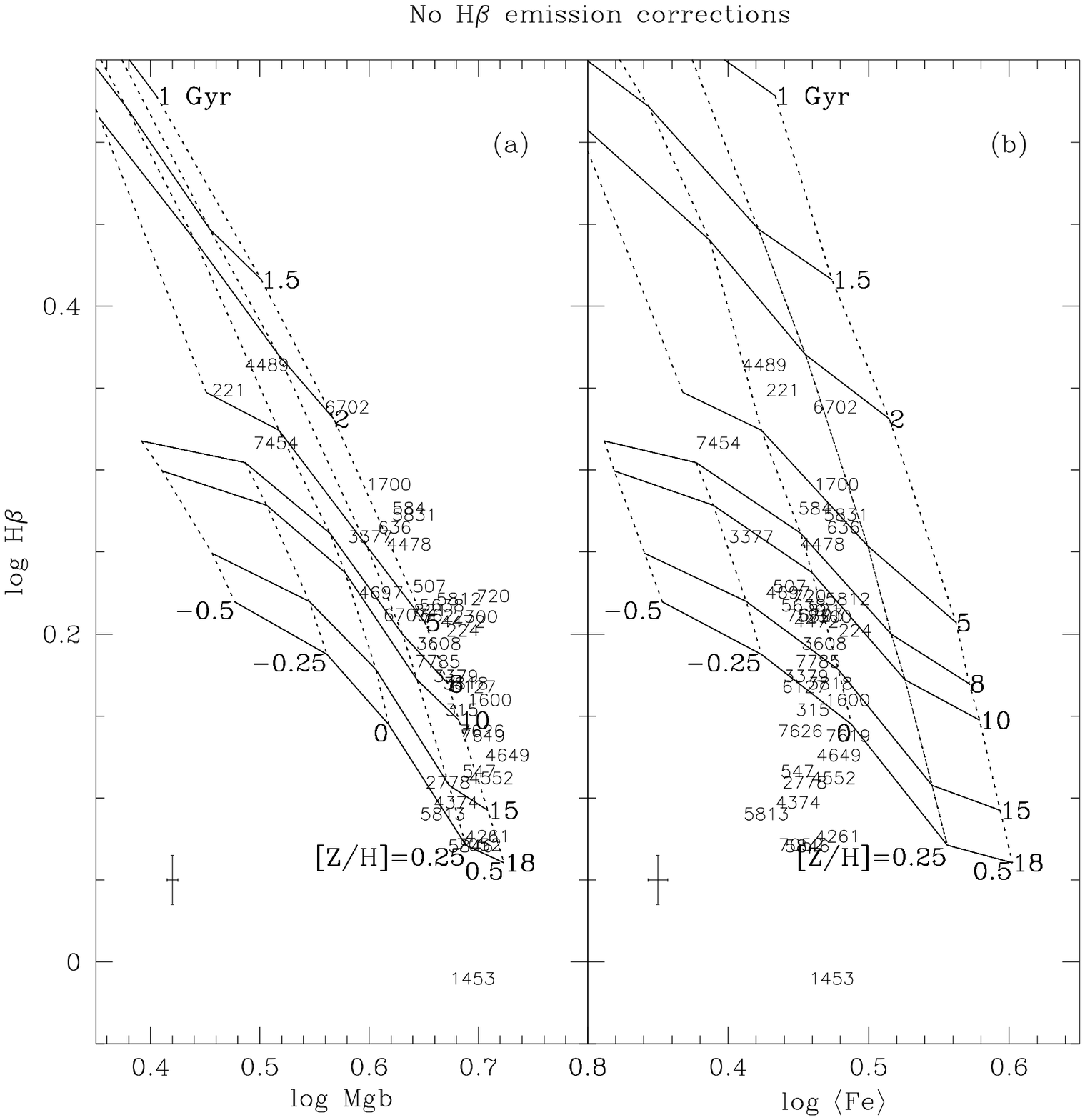}
\caption{Line strengths of early-type galaxies in the Gonz{\'a}lez
(1993) sample in the central \reo{8}\ aperture, but now with \hbeta\
emission corrections omitted (Section~\ref{sec:emission}).  W94 model
grids are overlaid as in Figure~\ref{fig:re8m}.  Galaxies on average
lie a little lower here than in Figure 1 but their relative parameters
(including ages) are little affected, showing that SSP parameters are
rather insensitive to the exact \hbeta\ correction
used.\label{fig:re8m_noem}}
\end{figure*}

\begin{figure*}
\plotone{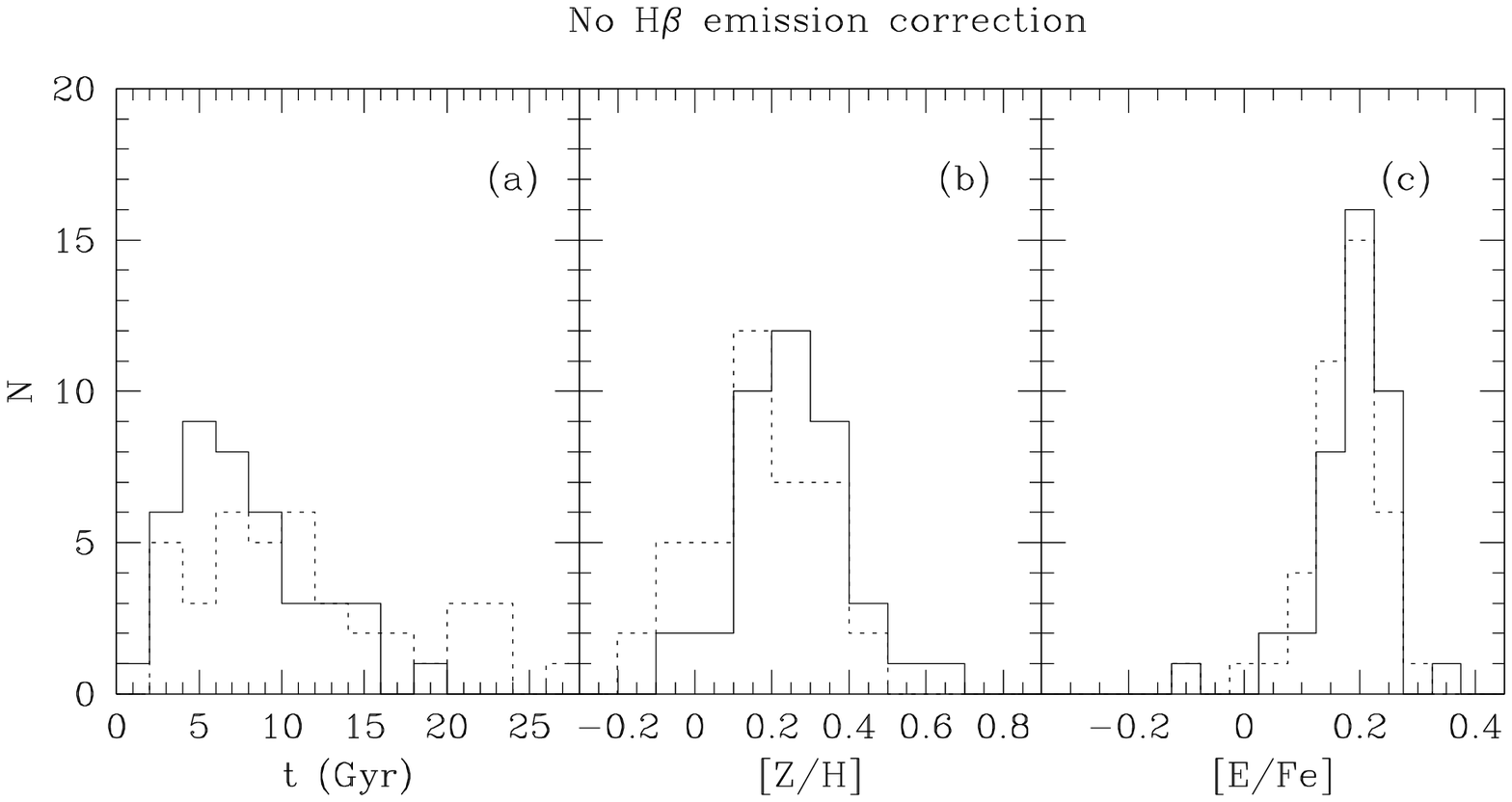}
\caption{The effect of \hbeta\ emission corrections on the SSP
parameter distributions of the Gonz{\'a}lez (1993) sample using
enhancement model 4 (central \reo{8}\ aperture).  Solid histograms use
the \hbeta\ emission fill-in correction (Section~\ref{sec:emission};
see Figure~\ref{fig:tza_hist_re8}); dotted histograms do not include
this correction.  Without the emission correction, galaxies are
typically slightly more metal-poor and older (NGC 1453 is $\sim26$ Gyr
old in this model), but their relative ranking in the various
parameters is little affected.
\label{fig:tza_hist_noem}}
\end{figure*}

Figure~\ref{fig:re8m_noem} presents the Balmer--metal--line diagrams
for the G93 galaxies through the $r_e/8$ aperture with the \hbeta\
correction omitted.  As expected from the additive nature of the
correction, galaxies now appear lower in the grid, and therefore older
and more metal-poor than in Figure~\ref{fig:re8m}.  Neglecting
emission corrections forces some galaxies to have unreasonably old
ages: for example, without corrections, NGC 1453, NGC 2778, NGC 4261,
NGC 4374, NGC 5813, NGC 5846, NGC 7052 have ages $\ga$ 20 Gyr.  Since
all of these galaxies have clear \hbeta\ emission (see Figs.~3.11 and
4.10 of G93), omitting the corrections makes no sense.  Furthermore,
careful checking reveals that a few galaxies (e.g., NGC 4552, NGC
4649, NGC 5813, NGC 5846, NGC 7052) actually appear to have \hbeta\ a
little \emph{stronger} than the standard ratio and are therefore
probably undercorrected in our standard treatment.  Fixing them would
move them up by a few hundredths in log \hbeta\ and decrease their
ages by $\la$20\%.  Since some of these are also objects that lie low
in the grid, this correction would improve their positions.

Figure~\ref{fig:re8m_noem} without \hbeta\ corrections looks
essentially like the original one---the large age spread and relative
parameter rankings of the galaxies are essentially the same.  This
point is reinforced in the histograms of
Figure~\ref{fig:tza_hist_noem}, which are nearly identical to those in
Figure~\ref{fig:tza_hist_re8}.  We conclude that emission corrections
are needed to derive the best age estimates for early-type galaxies,
but that their exact magnitude does not affect our broad conclusions.

\subsection{Velocity dispersion corrections}

\begin{figure*}
\plotone{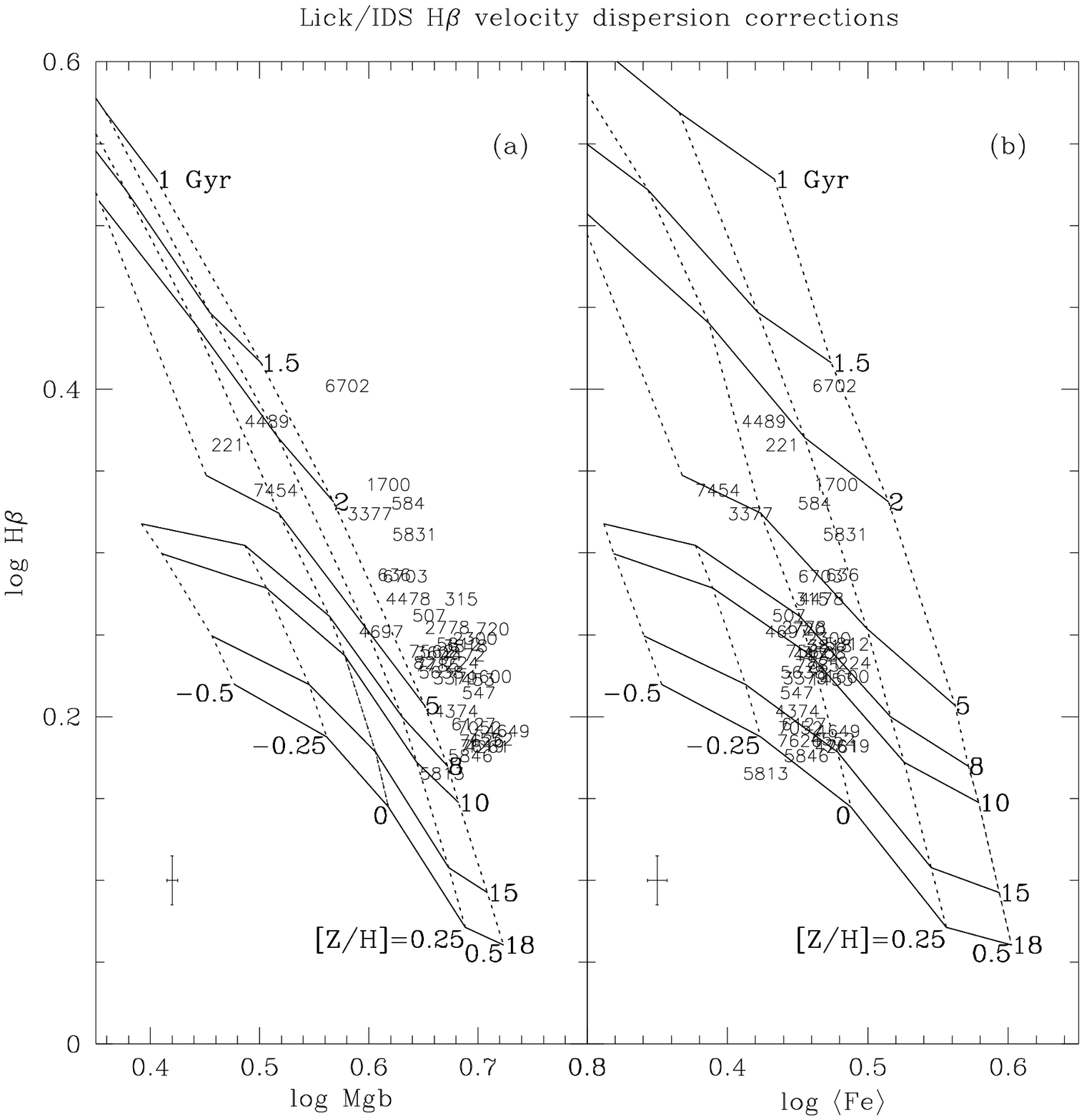}
\caption{Line strengths of early-type galaxies in the Gonz{\'a}lez
(1993) sample in the central \reo{8}\ aperture but now using the
Lick/IDS velocity dispersion corrections for
\hbeta~(Section~\ref{sec:sigmacorr}) rather than those of G93.  The
oldest galaxies lie slightly higher here than in
Figures~\ref{fig:re8m} and \ref{fig:re8c}, and their SSP-equivalent
ages are reduced by about 25\% (see Figure 16).  Overly old ages of
low-lying galaxies through the $r_e/2$ aperture are similarly reduced
(cf. Figure~\ref{fig:re2m}).
\label{fig:re8m_isc}}
\end{figure*}

\begin{figure*}
\plotone{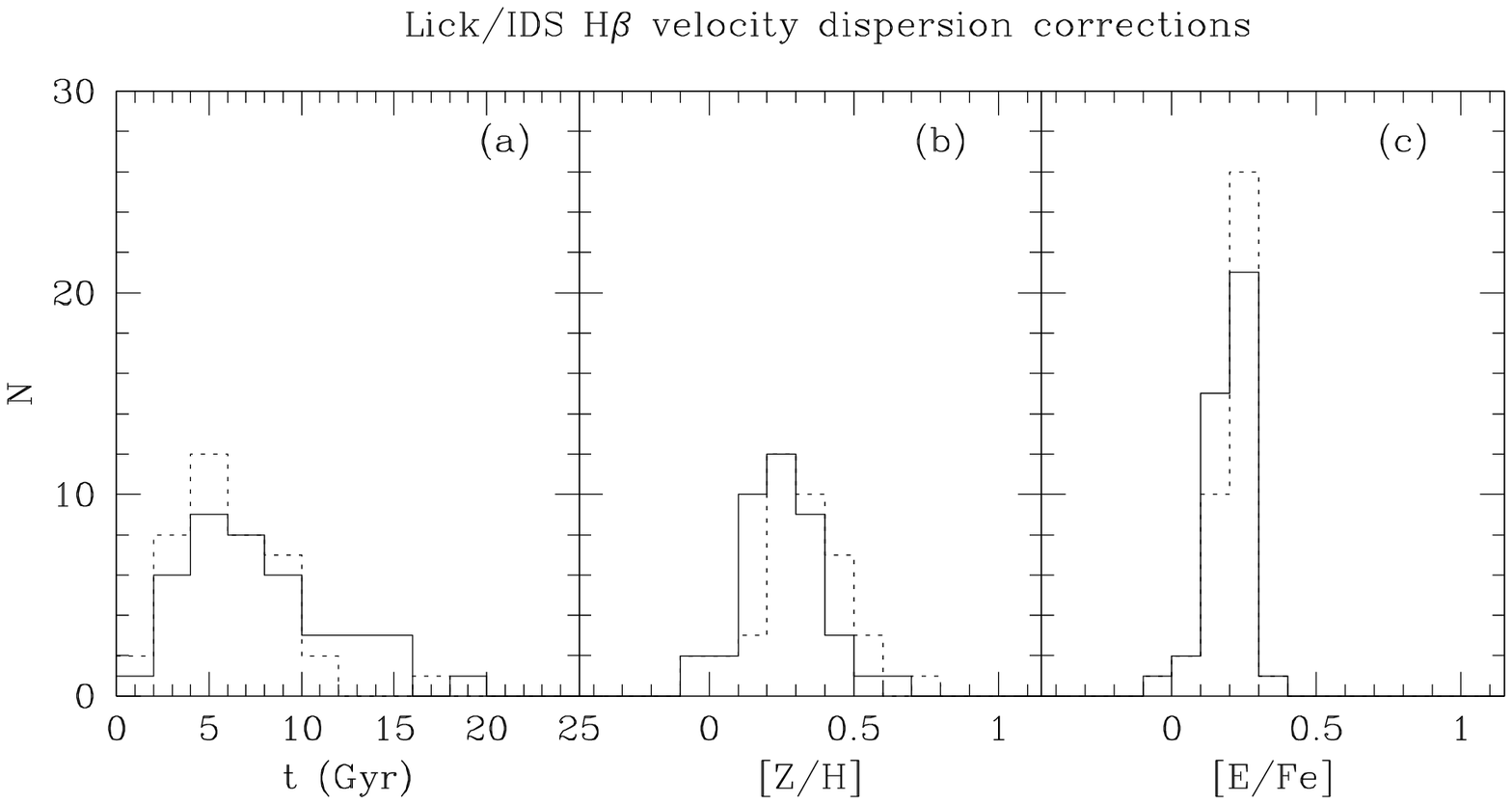}
\caption{The effect of \hbeta\ velocity corrections on the SSP
parameter distributions of the Gonz{\'a}lez (1993) sample using
enhancement model 4 (central \reo{8}\ aperture).  Solid histograms use
the G93 \hbeta\ velocity dispersion correction
(Section~\ref{sec:sigmacorr}; see Figure~\ref{fig:tza_hist_re8});
dotted histograms use the Lick/IDS velocity dispersion correction.
With the Lick/IDS correction, galaxies are younger, slightly more
metal-rich, and slightly more enhanced in $\alpha$-elements, but
retain their basic relative parameter
rankings.\label{fig:tza_hist_isc}}
\end{figure*}

This section derives a third set of population parameters using the
Lick/IDS velocity dispersion corrections of TWFBG98 for \hbeta\ rather
than the template-based corrections of G93.  As G93 does not provide
raw \hbeta\ strengths, we use his Figure 4.1 to estimate his velocity
corrections and use them to ``uncorrect'' the fully corrected line
strengths back to raw strengths (after removing the emission
correction discussed in Section~\ref{sec:emission}).  For most
galaxies, G93's velocity corrections are insignificant, but for
high-$\sigma$ galaxies they tend to be \emph{negative}.  After the G93
corrections are removed, we apply the \emph{positive} corrections
presented in TWFBG98 (their Figure~3) and then reapply the emission
corrections discussed in Section~\ref{sec:emission}.  The newly
corrected line strengths are plotted in Figure~\ref{fig:re8m_isc}, and
the resulting stellar population parameter histograms are presented in
Figure~\ref{fig:tza_hist_isc}.

The new corrections move only a few high-$\sigma$ galaxies, and
therefore the \emph{relative} age ranking of the sample is unaffected.
The affected galaxies again tend to lie at the bottom of the grid and
are again moved \emph{up} by the new corrections (by $\la$0.04 in log
\hbeta) so that galaxies that formerly lay below the grid at high ages
now tend to lie on it.  This correction, like the refined emission
corrections of the previous section, thus improves the ages of the
oldest objects.

The G93 \hbeta\ velocity corrections were based on a very high-S/N
stellar template fit to each galaxy, whereas the TWFBG98 corrections
are based on a statistical average over stellar spectral types whose
correction curves scattered widely.  Nevertheless, it is possible that
the TWFBG98 corrections are actually more acurate.  The G93 stellar
templates are a superb match to most of the spectrum \emph{except} at
\hbeta, where emission corrupted the data.  Hence the template match
(and correction) at \hbeta\ in particular may be poor.  The TWFBG98
corrections were selected to match a large number of stars with about
the same \hbeta\ strength as typical galaxies and could therefore be
better on average.

\end{document}